%% file: main.tex
 \documentclass[preprint,12pt]{clear2023} 

\title[Sparse Cox Subgroups]{Recovering Sparse and Interpretable Subgroups with Heterogeneous Treatment Effects with Censored Time-to-Event Outcomes}
\usepackage{times}
\usepackage[utf8]{inputenc}
\usepackage{graphicx}
\usepackage{enumerate}
\usepackage{natbib}
\usepackage{url} 
\usepackage{hyperref}
\hypersetup{colorlinks=true,
    linkcolor=mydarkblue,
    citecolor=citecol,
    filecolor=mydarkblue,
    urlcolor=mydarkblue}

\usepackage{listings}
\usepackage{inconsolata}
\usepackage{xcolor}
\usepackage{enumitem}
\usepackage{oubraces}
\usepackage{booktabs}
\usepackage{nicefrac}
\usepackage{cancel}

\usepackage{amsmath}
\usepackage{bbm}

\definecolor{red}{rgb}{1.0,0.0,0.0}
\definecolor{mydarkblue}{rgb}{0,0.08,0.45}
\definecolor{citecol}{rgb}{0.45,0.0,0.0}
\definecolor{filcol}{rgb}{0.45,0.,0.45}
\definecolor{string}{RGB}{200, 170, 0}
\definecolor{comment}{RGB}{117, 113, 94}
\definecolor{normal}{RGB}{0, 0, 0}
\definecolor{identifier}{RGB}{166, 226, 46}
\definecolor{number}{RGB}{166, 0, 166}

\newtheorem{asmp}{Assumption}
\newtheorem{result}{Result}

\include{math_commands}

\usepackage[format=plain]{caption}
\clearauthor{\Name{Chirag Nagpal} \Email{chiragn@cs.cmu.edu}\\
 \Name{Vedant Sanil} \Email{vsanil@andrew.cmu.edu}\\
 \Name{Artur Dubrawski} \Email{awd@cs.cmu.edu}\\
 \addr Auton Lab, Carnegie Mellon}

\begin{document}

\maketitle

\begin{abstract}%
    Studies involving both randomized experiments as well as observational data typically involve time-to-event outcomes such as time-to-failure, death or onset of an adverse condition. Such outcomes are typically subject to censoring due to loss of follow-up and established statistical practice involves comparing treatment efficacy in terms of hazard ratios between the treated and control groups. In this paper we propose a statistical approach to recovering sparse phenogroups (or subtypes) that demonstrate differential treatment effects as compared to the study population. Our approach involves modelling the data as a mixture while enforcing parameter shrinkage through structured sparsity regularization. We propose a novel inference procedure for the proposed model and demonstrate its efficacy in recovering sparse phenotypes across large landmark real world clinical studies in cardiovascular health.
\end{abstract}


\section{Introduction}

Data driven decision making across multiple disciplines including healthcare, epidemiology, econometrics and prognostics often involves establishing efficacy of an intervention when outcomes are measured in terms of the time to an adverse event, such as death, failure or onset of a critical condition. Typically the analysis of such studies involves assigning a patient population to two or more different treatment arms often called the `treated' (or `exposed') group and the `control' (or `placebo') group and observing whether the populations experience an adverse event (for instance death or onset of a disease) over the study period at a rate that is higher (or lower) than for the control group. Efficacy of a treatment is thus established by comparing the relative difference in the rate of event incidence between the two arms called the hazard ratio. However, not all individuals benefit equally from an intervention. Thus, very often potentially beneficial interventions are discarded even though there might exist individuals who benefit, as the population level estimates of treatment efficacy are inconclusive.  

In this paper we assume that patient responses to an intervention are typically heterogeneous and there exists patient subgroups that are unaffected by (or worse, \textbf{harmed}) by the intervention being assessed. The ability to discover or phenotype these patients is thus clinically useful as it would allow for more precise clinical decision making by identifying individuals that actually benefit from the intervention being assessed.

Towards this end, we propose \textbf{Sparse Cox Subgrouping}, (SCS) a latent variable approach to model patient subgroups that demonstrate heterogeneous effects to an intervention. As opposed to existing literature in modelling heterogeneous treatment effects with censored time-to-event outcomes our approach involves structured regularization of the covariates that assign individuals to subgroups leading to parsimonious models resulting in phenogroups that are interpretable. We release a \texttt{python} implementation of the proposed SCS approach as part of the \texttt{auton-survival} package \citep{nagpal2022auton} for survival analysis: 

\centerline{\url{https://autonlab.github.io/auton-survival/}}

\section{Related Work}

Large studies especially in clinical medicine and epidemiology involve outcomes that are time-to-events such as death, or an adverse clinical condition like stroke or cancer.  Treatment efficacy is typically estimated by comparing event rates between the treated and control arms using the Proportional Hazards \citep{cox1972regression} model and its extensions.

Identification of subgroups in such scenarios has been the subject of a large amount of traditional statistical literature. Large number of such approaches involve estimation of the factual and counterfactual outcomes using separate regression models (T-learner) followed by regressing the difference between these estimated potential outcomes. Within this category of approaches, \cite{subgrouplipkovich} propose the subgroup identification based on differential effect search (SIDES) algorithm, \cite{subgroupsu} propose a recursive partitioning method for subgroup discovery, \cite{subgroupelise} propose the qualitative interaction trees (QUINT) algorithm, and \cite{foster2011subgroup} propose the virtual twins (VT) method for subgroup discovery involving decision tree ensembles. We include a parametric version of such an approach as a competing baseline.

Identification of heterogeneous treatment effects (HTE) is also of growing interest to the machine learning community with multiple approaches involving deep neural networks with balanced representations \citep{tesontag, johansson2020generalization}, generative models \cite{louizos2017causal} as well as Non-Parametric methods involving random-forests \citep{wager2018estimation} and Gaussian Processes \citep{alaa2017bayesian}. There is a growing interest in estimating HTEs from an interpretable and trustworthy standpoint \citep{lee2020causal, nagpal2020interpretable, morucci2020adaptive, wu2022interpretable, crabbe2022benchmarking}. \cite{wang2022causal} propose a sampling based approach to discovering interpretable rule sets demonstrating HTEs.

However large part of this work has focused extensively on outcomes that are binary or continuous. The estimation of HTEs in the presence of censored time-to-events has been limited. \cite{xu2022treatment} explore the problem and describe standard approaches to estimate treatment effect heterogeneity with survival outcomes. They also describe challenges associated with existing risk models when assessing treatment effect heterogeneity in the case of cardiovascular health.

There has been some initial attempt to use neural network for causal inference with censored time-to-event outcomes.  \cite{curth2021survite} propose a discrete time method along with regularization to match the treated and control representations. \cite{chapfuwa2021enabling}'s approach is related and involves the use of normalizing flows to estimate the potential time-to-event distributions under treatment and control. While our contributions are similar to \cite{nagpal2022counterfactual}, in that we assume treatment effect heterogeneity through a latent variable model, our contribution differs in that 1) Our approach is free of the expensive Monte-Carlo sampling procedure and 2) Our generalized EM inference procedure allows us to naturally incorporate structured sparsity regularization, which helps recovers phenogroups that are parsimonious in the features they recover that define subgroups. 

Survival and time-to-event outcomes occur pre-eminently in areas of cardiovascular health. One such area is reducing combined risk of adverse outcomes from atherosclerotic disease\footnote{A class of related clinical conditions from increasing deposits of plaque in the arteries, leading to Stroke, Myorcardial Infarction and other Coronary Heart Diseases.} \citep{herrington2016epidemiology, furberg2002major, bari2009randomized, buse2007action} The ability of recovering groups with differential benefits to interventions can thus lead to improved patient care through framing of optimal clinical guidelines.

\section{Proposed Model: Sparse Cox Subgrouping}

\begin{figure*}[!htbp]
    \begin{minipage}{0.5\textwidth}
    \centering
    \textbf{Case 1: $Z \in \{ 0, +1\}$}
    \includegraphics[width=\textwidth]{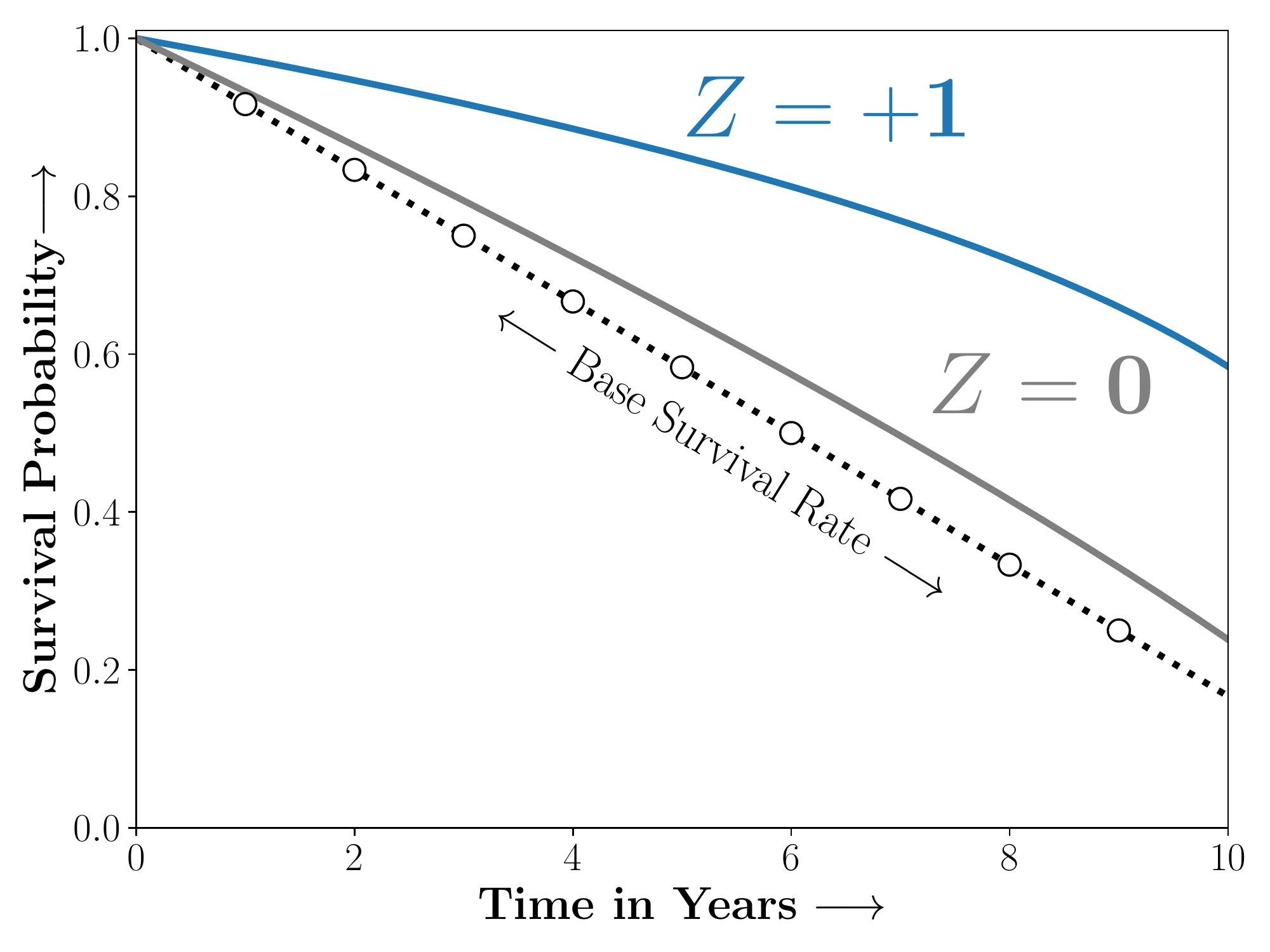}
    \end{minipage}\hfill
    \begin{minipage}{0.5\textwidth}
    \centering
    \textbf{Case 2: $Z \in \{ 0, +1, -1\}$ }
    \includegraphics[width=\textwidth]{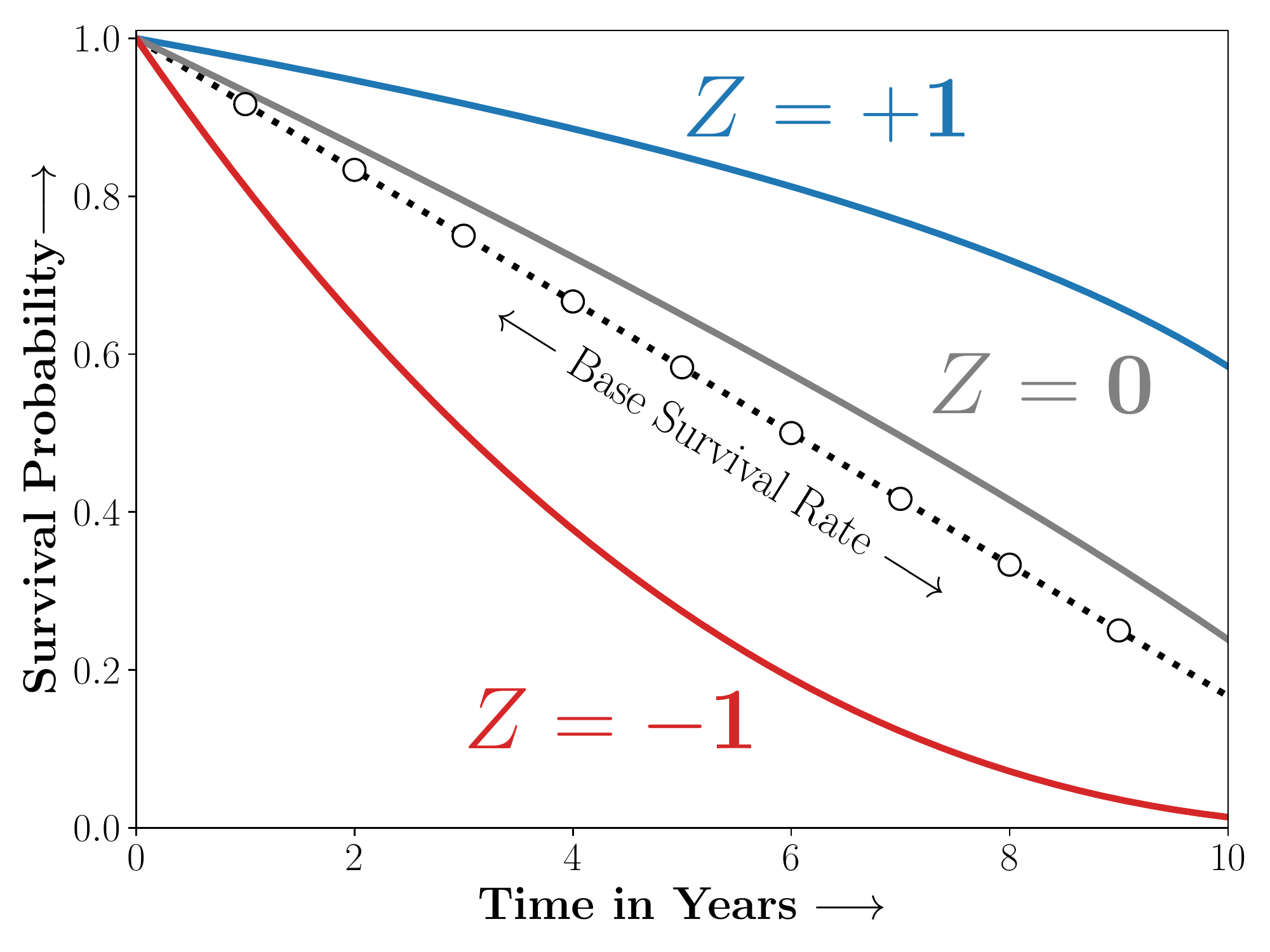}
    \end{minipage}
    \captionof{figure}{Potential outcome distributions under the assumptions of treatment 
    effect heterogeneity. \textbf{Case 1}: Amongst the treated population, conditioned on the latent $Z$, there are two subgroups that \textbf{benefit} and are \textbf{unaffected} by the intervention.  \textbf{Case 2}: There is an additional latent subgroup conditioned on which, the treated population is \textbf{harmed} with a worse survival rate. }
    \label{fig:identifiability}
\end{figure*}

\begin{description}[leftmargin=*]

\item[Notation] As is standard in survival analysis, we assume that we either observe the true time-to-event or the time of censoring $U = \mathrm{min}\{ T, C\}$ indicated by the censoring indicator defined as $\Delta = \bm{1} \{ T< C\}$. We thus work with a dataset of right censored observations in the form of 4-tuples, $\mathcal{D} = \{ (\vx_i, \delta_i, \vu_i, \va_i ) \}_{i=1}^{n}$, where $\vu_i \in \mathbb{R}^+$ is the time-to-event or censoring as indicated by $\delta_i\in\{0, 1\}$, $\va_i\in \{0, 1 \}$ is the indicator of treatment assignment, and $\vx_i$ are individual covariates that confound the treatment assignment and the outcome. 

\begin{asmp}[Independent Censoring]
The time-to-event $T$ and the censoring distribution $C$ are independent conditional on the covariates $X$ and the intervention $A$.
\label{asmp:censoring}
\end{asmp}
\item[Model] Consider a maximum likelihood approach to model the data $\mathcal{D}$ the set of parameters $\bm{\Omega}$. Under Assumption \ref{asmp:censoring} the likelihood of the data $\mathcal{D}$ can be given as,
\begin{align}
    \mathcal{L}( \bm{\Omega} ;\mathcal{D}) &\propto  \prod_{i=1}^{|\mathcal{D}|} \bm{\lambda}(u_i|X=\vx_i,A=\va_i)^{\delta_i} \bm{S}(u_i|X=\vx_i, A=\va_i),
    \label{eqn:likelihood}
\end{align}
here $\bm{\lambda}(t) = \mathop{\lim}\limits_{\Delta t\to 0}\frac{\mathbb{P}(t\leq T<t+\Delta t| T \geq t)}{\Delta t}$ is the hazard and $\bm{S}(t) = \mathbb{P}(T>t)$ is the survival rate. 

\begin{asmp}[PH]
The distribution of the  time-to-event $T$ conditional on the covariates and the treatment assignment obeys proportional hazards.
\label{asmp:ph}
\end{asmp}

From Assumption \ref{asmp:ph} (Proportional Hazards), an individual with covariates $(X=\vx)$ under intervention $(A=\va)$ under a Cox model with parameters $\beta$ and treatment effect $\omega$ is given as 
\begin{align}
    \bm\lambda(\vt|A=\va, X=\vx) = \bm{\lambda}_0(t) \exp \big( \bm{\beta}^{\top}\vx + \bm{ \omega} \cdot \bm{a} \big),
\end{align}
Here, $\bm{\lambda}_0(\cdot)$ is an infinite dimensional parameter known as the base survival rate. In practice in the Cox's model the base survival rate is a nuisance parameter and is estimated non-parametrically. In order to model the heterogeneity of treatment response. We will now introduce a latent variable $Z \in \{0, 1, -1 \}$ that mediates treatment response to the model,
\begin{align}
\nonumber \bm\lambda(\vt|A=\va, X=\vx, Z=\vk) &= \bm{\lambda}_0(t) \exp(\beta^\top \vx ) \exp (\bm{\omega} )^{\bm{ka}},\\
\text{and, }\; \; \mathbb{P}(Z &= \vk | X=\vx) = \frac{\exp(\bm{\theta}_k^{\top} \vx)}{\sum_j \exp(\bm{\theta}_j^{\top} \vx)}.
\label{eqn:model}
\end{align}

Here, $\bm{\omega} \in \mathbb{R}$ is the treatment effect, and $\bm{\theta} \in \mathbb{R}^{k \times d}$ is the set of parameters that mediate assignment to the latent group $Z$ conditioned on the confounding features $\vx$. Note that the above choice of parameterization naturally enforces the requirements from the model as in Figure \ref{fig:identifiability}. Consider the following scenarios,

\noindent \textbf{Case 1}: The study population consists of two sub-strata ie. $Z\in \{0, +1\}$, that are benefit and are unaffected by treatment respectively. 

\noindent \textbf{Case 2}: The study population consists of three sub-strata ie. $Z\in \{0, +1, -1\}$, that benefit, are harmed or unaffected by treatment respectively.

Following from Equations \ref{eqn:likelihood} \& \ref{eqn:model}, the complete likelihood of the data $\mathcal{D}$ under this model is,
\begin{align}
\nonumber \mathcal{L}(\bm{\Omega}; \mathcal{D})  = \prod_{i=1}^{|\mathcal{D}|} \sum_{k\in Z}  \bigg( \bm\lambda_0(u_i) \bm{h}(\vx, \va, \vk ) \bigg)^{\delta_i} &\mS_0(u_i) ^{ \bm{h}(\vx, \va, \vk ) } \mathbb{P}(Z=k|X=\vx_i) \\
\text{ where, } \ln  \bm{h}(\vx, \va, \vk) = \bm{\beta}^{\top} \vx + \vk \cdot \bm{a} \cdot \bm{w} \text{ and } \ln  &\mS_0(\cdot) = -\bm{\Lambda}_0(\cdot),
\label{eqn:completelikelihood}
\end{align}
Note that $\bm{\Lambda}_0(\cdot) = \int_{0}^{t} \bm{\lambda}_0(\cdot) $ is the infinite dimensional cumulative hazard and is inferred when learning the model. We will notate the set of all learnable parameters as $\bm{\Omega} = \{  \bm{\theta},  \bm{\beta}, \bm{w},  \bm{\Lambda}_0 \} $.



\item[Shrinkage] In retrospective analysis to recover treatment effect heterogeneity a natural requirement is parsimony of the recovered subgroups in terms of the covariates to promote model interpretability. Such parsimony can be naturally enforced through appropriate shrinkage on the coefficients that promote sparsity. We want to recover phenogroups that are `sparse' in $\bm{\theta}$. We enforce sparsity in the parameters of the latent $Z$ gating function via a group $\ell_1$ (Lasso) penalty. The final loss function to be optimized including the group sparsity regularization term is, 
\begin{align}
   \nonumber \mathcal{L}(\bm{\Omega}; \mathcal{D}) + & \bm{\epsilon} \cdot \mathcal{R}(\bm{\theta})    \text{ where, } \mathcal{R}(\bm{\theta}) = \sum_d \sqrt{\sum_{k\in \mathcal{Z}} \big( \bm{\theta}^{k}_d \big)^{2}} \\
    \text{ and } & \bm{\eps}>0 \text{ is the strength of the shrinkage parameter}. 
\end{align}
\item[Identifiability] Further, to ensure identifiability we restrict the gating parameters for the $(Z=0)$ to be $0$. Thus  $\bm{\theta}_1= 0$.


\item[Inference] We will present a variation of the \textbf{Expectation Maximization} algorithm to infer the parameters in \hyperref[eqn:model]{Equation \ref*{eqn:model}}. Our approach differs from \cite{nagpal2022counterfactual, nagpal2021deep} in that it does not require storchastic Monte-Carlo sampling. Further, our generalized EM inference allows for incorporation of the structured sparsity in the \textbf{M-Step}.



\item[A Semi-Parametric $Q(\cdot)$] Note that the likelihood in Equation \ref{eqn:model} is semi-parametric and consists of parametric components and the infinite dimensional base hazard $\bm{\Lambda}(\cdot)$. We define the $Q(\cdot)$ as:
\begin{align}
\nonumber Q(\bm{\Omega}; \mathcal{D}) &= \sum_{i=1}^{n} \sum_{k\in \mathcal{Z}} \bm{\gamma}^{k}_i \bigg(  \ln  \bm{p}_{\bm{\theta}}(Z=k|X=\vx_i)  + \ln  \bm{p}_{\bm{w}, \bm{\beta}, \bm{\Lambda}}(T| Z=k, X=\vx_i) \bigg) + \mathcal{R} (\bm{\theta}) 
\end{align}

\item[The E-Step] Requires computation of the posteriors counts $\bm{\gamma}:= \bm{p}(Z={k} |T, X=\bm{x}, A=\va).$

\begin{result}[Posterior Counts] The posterior counts $\bm{\gamma}$ for the latent $Z$ are estimated as,
\begin{align}
 \nonumber \bm{\gamma}^{k} &= \widehat{\mathbb{P}}(Z=k|X=\vx, A=\va, \vu) \\
 \nonumber &= \frac{\mathbb{P}(\vu|Z=\vk, X=\vx, A=\va)\mathbb{P}(Z=\vk|X=\vx)}{\sum_k \mathbb{P}(\vu|Z=\vk, X=\vx, A=\va)\mathbb{P}(Z=\vk|X=\vx)}\\
 &= \frac{  {\vh(\vx, \va, \vk)}^{\delta_i} \widehat{\mS}_0( \vu )^{\vh(\vx, \va, \vk)} \exp(\bm{\theta_k}^\top \bm{x})}{\mathop{\sum}_{j\in \mathcal{Z}} {\vh(\vx, \va, \vj)}^{\delta_i}\widehat{\mS}_0( \vu )^{\vh(\vx, \va, \vj)} \exp(\bm{\theta_j}^\top \bm{x})} \label{eqn:posterior}.
\end{align}
\label{result:censoring}
\end{result}
For a full discussion on derivation of the $Q(\cdot)$ and the posterior counts please refer to Appendix \ref{apx:posterior}

\item[The M-Step]

%
Involves maximizing the $Q(\cdot)$ function. Rewriting the $Q(\cdot)$ as a sum of two terms,
\begin{align}
\nonumber  Q(\bm{\Omega} ) 
= \nonumber  \underbrace{\sum_{i=1}^{n} \sum_{k\in \mathcal{Z}}   \bm{\gamma}^{k}_i \ln  \bm{p}_{\bm{w}, \bm{\beta}, \bm{\Lambda}_0}(T| Z=k, X=\vx_i, A=\va_i)}_{\mA(\bm{w}, \bm{\beta}, \bm{\Lambda}_0)}  + \underbrace{ \sum_{i=1}^{n} \sum_{k\in \mathcal{Z}} \bm{\gamma}^{k}_i \ln  \bm{p}_{\bm{\theta}}(Z=k|X=\vx_i) + \mathcal{R} (\bm{\theta}) }_{\mB(\bm{\theta})}
\end{align}

\begin{result}[Weighted Cox model]  The term $\mA$ can be rewritten as a weighted Cox model and thus optimized using the corresponding `\textit{partial likelihood}',
\end{result}


\noindent \textbf{Updates for $\{ \bm{\beta}, \bm{\omega} \}$}: The partial-likelihood, $\mathcal{PL(\cdot)}$ under sampling weights \citep{binder1992fitting} is
\begin{align}
\mathcal{PL}(\bm{\Omega}; \mathcal{D}) = 
\sum_{\substack{i=1,\delta_i=1}}^{n} \sum_{k\in \mathcal{Z}}\bm{\gamma}^{k}_{i}  \bigg(  \ln \bm{h}_k(\vx_i,\bm{a}_i, \vk)  - \ln &\sum_{j\in{\mathsf{RiskSet}(u_i)}} \sum_{k\in \mathcal{Z}} \bm{\gamma}^{k}_j \bm{h}_k(\vx_j,\bm{a}_j, \vk)  \bigg) \bigg]
\label{eqn:pll}
\end{align}
Here $\mathsf{RiskSet}(\cdot)$ is the \textit{`risk set'} or the set of all individuals who haven't experienced the event till the corresponding time, i.e. $\mathsf{RiskSet}(t) := \{ i: u_i > t\}$. $\mathcal{PL(\cdot)}$ is convex in  $\{ \bm{\beta}, \bm{\omega} \}$ and we update these with a gradient step.

\noindent \textbf{Updates for $\bm{\Lambda}_0$}: The base hazard $\bm{\Lambda}_0$ are updated using a weighted Breslow's estimate \citep{breslow1972contribution, lin2007breslow} assuming the posterior counts $\bm{\gamma}$ to be sampling weights \citep{chen2009weighted},
\begin{align}
\widehat{\bm{\Lambda}}_0(t)^{+} = \sum_{i=1}^{n} \sum_{k\in\mathcal{Z}} \bm{1}\{ u_i < t\}  \frac{\bm{\gamma}^{k}_i \cdot \delta_i }{ \mathop{\sum}\limits_{j\in \mathsf{RiskSet}(u_i)} \mathop{\sum}\limits_{k\in \mathcal{Z}} \bm{\gamma}^{k}_j  \bm {h}_k(\vx_j,\bm{a}_j, \vk)}
\label{eqn:breslow}
\end{align}

\noindent Term $\mB$ is a function of the gating parameters $\bm{\theta}$ that determine the latent assignment $Z$ along with sparsity regularization. We update $\mB$ using a Proximal Gradient update as is the case with Iterative Soft Thresholding (ISTA) for group sparse $\ell_1$ regression.

\noindent \textbf{Updates for $\bm{\theta}$}: The proximal update for $\bm{\theta}$ including the group regularization \citep{friedman2010note}  term $\mathcal{R}(\cdot)$ is,
\begin{align}
   \widehat{\bm{\theta}}^{+} =  \mathsf{prox}_{\eta\eps} \bigg( \bm{\theta} - \frac{d}{d\bm{\theta}} \mB (\bm{\theta}) \bigg), \; \; \; \text{ where } \mathsf{prox}_{\eta\eps}(\vy) = \frac{\vy}{||\vy||_2}\mathrm{max}\{ 0, ||\bm{y}||_2 - \eta\bm{\epsilon}\} .
\end{align}



\end{description}

\noindent All together the inference procedure is described in Algorithm \ref{alg:learning}. 

\begin{algorithm}[!h]
\caption{\textbf{ Parameter Learning for SCS with a Generalized EM}} 
\SetAlgoLined
  \SetKwInOut{Input}{Input}
  \SetKwInOut{return}{Return}
    \Input{Training set, $\mathcal{D} = \{ (\vx_{i}, u_i, a_i, \delta_i)_{i=1}^{n}  \}$; maximum EM iterations, $B$, step size $\eta$} \hrulefill\\

\While{\texttt{<not converged>}}{
  \For{$b \in \{1, 2, ...,  B \} $ }{
    \textbf{\textsc{{E-Step}}} \dotfill
  
    \vspace{1em}  
    $\bm{\gamma}_i^{k} =  \frac{  {\vh(\vx, \va, \vk)}^{\delta_i} \widehat{\mS}_0( \vu )^{\vh(\vx, \va, \vk)} \exp(\bm{\theta_k}^\top \bm{x})}{\mathop{\sum}_{j\in \mathcal{Z}} {\vh(\vx, \va, \vj)}^{\delta_i}\widehat{\mS}_0( \vu )^{\vh(\vx, \va, \vj)} \exp(\bm{\theta_j}^\top \bm{x})} $ $\hfill \triangleright$ Compute posterior counts (Equation \ref{eqn:posterior}).
    \vspace{1em}  
    
    \textbf{\textsc{M-Step}} \dotfill
    \vspace{1em}
    
    $\widehat{\bm{\beta}}^{+} \gets \widehat{\bm{\beta}} - \eta \cdot \nabla_{\bm{\beta}}  \mathcal{PL}(\bm \beta, \vw; \mathcal{D})$\\

    $\widehat{\bm{w}}^{+} \gets \widehat{\bm{w}} - \eta \cdot \nabla_{\bm{w}}  \mathcal{PL}(\bm \beta, \vw; \mathcal{D})$  \hfill $\triangleright$  Gradient descent update.\\

    $\widehat{\bm{\Lambda}}_0(t)^{+} \gets \mathop{\sum}_{i=1}^{n} \sum_{k\in\mathcal{Z}} \bm{1}\{ u_i < t\}  \frac{\bm{\gamma}^{k}_i \cdot \delta_i }{ \mathop{\sum}\limits_{j\in \mathsf{RiskSet}(u_i)} \mathop{\sum}\limits_{k\in \mathcal{Z}} \bm{\gamma}^{k}_j  \bm {h}_k(\vx_j,\bm{a}_j, \vk)}$ \hfill$ \hfill \triangleright$\citet{breslow1972contribution}'s estimator.\\ 

    $\widehat{\bm{\theta}}^{+} \gets \widehat{\bm{\theta}} - \eta \cdot \nabla_\theta \mB(\theta)$     \hfill $\triangleright$ Update $\bm{\theta}$ with gradient of $\widehat{Q}$.\\ 

    $\widehat{\bm{\theta}}^{+} \gets \mathsf{prox}_{\eps\eta} (\widehat{\bm\theta})$     \hfill $\triangleright$ Proximal update.\\ 
  }
}
\hrulefill\\
\return{learnt parameters $\bm{\Omega}$;}
\label{alg:learning}
\end{algorithm}

\section{Experiments}

In this section we describe the experiments conducted to benchmark the performance of SCS against alternative models for heterogenous treatment effect estimation on multiple studies including a synthetic dataset and multiple large landmark clinical trials for cardiovascular diseases.

\subsection{Simulation}

\begin{figure}[!ht]
    \begin{minipage}{0.33\textwidth}
    \centering
    a) The Time-to-Event
    \end{minipage}%
    \begin{minipage}{0.33\textwidth}
    \centering
      b) Learnt Decision Boundary
    \end{minipage} %
    \begin{minipage}{0.33\textwidth}
    \centering
    c) ROC Curves
    \end{minipage}
    \includegraphics[width=.33\textwidth]{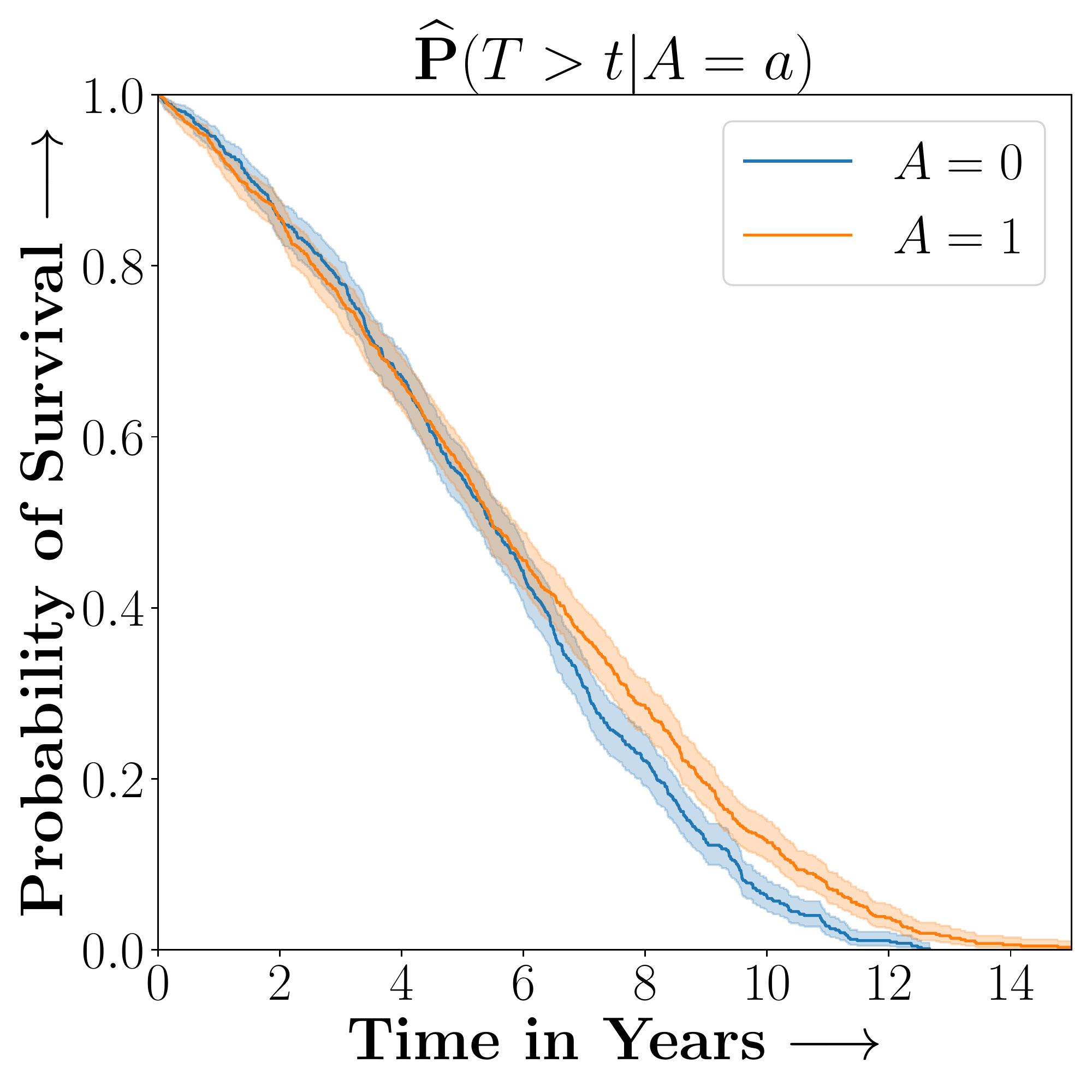} \hfill
    \includegraphics[width=.33\textwidth]{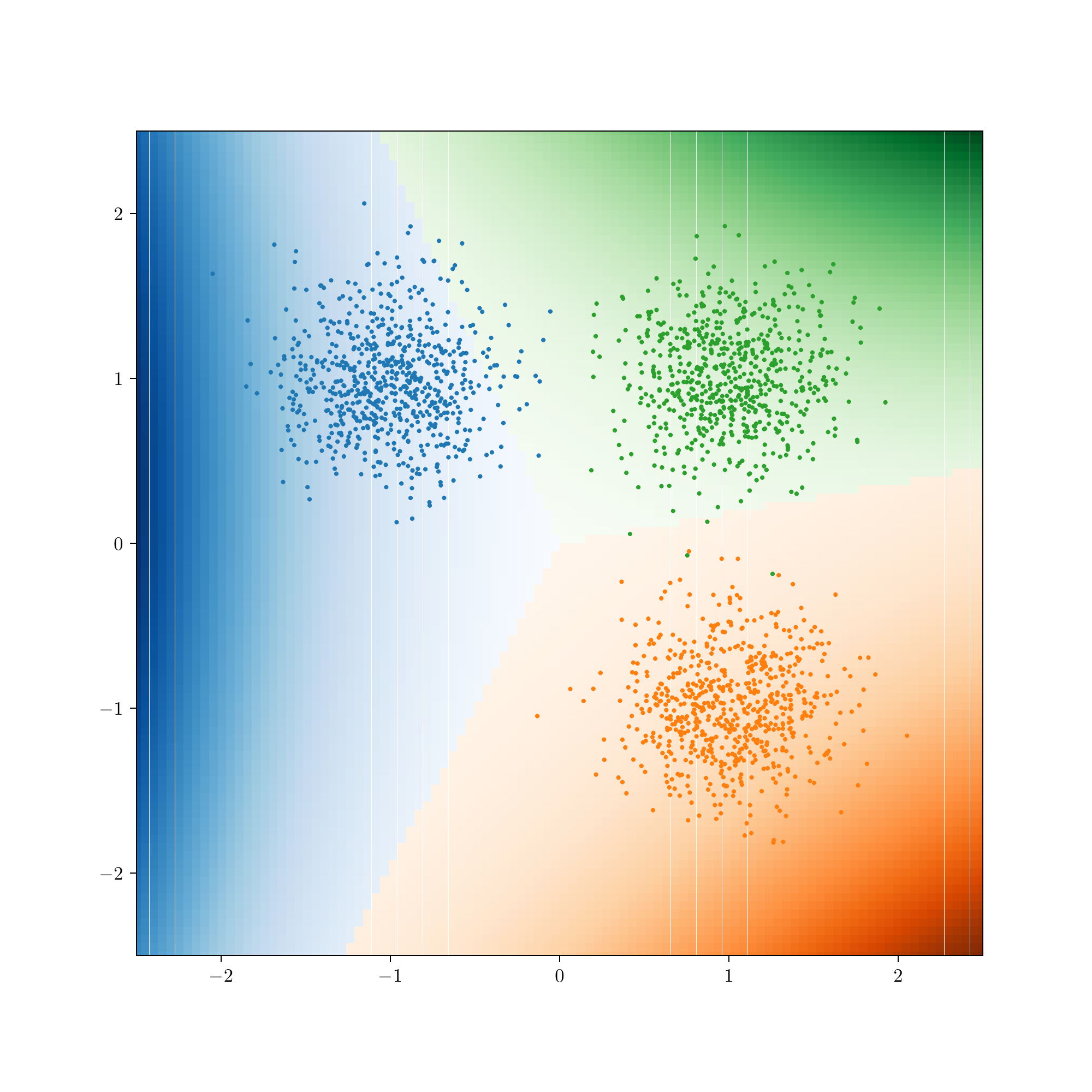}\hfill
    \includegraphics[width=.33\textwidth]{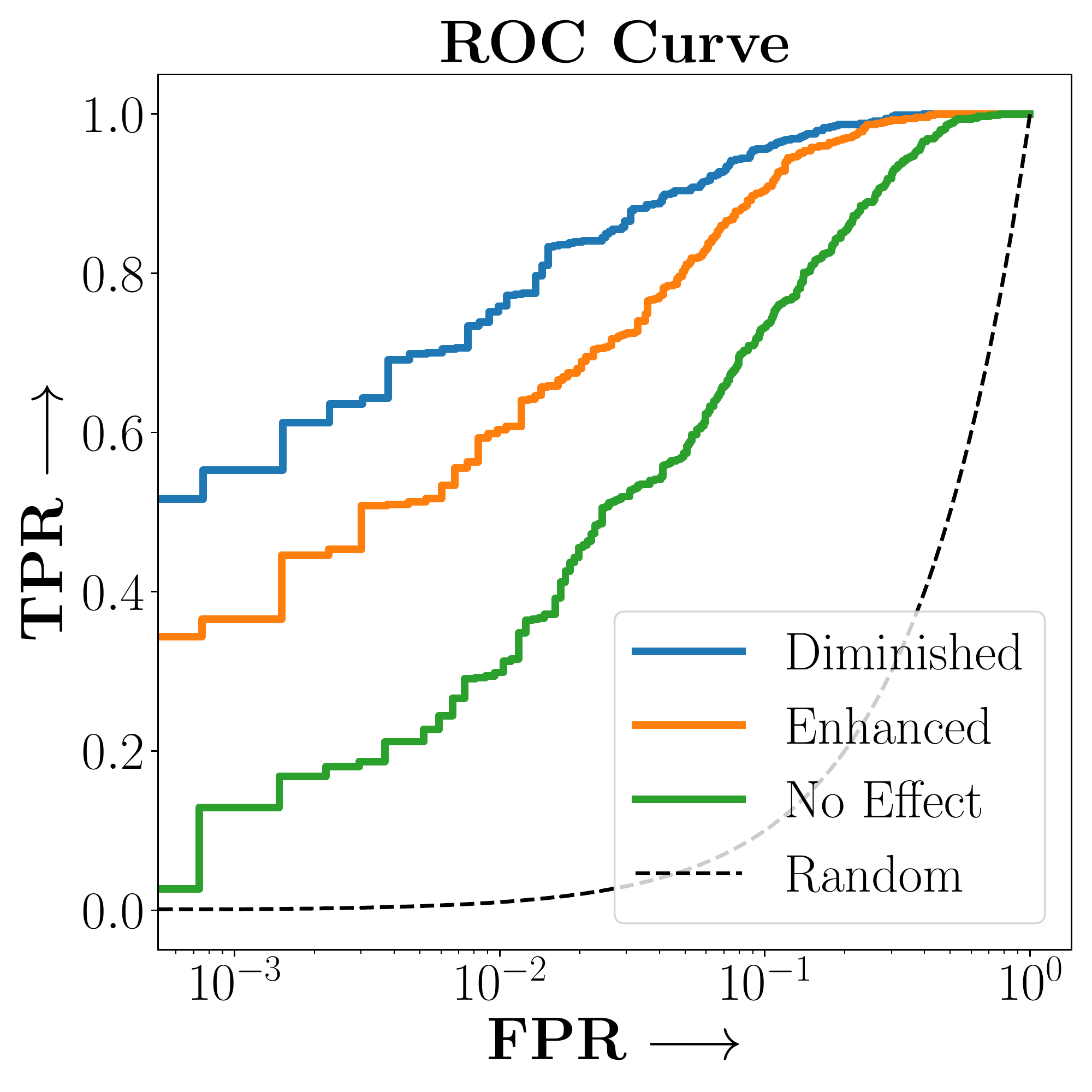}%
    \caption{a) Population level Kaplan-Meier Estimates of the Survival Distribution stratified by the treatment assignment. b) Distribution of the Latent $Z$ in $X$ and the recovered decision boundary by SCS. c) Receiver Operator Characteristics of SCS in recovering the true phenotype.}
    \label{fig:synthetic}
\end{figure}

\begin{figure}[!ht]
    \centering
    \includegraphics[width=.325\textwidth]{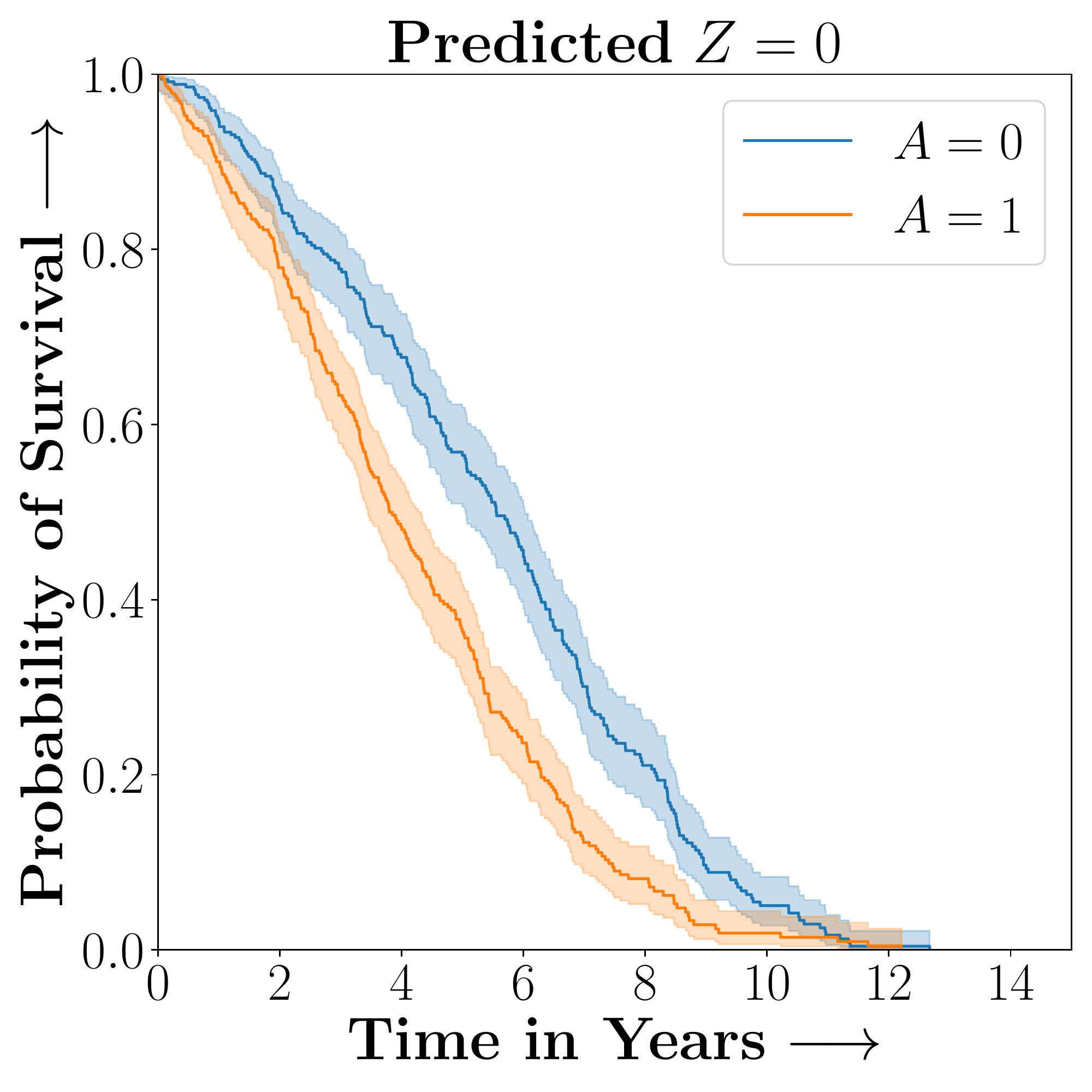}%
    \includegraphics[width=.325\textwidth]{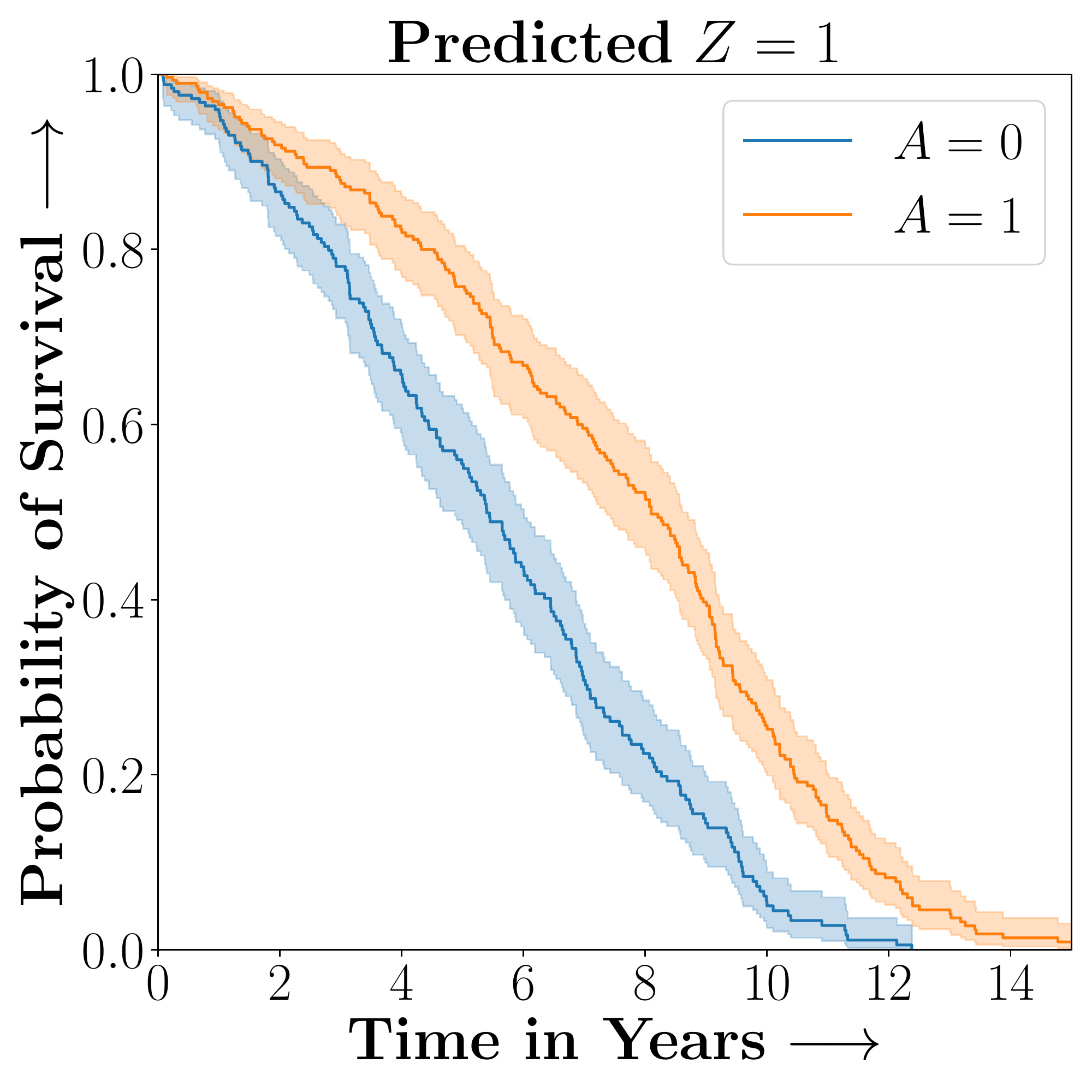}%
    \includegraphics[width=.325\textwidth]{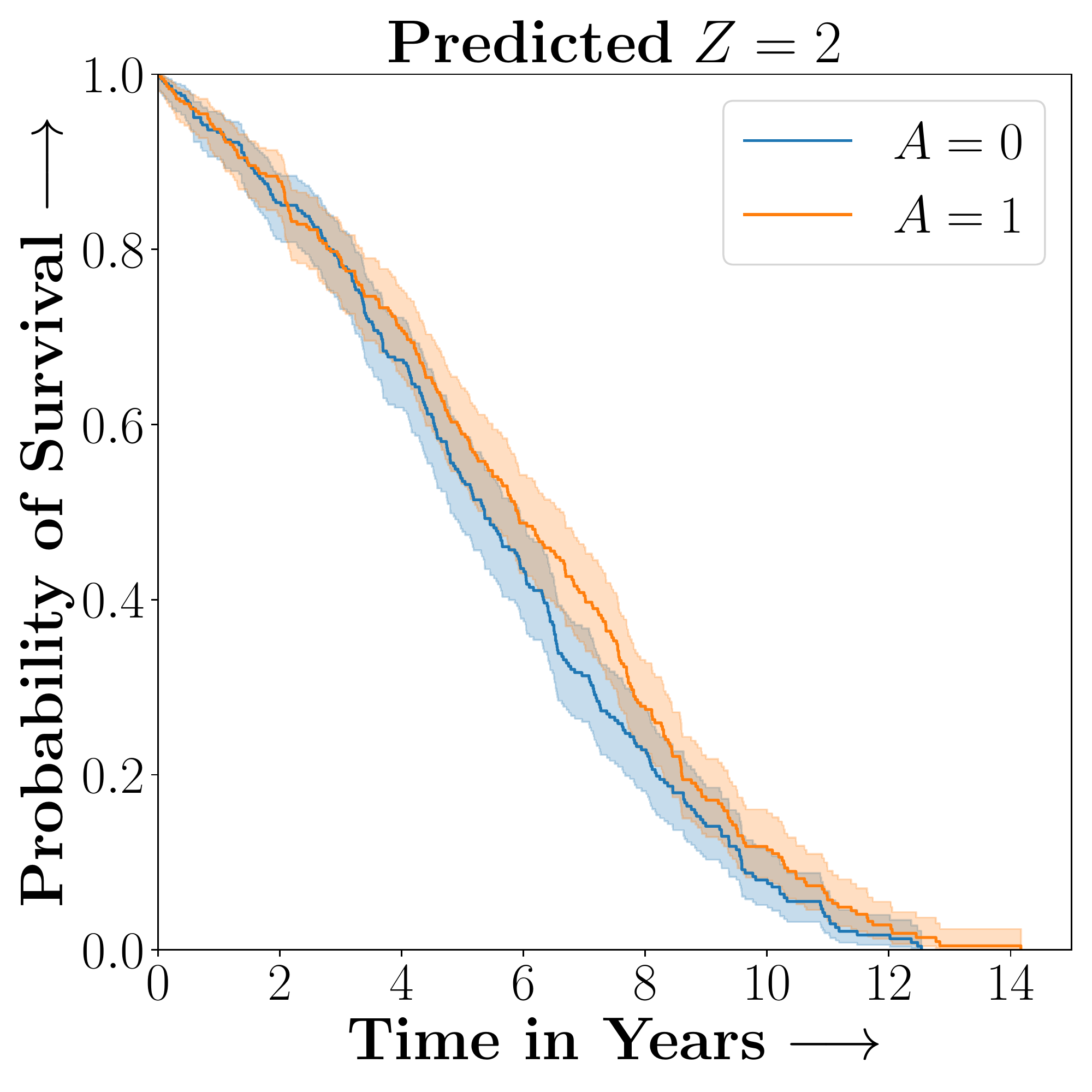}
    \caption{The phenotypes recovered with Sparse Cox Subgrouping on the Synthetic Data. As expected, the recovered phenotypes conform to the modelling assumptions as in \href{fig:studies}{Figure \ref*{fig:studies}.}}.
    \label{fig:pred_synthetic}
\end{figure}

In this section we first describe the performance of the proposed Sparse Cox Subgrouping approach on a synthetic dataset designed to demonstrate heterogeneous treatment effects. We randomly assign individuals to the treated or control group. The latent variable $Z$ is drawn from a uniform categorical distribution that determines the subgroup,
$$ A\sim \mathrm{Bernoulli}(\nicefrac{1}{2}), \quad Z \sim \mathrm{Categorical}(\nicefrac{1}{3})$$
Conditioned on $Z$ we sample $X_{1:2} \sim \textrm{Normal}(\bm{\mu}_z, \bm{\sigma}_{z})$ as in Figure \ref{fig:synthetic} that determine the conditional Hazard Ratios $\textrm{HR}(k)$, and randomly sample noisy covariates $X_{3:6} \sim \text{Uniform}(-1,1)$ . The true time-to-event $T$ and censoring times $C$ are then sampled as, 
\begin{align*}
T | (X=\vx, A=\va, Z=\vk ) &\sim \mathrm{Gompertz}(\beta=1, \eta=0.25 \cdot \text{HR}(k)^{\bm{a}}), \quad C|T\sim \text{Uniform}(0, T)   
\end{align*}

\noindent Finally we sample the censoring indicator $\Delta \sim \mathrm{Bernoulli}(0.8)$ and set the observed time-to-event,
$$U = T \text{ if }  \Delta = 1 \text{, else we set } U = C.$$
\noindent Figure \ref{fig:synthetic} presents the ROC curves for SCS's ability to identify the groups with enhanced and diminished treatment effects respectively. In Figure \ref{fig:pred_synthetic} we present Kaplan-Meier estimators of the Time-to-Event distributions conditioned on the predicted $Z$ by SCS. Clearly, SCS is able to identify the phenogroups corresponding to differential benefits.


\subsection{Recovering subgroups demonstrating Heterogeneous Treatment Effects from Landmark studies of Cardiovascular Health}

\begin{figure}[!h]
    \centering
    \includegraphics[width=0.5\textwidth]{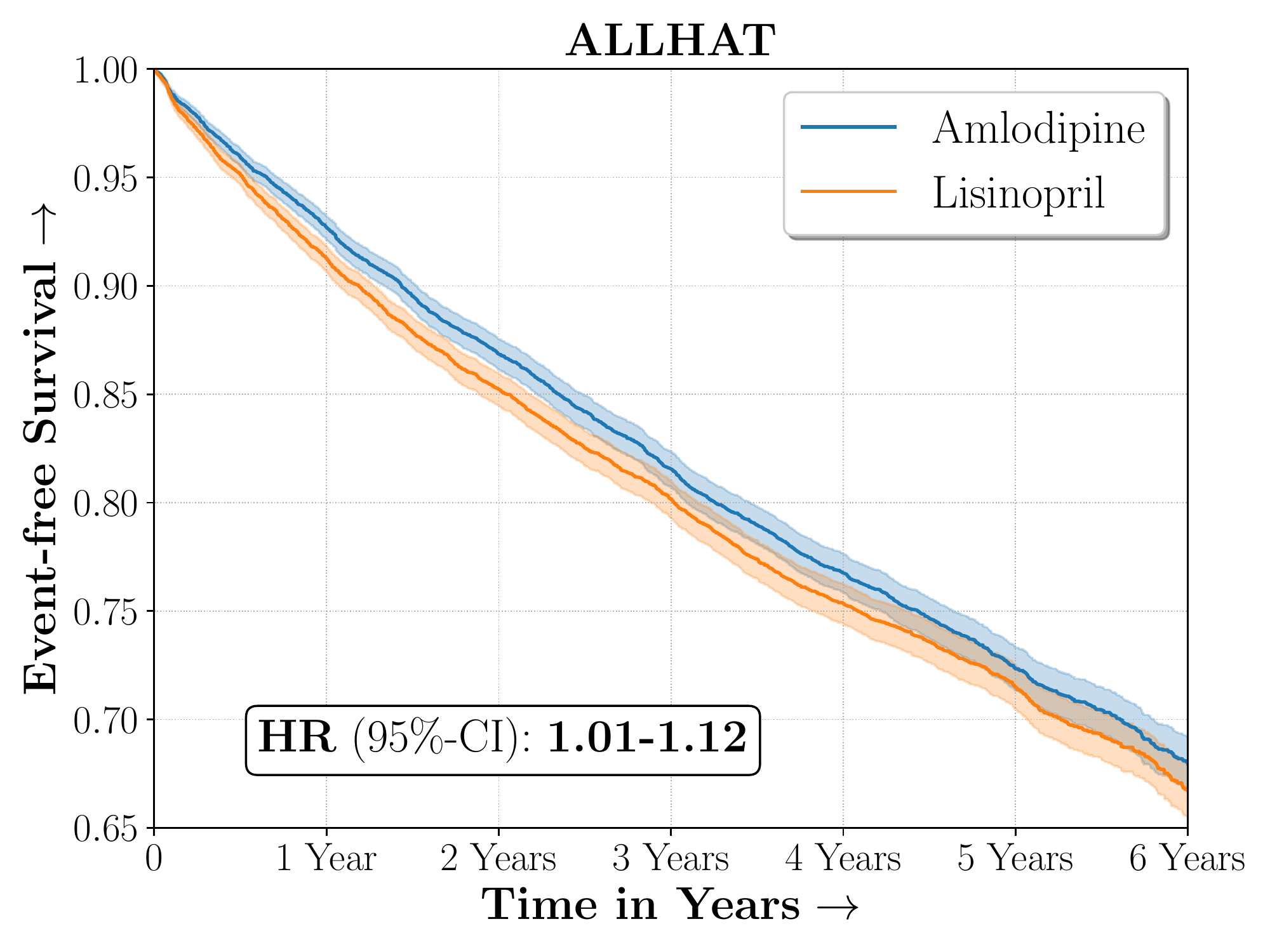}%
    \includegraphics[width=0.5\textwidth]{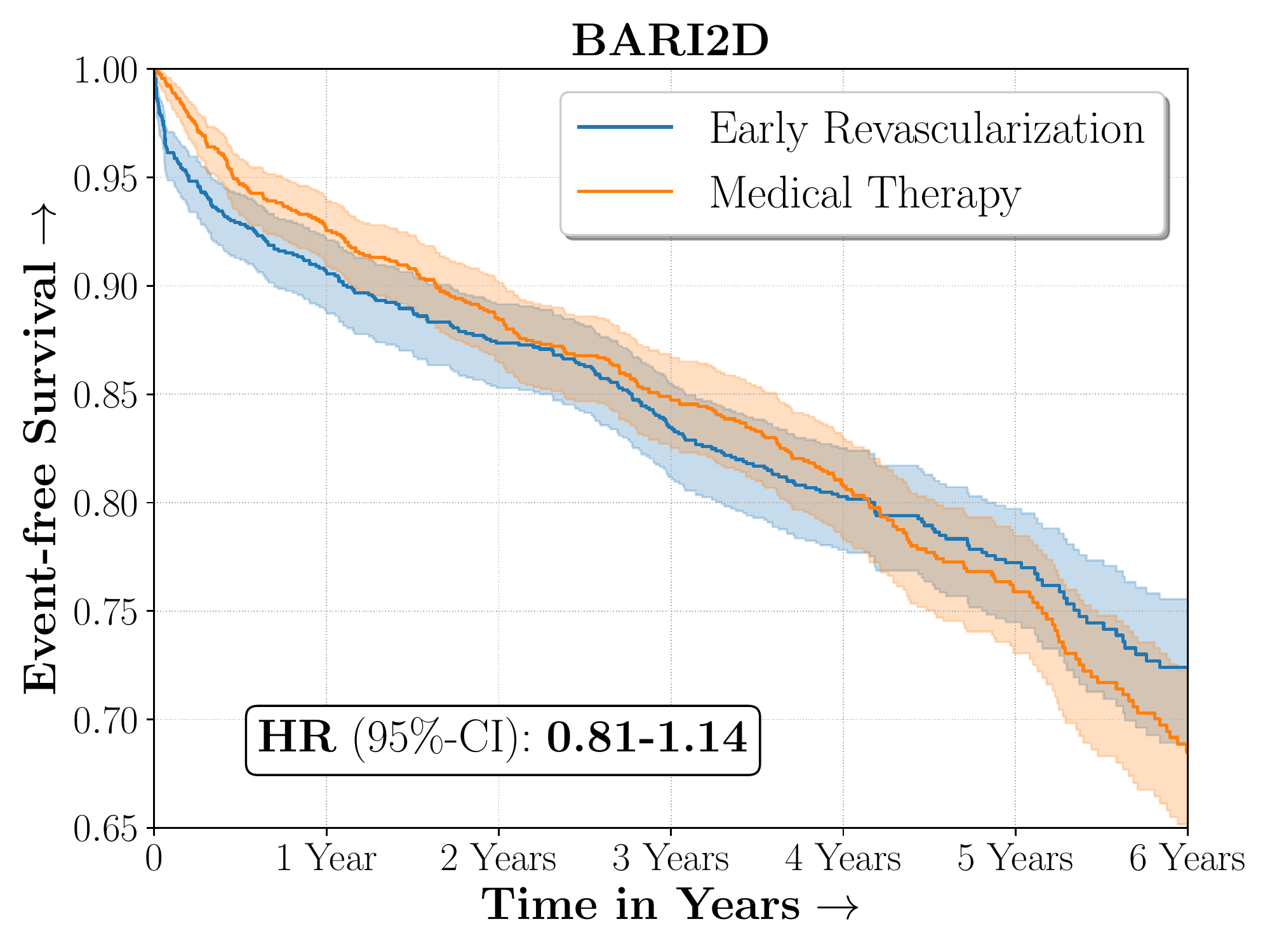}%
    
    \begin{minipage}{0.5\textwidth}
    \centering
    \begin{tabular}{lr}
    \toprule \hline
    \multicolumn{2}{c}{\textbf{ALLHAT}}\\ \hline
        Size &  18,102\\
        Outcome & Combined CVD\\
        Intervention & Lisinopril\\
        Control & Amlodipine\\
        Hazard Ratio & 1.06, (1.01, 1.12)\\
        5-year RMST & -24.86, (-37.35, -8.89)\\ \hline \bottomrule

    \end{tabular}
    
    \end{minipage}%
    \begin{minipage}{0.5\textwidth}
    \centering
    \begin{tabular}{lr}
    \toprule \hline 
    \multicolumn{2}{c}{\textbf{BARI2D}}\\ \hline
        Size &  2,368\\
        Outcome & Death, MI or Stroke\\
        Intervention & Medical Therapy\\
        Control & Early Revascularization\\
        Hazard Ratio & 1.02, (0.81, 1.14)\\
        5-year RMST & 23.26, (-27.01, 64.84)\\ \hline \bottomrule
    \end{tabular}
    \end{minipage}

    \captionof{figure}{Event-free Kaplan-Meier survival curves stratified by the treatment assignment and summary statistics for the \textbf{ALLHAT} and \textbf{BARI2D} studies. (Combined CVD: Coronary Heart Disease, Stroke, other treated angina, fatal or non-fatal Heart Failure, and Peripheral Arterial Disease.)}
    \label{fig:studies}
\end{figure}

\begin{description}[leftmargin=*]

\item[Antihypertensive and Lipid-Lowering Treatment to Prevent Heart Attack] \citep{furberg2002major}

The ALLHAT study was a large randomized experiment conducted to assess the efficacy of multiple classes of blood pressure lowering medicines for patients with hypertension  in reducing risk of adverse cardiovascular conditions. We considered a subset of patients from the original \textbf{ALLHAT} study who were randomized to receive either Amlodipine (a calcium channel blocker) or Lisinopril (an Angiotensin-converting enzyme inhibitor). Overall, Amlodipine was found to be more efficacious than Lisinopril in reducing combined risk of cardio-vascular disease. 
\item[Bypass Angioplasty Revascularization Investigation in Type II Diabetes] \citep{bari2009randomized}

Diabetic patients have been traditionally known to be at higher risk of cardiovascular disease however appropriate intervention for diabetics with ischemic heart disease between surgical coronary revascularization or management with medical therapy is widely debated. The \textbf{BARI2D} was a large landmark experiment conducted to assess efficacy between these two possible medical interventions. Overall \textbf{BARI2D} was inconclusive in establishing the appropriate therapy between Coronary Revascularization or medical management for patients with Type-II Diabetes. 

\end{description}

\noindent \href{fig:studies}{Figure \ref*{fig:studies}} presents the event-free survival rates as well as the summary statistics for the studies. In our experiments, we included a large set of confounders collected at baseline visit of the patients which we utilize to train the proposed model. A full list of these features are in \href{apx:confounders}{Appendix \ref*{apx:confounders}}.


\subsection{Baselines}

\begin{description}[leftmargin=*]
\item[Cox PH with $\ell_1$ Regularized Treatment Interaction (\textsc{cox-int})]\phantom{a}

We include treatment effect heterogeneity via interaction terms that model the time-to-event distribution using a proportional hazards model as in \cite{kehl2006responder}. Thus, 
\begin{align}
    \bm{\lambda}(t|X=\vx, A=\va) = \bm{\lambda}_0(t)\exp\big( \bm{\beta}^{\top}\vx + \va\cdot \bm{\theta}^{\top}\vx \big)
\end{align}
The interaction effects $\bm{\theta}$ are regularized with a lasso penalty in order to recover a sparse phenotyping rule defined as $G(\bm{x}) = \bm{\theta}^{\top} \bm{x}$.

\item[Binary Classifier with $\ell_1$ Regularized Treatment Interaction (\textsc{bin-int})] \phantom{a}

Instead of modelling the time-to-event distribution we directly model the thresholded survival outcomes $Y=\bm{1}\{T<t\}$ at a five year time horizon using a log-linear parameterization with a logit link function. As compared to \textsc{cox-int}, this model ignores the data-points that were right-censored prior to the thresholded time-to-event, however it is not sensitive to the strong assumption of Proportional Hazards.
\begin{align}
  \nonumber \mathbb{E}[T>t|X=\vx, A=\va]  = \sigma( \bm{\beta}^{\top}\vx + \bm{\beta}_0 + \va\cdot \bm{\theta}^{\top}\vx),\\
  \text{and, }\sigma(\cdot) \text{ is the logistic link function.} 
\end{align}

\item[Cox PH T-Learner with $\ell_1$ Regularized Logistic Regression (\textsc{cox-tlr})] \phantom{a}

\noindent We train two separate Cox Regression models on the treated and control arms (T-Learner) to estimate the potential outcomes under treatment $(A=1)$ and control $(A=0)$. Motivated from the \textit{`Virtual Twins'} approach as in \cite{foster2011subgroup}, a logistic regression with an $\ell_1$ penalty is trained to estimate if the risk of the potential outcome under treatment is higher than under control. This logistic regression is then employed as the phenotyping function $G(\cdot)$ and is given as,
\begin{align}
 \nonumber   G(\vx) &= \mathbb{E}[\bm{1}\{f_1(\vx, t) > f_0(\vx, t)   \} | X =\vx]\\
    \text{where, } f_{\bm{a}} (\vx, t) &= \mathbb{P}(T>t| \text{do}(A = \bm{a}), X=\vx).
\end{align}

The above models involving sparse $\ell_1$ regularization were trained with the \texttt{glmnet} \citep{friedman2009glmnet} package in \texttt{R}.

\item[The ACC/AHA Long term Atheroscleoratic Cardiovascular Risk Estimate]\footnote{\url{https://tools.acc.org/ascvd-risk-estimator-plus/}}   \phantom{a}

\noindent The American College of Cardiology and the American Heart Association model for estimation of risk of Atheroscleratic disease risk \citep{goff20142013} involves pooling data from multiple observational cohorts of patients followed by modelling the 10-year risk of an adverse cardiovascular condition including death from coronary heart disease, Non-Fatal Myocardial Infarction or Non-fatal Stroke. While the risk model was originally developed to assess factual risk in the observational sense, in practice it is also employed to assess risk when making counterfactual decisions.

\end{description}

\begin{figure}[!htbp]
    \centering
    \textbf{Amlodipine versus Lisinopril in the ALLHAT Study}\\
    
    \includegraphics[width=0.5\textwidth]{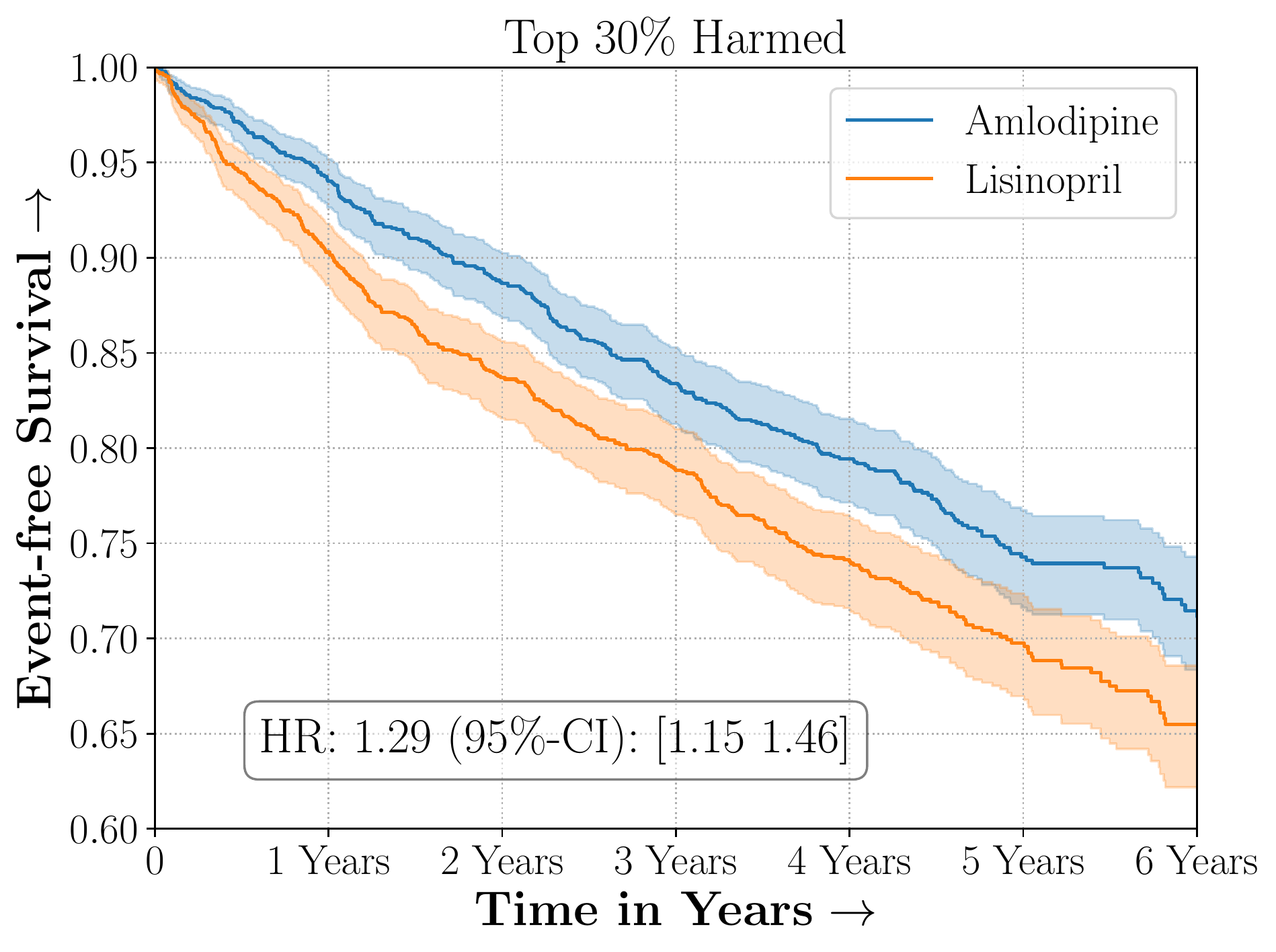}%
    \includegraphics[width=0.5\textwidth]{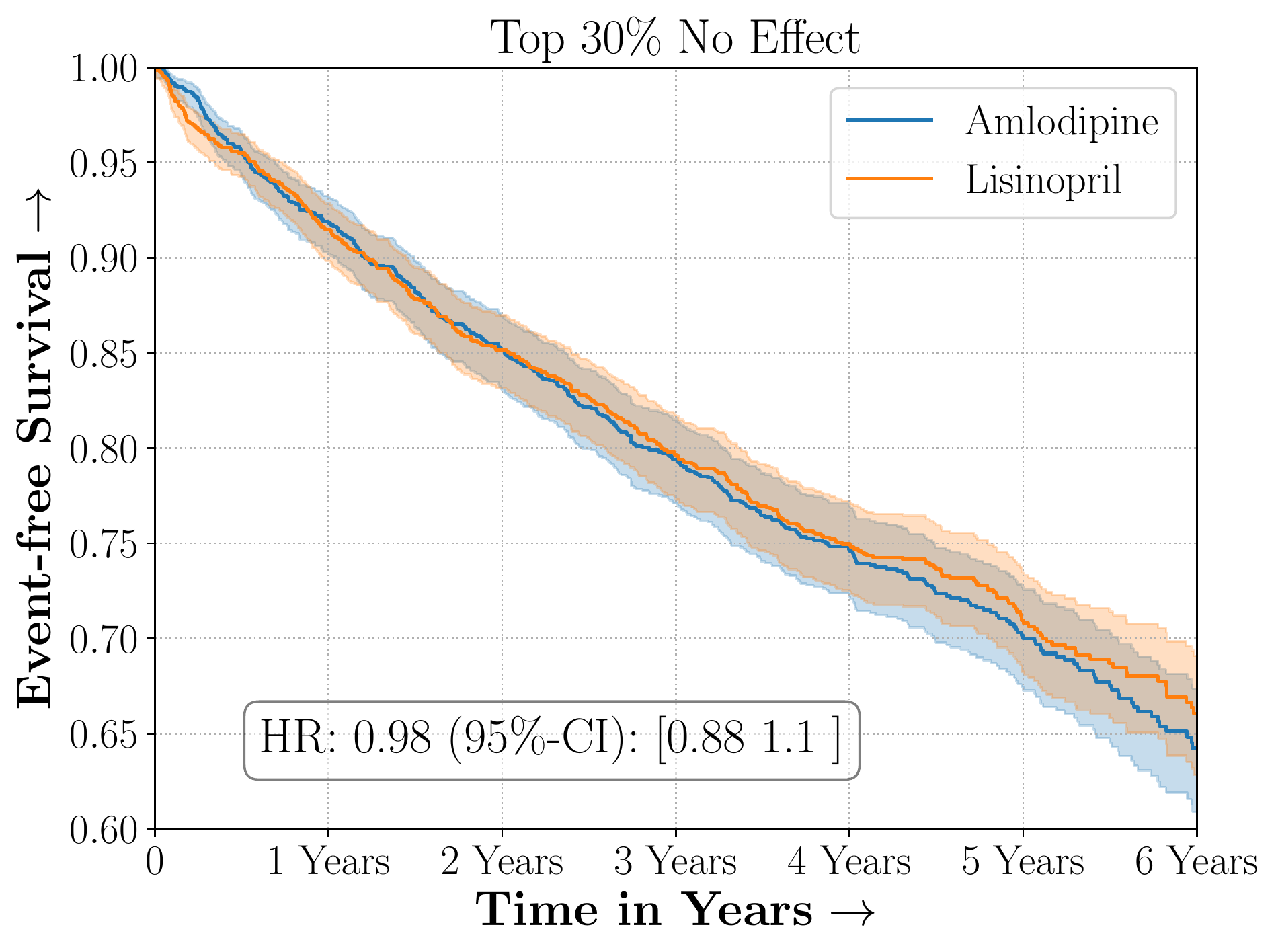}
    
    \textbf{$||\bm{\theta}||_0 \leq 5$}\\
    \includegraphics[width=0.5\textwidth]{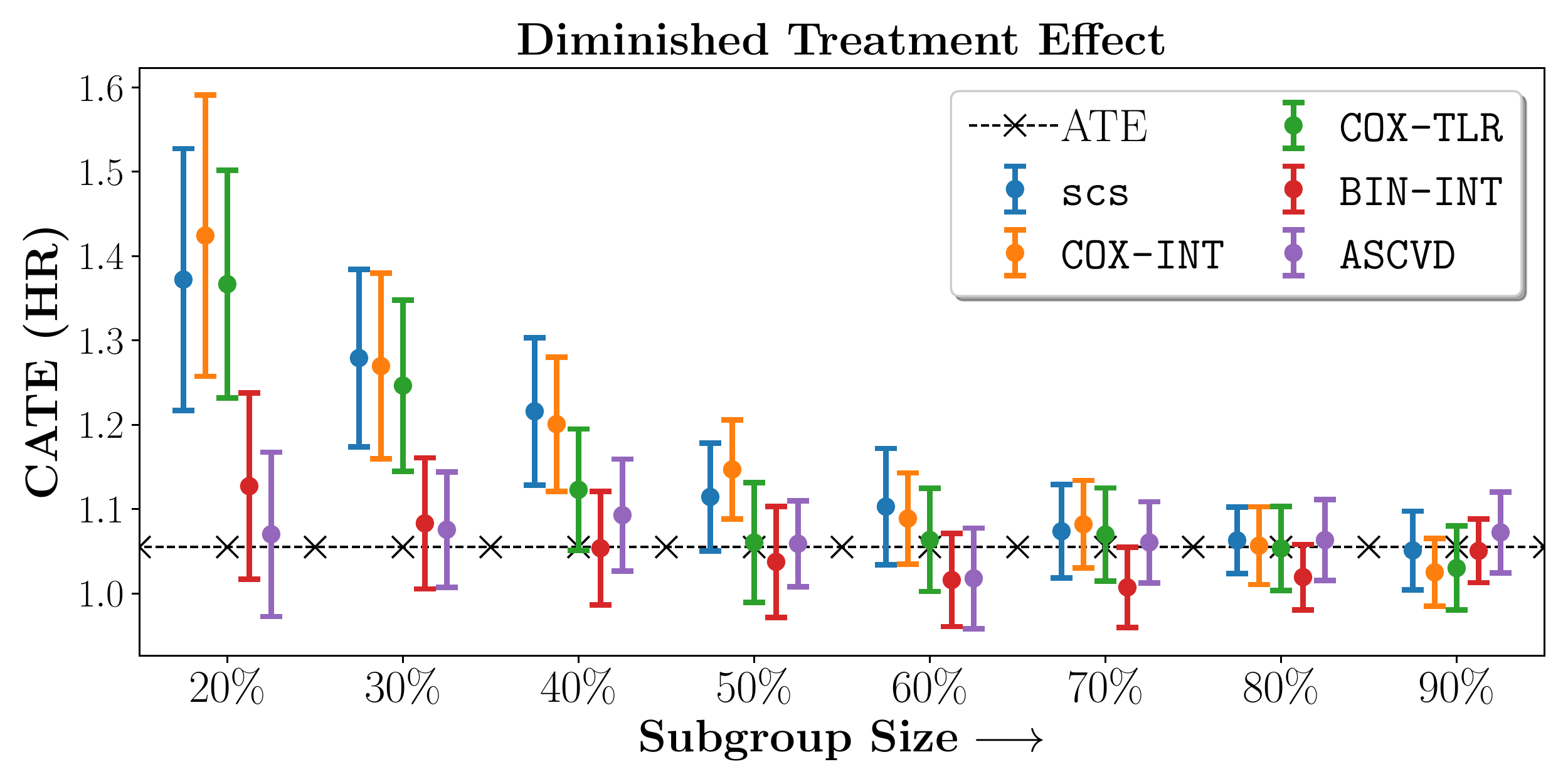}%
    \includegraphics[width=0.5\textwidth]{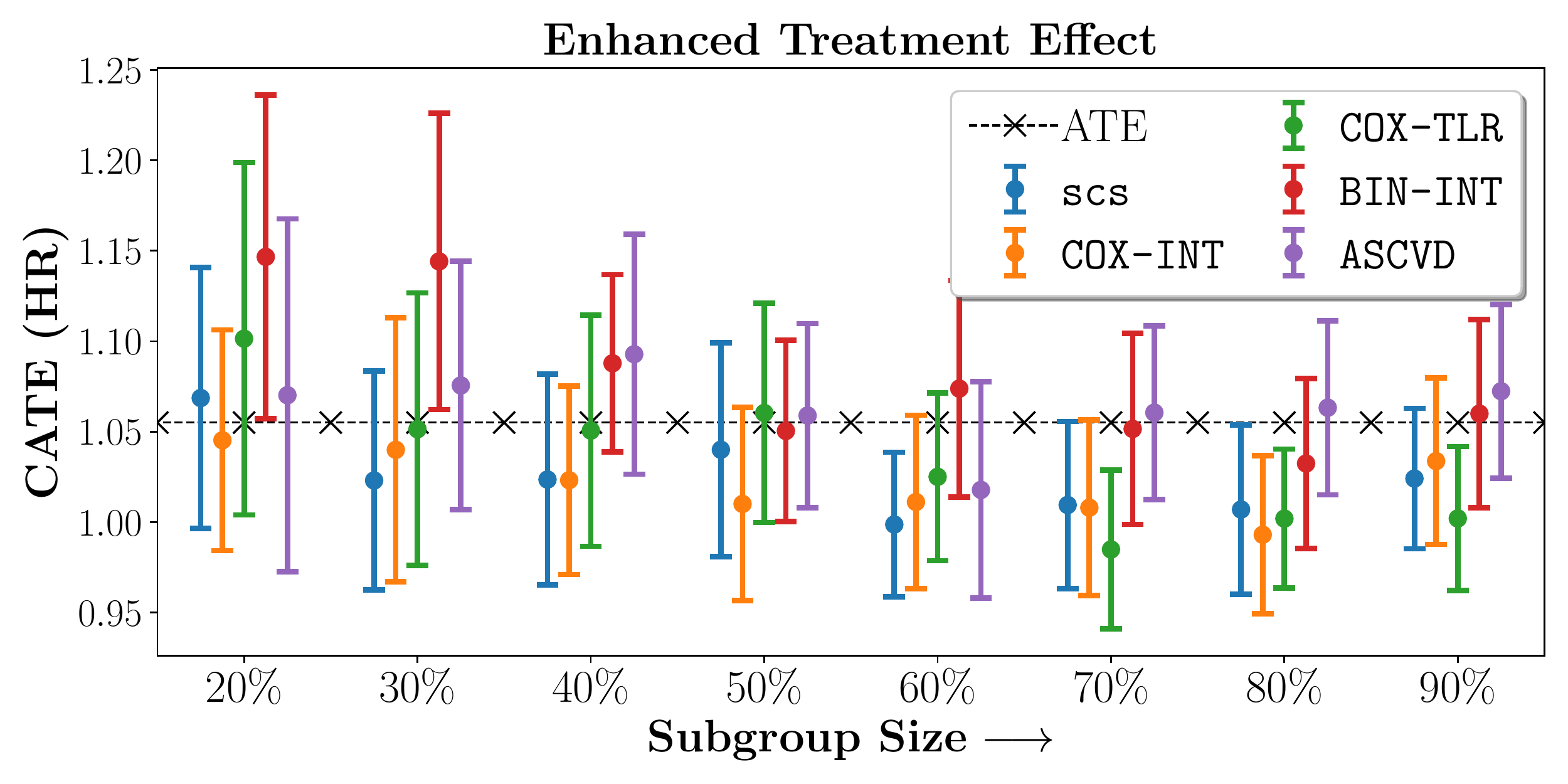}
    \textbf{$||\bm{\theta}||_0 \leq 10$}
    \includegraphics[width=0.5\textwidth]{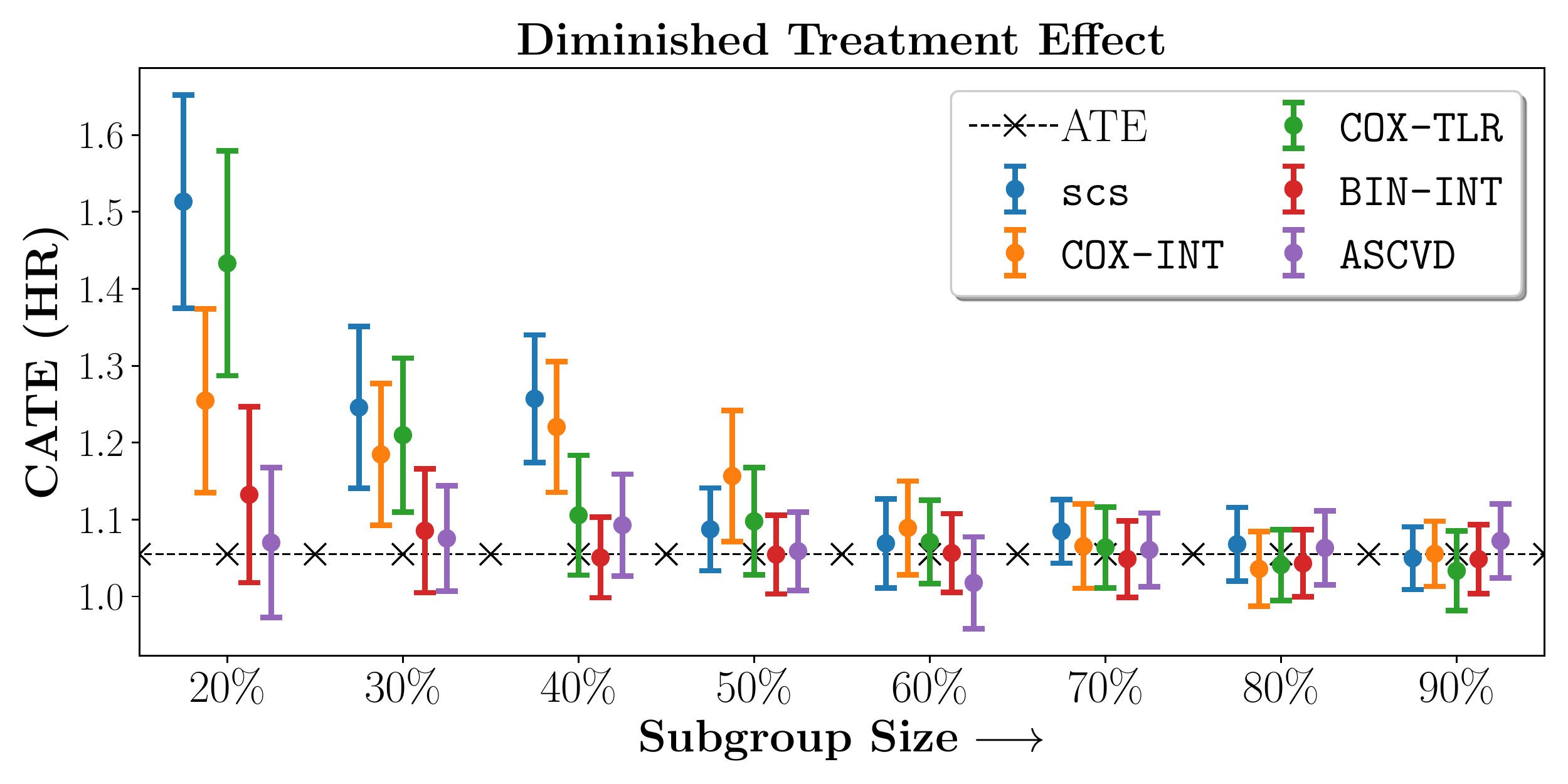}%
    \includegraphics[width=0.5\textwidth]{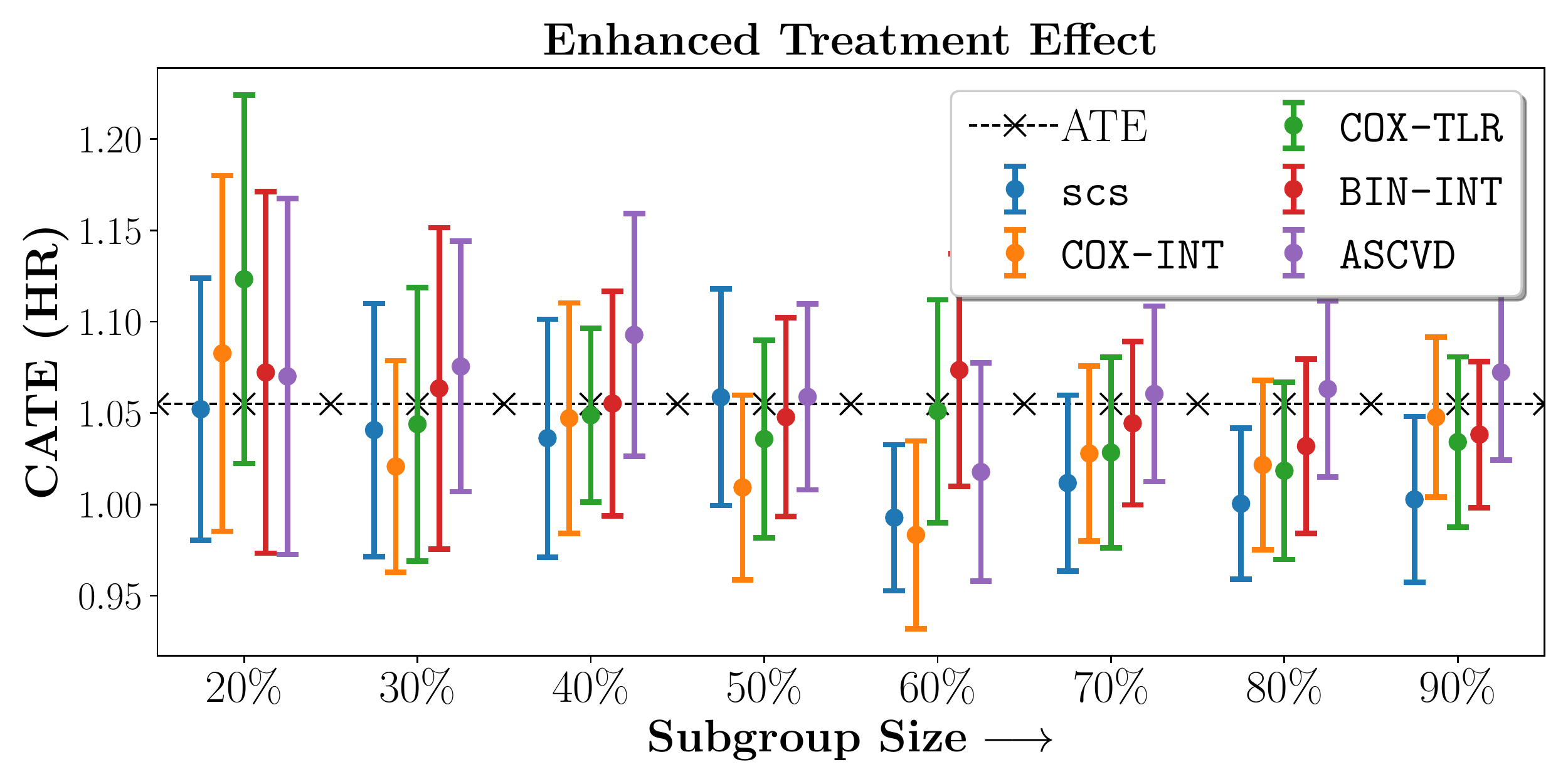}
    No Sparsity\\
    \includegraphics[width=0.5\textwidth]{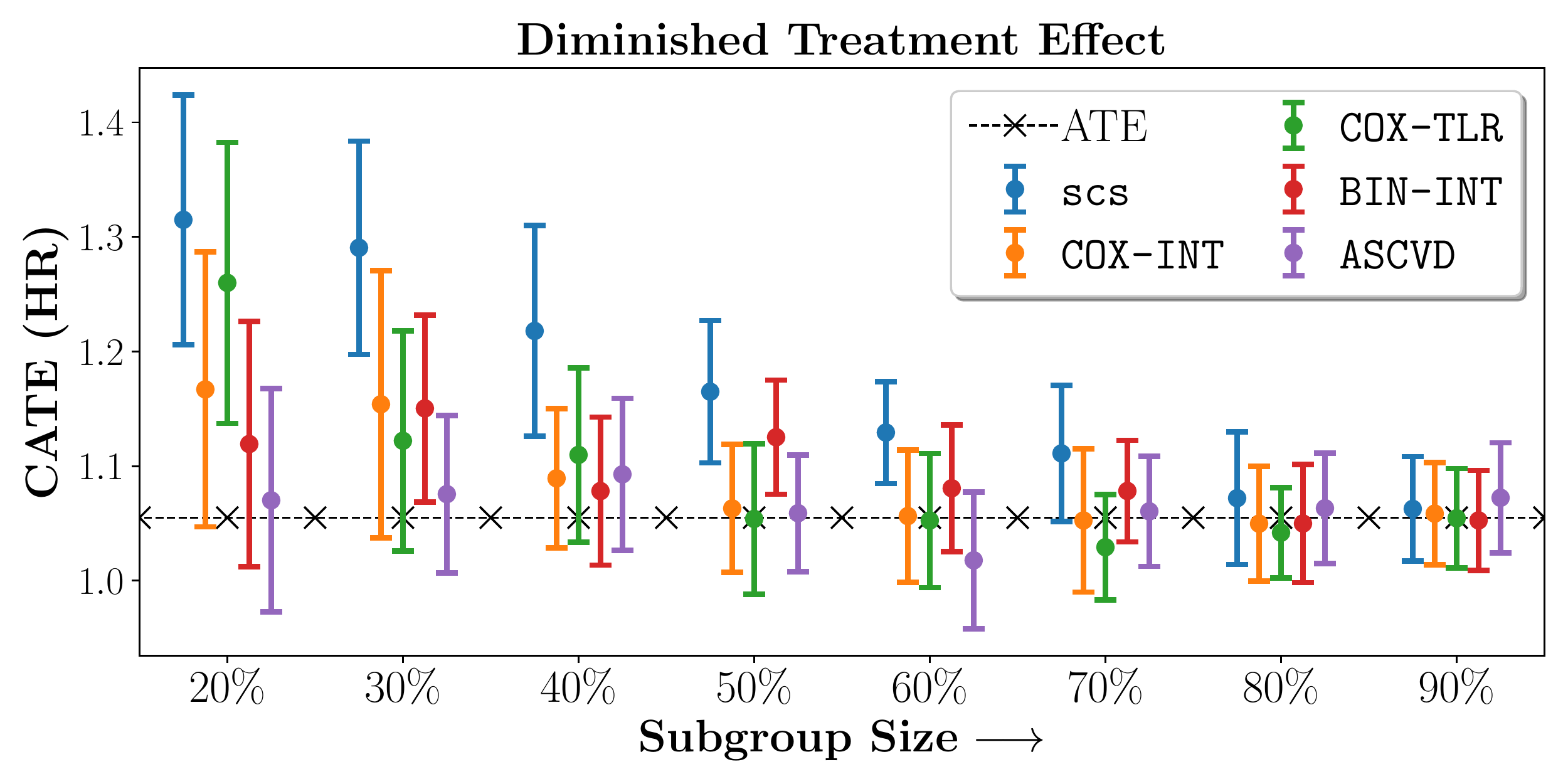}%
    \includegraphics[width=0.5\textwidth]{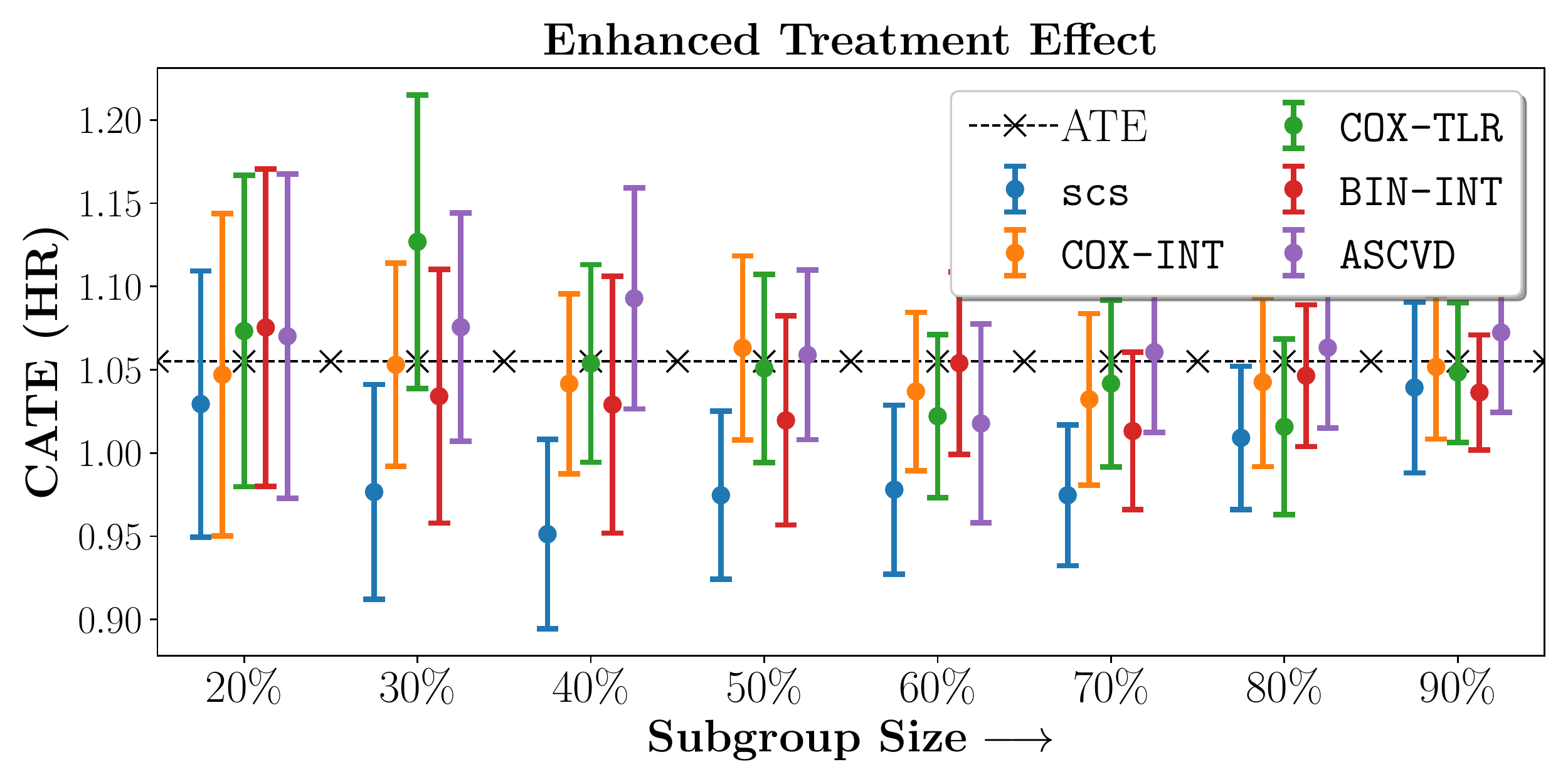}
    \caption{Conditional Average Treatment Effect in Hazard Ratio versus subgroup size for the latent phenogroups extracted from the \textbf{ALLHAT} study.}
    \label{fig:allhat}
\end{figure}

\begin{figure}
    \centering
    \textbf{Early Revascularization versus Medical Therapy in the BARI2D Study}\\
    
    \includegraphics[width=0.5\textwidth]{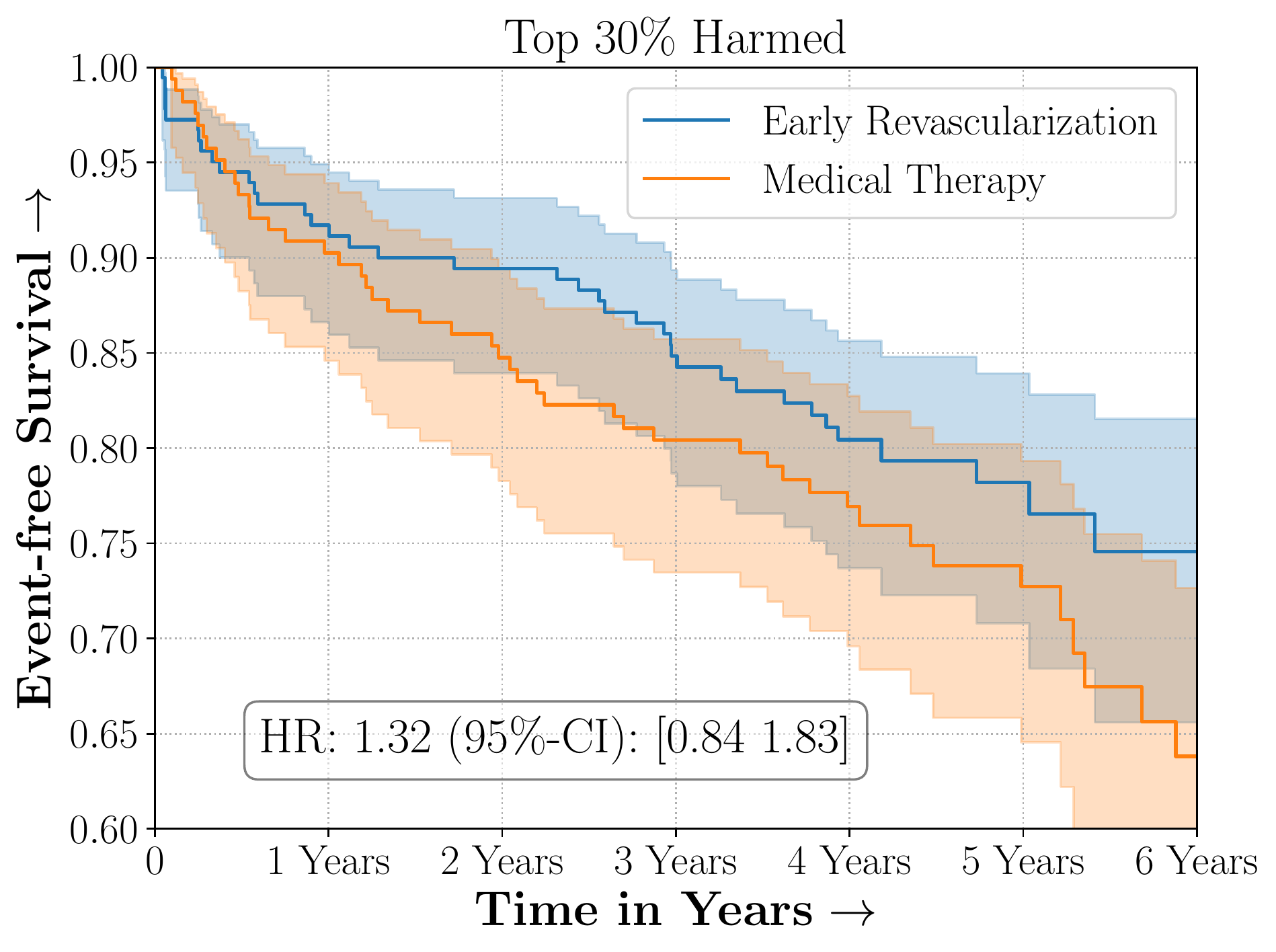}%
    \includegraphics[width=0.5\textwidth]{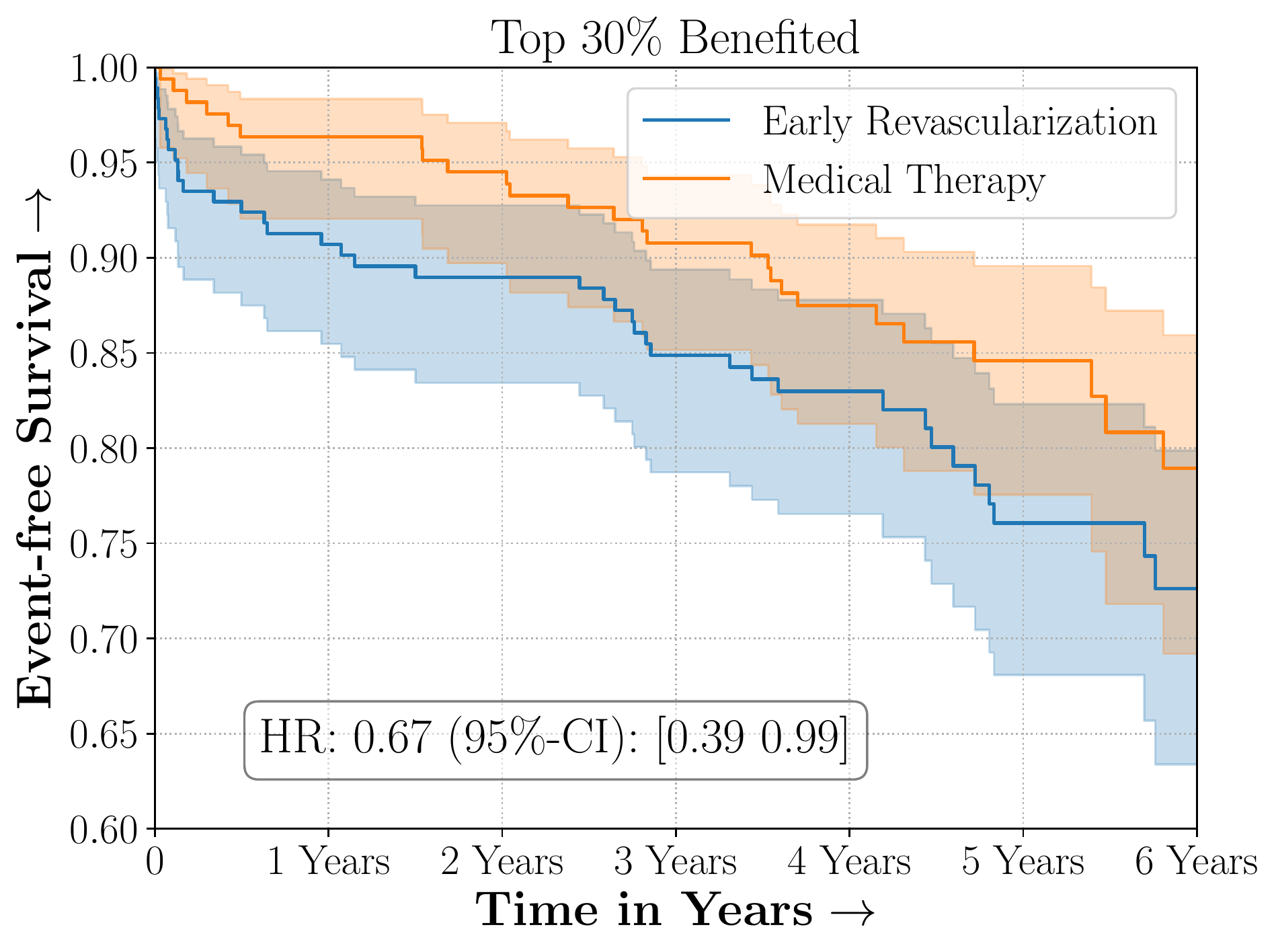}
    
    \textbf{$||\bm{\theta}||_0 \leq 5$}
    \includegraphics[width=0.5\textwidth]{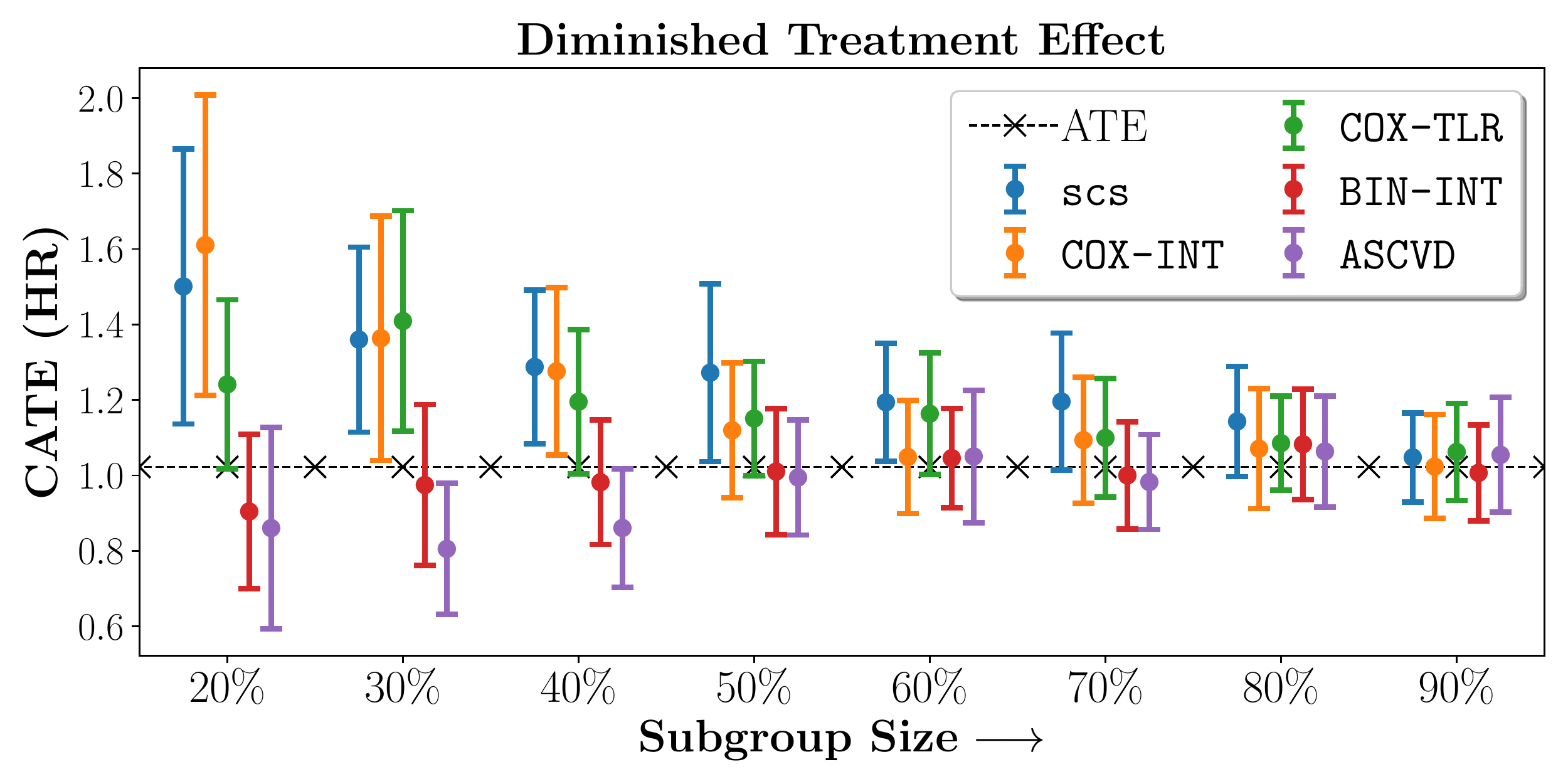}%
    \includegraphics[width=0.5\textwidth]{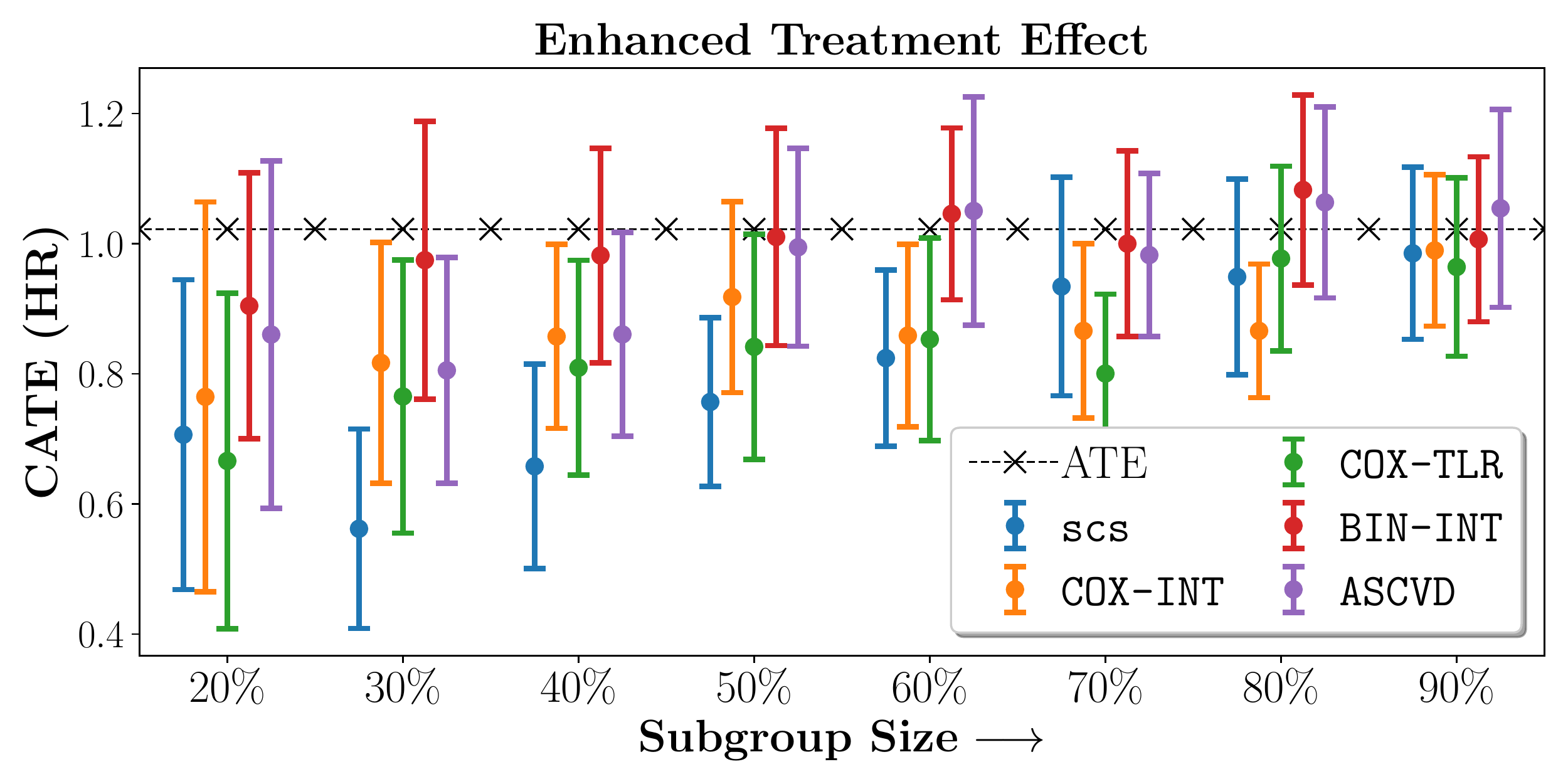}
    \textbf{$||\bm{\theta}||_0 \leq 10$}
    \includegraphics[width=0.5\textwidth]{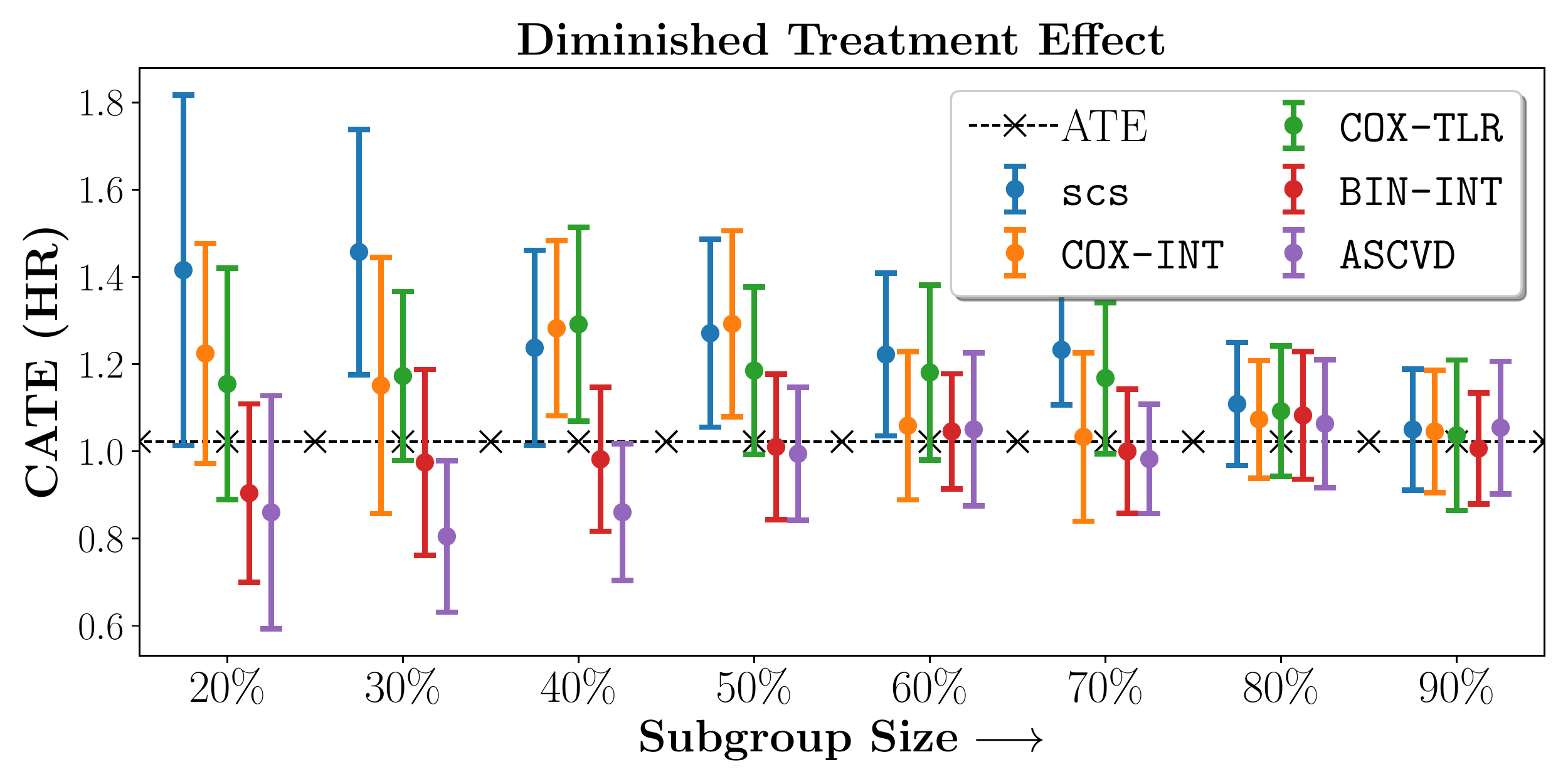}%
    \includegraphics[width=0.5\textwidth]{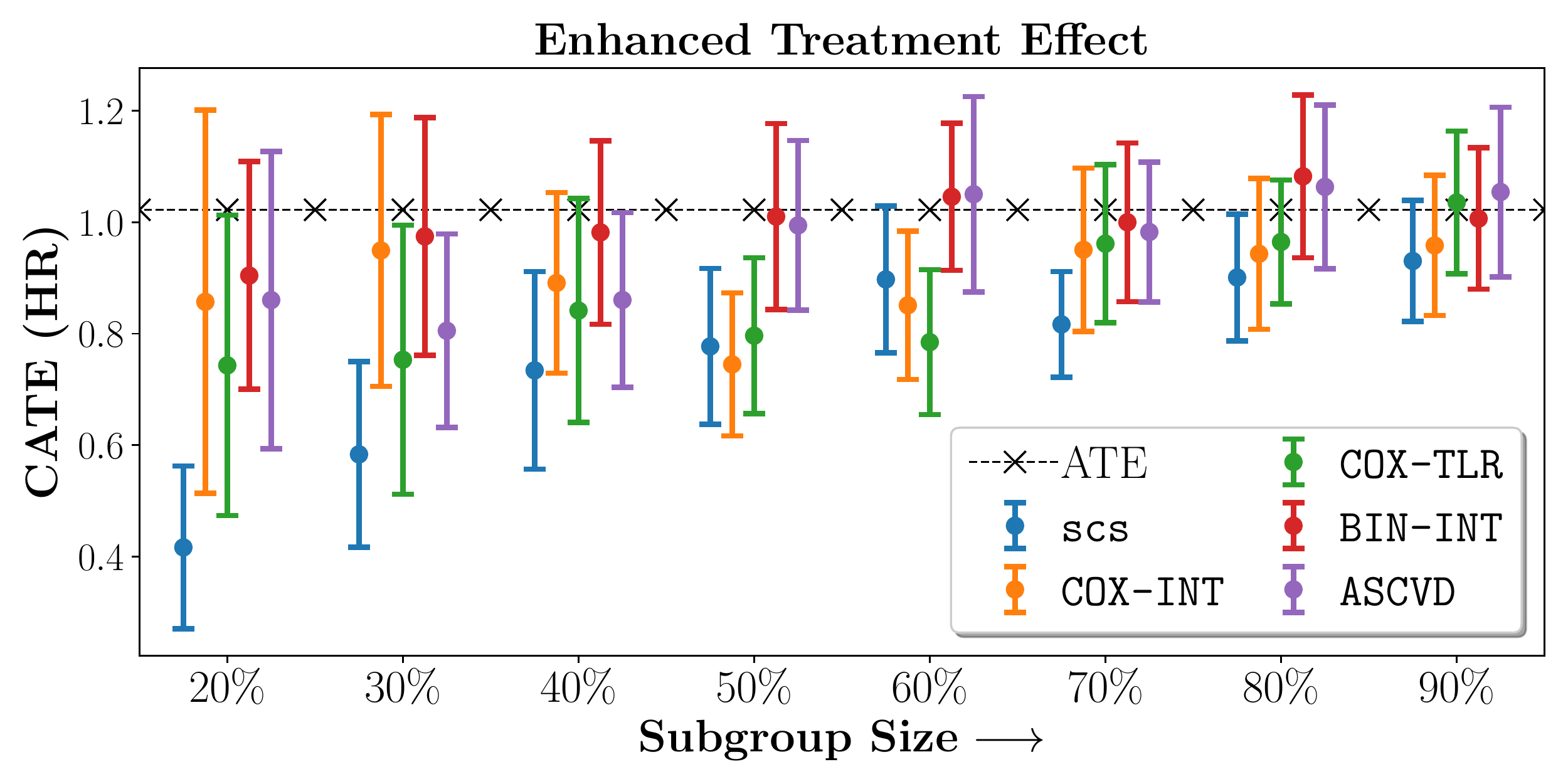}
    No Sparsity\\
   \includegraphics[width=0.5\textwidth]{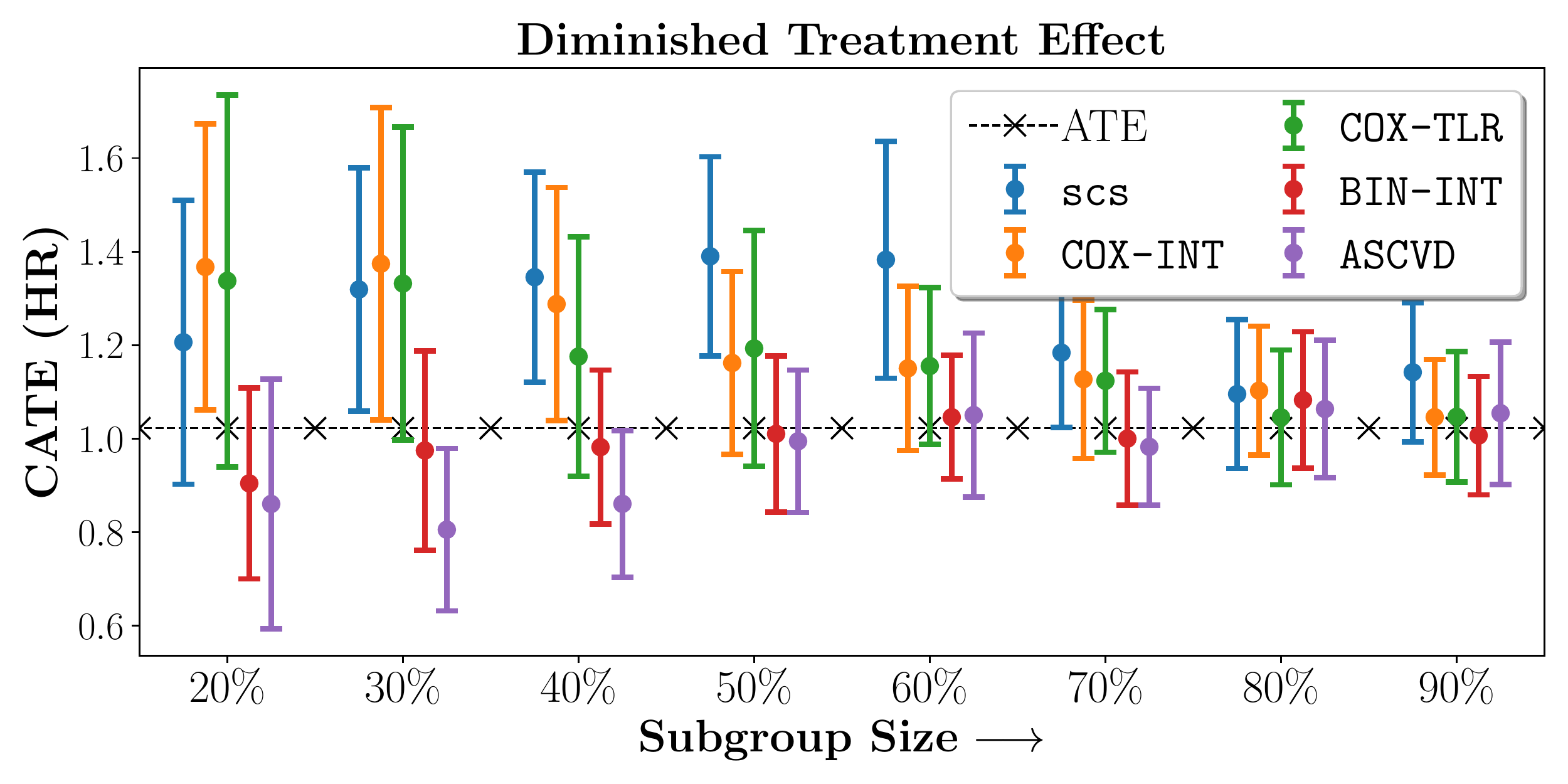}%
    \includegraphics[width=0.5\textwidth]{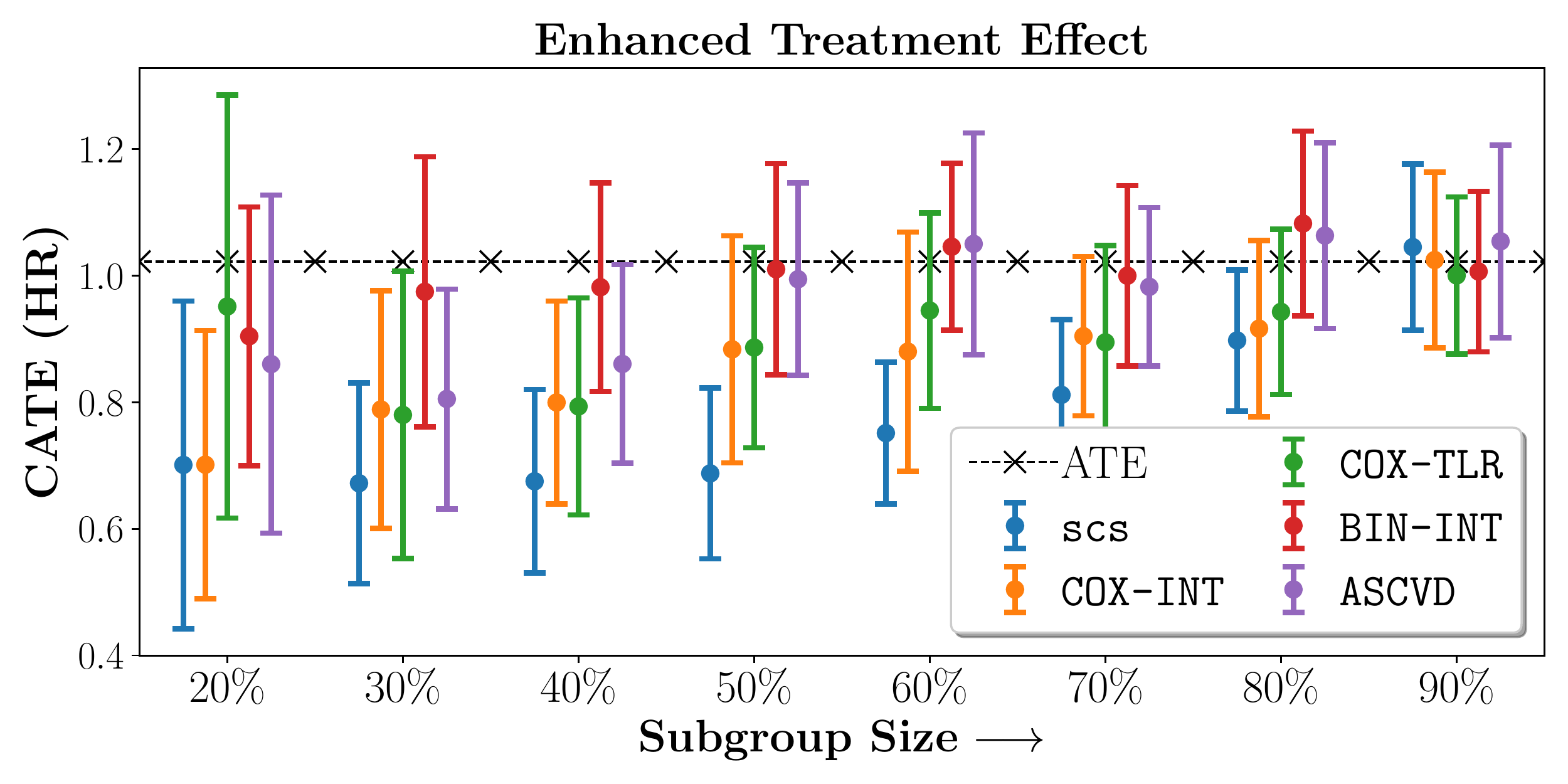}
    \caption{Conditional Average Treatment Effect in Hazard Ratio versus subgroup size for the latent phenogroups extracted from the \textbf{BARI2D} study.}
    \label{fig:bari2d}
\end{figure}






\subsection{Results and Discussion}

\begin{description}[leftmargin=*]

\item[Protocol] We compare the performance of SCS and the corresponding competing methods in recovery of subgroups with enhanced (or diminished treatment effects). For each of these studies we stratify the study population into equal sized sets for training and validation while persevering the proportion of individuals that were assigned to treatment and experienced the outcome in the follow up period. The models were trained on the training set and validated on the held-out test set. For each of the methods we experiment with models that do not enforce any sparsity ($\bm{\epsilon }=0$) as well as tune the level of sparsity to recover phenotyping functions that involve $5$ and $10$ features. The subgroup size are varied by controlling the threshold at which the individual is assigned to a group. Finally, the treatment effect is compared in terms of Hazard Ratios, Risk Differences as well as Restricted Mean Survival Time over a 5 Year event period.

\item[Results] We present the results of SCS versus the baselines in terms of Hazard Ratios on the \textbf{ALLHAT} and \textbf{BARI2D} datasets in Figures \ref{fig:allhat} and \ref{fig:bari2d}. In the case of \textbf{ALLHAT}, SCS consistently recovered phenogroups with more pronounced (or diminished) treatment effects. On external validation on the heldout dataset, we found a subgroup of patients that had similar outcomes whether assigned to Lisinopril or Amlodipine, whereas the other subgroup clearly identified patients that were harmed with Lisinopril. The group harmed with Lisinopril had higher Diastolic BP. On the other hand, patients with Lower kidney function did not seem to benefit from Amlodipine.

In the case of \textbf{BARI2D}, SCS recovered phenogroups that were both harmed as well as benefitted from just medical therapy without revascularization. The patients who were harmed from Medical therapy were typically older, on the other hand the patients who benefitted primarily included patients who were otherwise assigned to receive PCI instead of CABG revascularization, suggesting PCI to be harmful for diabetic patients. 

Tables \ref{tab:5feats} and \ref{tab:10feats} present the features that were selected by the proposed model for the studies. Additionally, we also report tabulated results involving metrics like risk difference and restricted mean survival time in the Appendix \ref{apx:addresults}. 
\end{description}

\section{Concluding Remarks}

We presented Sparse Cox Subgrouping (SCS) a latent variable approach to recover subgroups of patients that respond differentially to an intervention in the presence of censored time-to-event outcomes. As compared to alternative approaches to learning parsimonious hypotheses in such settings, our proposed model recovered hypotheses with more pronounced treatment effects which we validated on multiple studies for cardiovascular health. 

While powerful in its ability to recover parsimonious subgroups there exists limitations in SCS in its current form. The model is sensitive to proportional hazards and may be ill-specified when the proportional hazards assumptions are violated as is evident in many real world clinical studies \citep{maron2018international, bretthauer2022effect}. Another limitation is that SCS in its current form looks at only a single endpoint (typically death, or a composite of multiple adverse outcome). In practice however real world studies typically involve multiple end-points. We envision that extensions of SCS would allow patient subgrouping across multiple endpoints, leading to discovery of actionable sub-populations that similarly benefit from the intervention under assessment.





\small
\bibliography{ref}

\appendix

\newpage

\normalsize

\section{Additional Details on the ALLHAT and BARI 2D Case Studies}
\label{apx:confounders}

Tables \ref{tab:ALLHAT-feats} and \ref{tab:BARI2D-feats} represent additional confounding variables found in the \textbf{ALLHAT} and \textbf{BARI2D} trials respectively.

\begin{minipage}{0.5\textwidth}
\small
    \centering
    \resizebox{1\textwidth}{!}{
    \begin{tabular}{|r|l|}
    \hline
    \textbf{Name} & \textbf{Description}\\
    \hline
    \textbf{\texttt{ETHNIC}}& Ethnicity\\
    \textbf{\texttt{SEX}}& Sex of Participant\\
    \textbf{\texttt{ESTROGEN}}& Estrogen supplementation \\
    \textbf{\texttt{BLMEDS}}& Antihypertensive treatment\\
    \textbf{\texttt{MISTROKE}}&History of Stroke\\
    \textbf{\texttt{HXCABG}}&History of coronary artery bypass\\
    \textbf{\texttt{STDEPR}}& Prior ST depression/T-wave inversion\\
    \textbf{\texttt{OASCVD}}& Other atherosclerotic cardiovascular disease\\
    \textbf{\texttt{DIABETES}}&Prior history of Diabetes\\
    \textbf{\texttt{HDLLT35}} & HDL cholesterol \textless 35mg/dl; 2x in past 5 years\\
    \textbf{\texttt{LVHECG}}& LVH by ECG in past 2 years\\
    \textbf{\texttt{WALL25}}&LVH by ECG in past 2 years\\
    \textbf{\texttt{LCHD}}& History of CHD at baseline\\
    \textbf{\texttt{CURSMOKE}}& Current smoking status.\\
    \textbf{\texttt{ASPIRIN}}& Aspirin use\\
    \textbf{\texttt{LLT}}&Lipid-lowering trial\\
    \textbf{\texttt{AGE}}& Age upon entry \\
    \textbf{\texttt{BLWGT}}& Weight upon entry\\
    \textbf{\texttt{BLHGT}}& Height upon entry\\
    \textbf{\texttt{BLBMI}}& Body Mass Index upon entry\\
    \textbf{\texttt{BV2SBP}}&Baseline SBP\\
    \textbf{\texttt{BV2DBP}}&Baseline DBP\\
    \textbf{\texttt{APOTAS}}& Baseline serum potassium\\
    \textbf{\texttt{BLGFR}}& Baseline est glomerular filtration rate\\
    \textbf{\texttt{ACHOL}}& Total Cholesterol\\
    \textbf{\texttt{AHDL}}& Baseline HDL Cholesterol\\
    \textbf{\texttt{AFGLUC}}& Baseline fasting serum glucose\\
    \hline
    \end{tabular}}
    \captionof{table}{\small List of confounding variables used for experiments involving the \textbf{ALLHAT} dataset.}
    \label{tab:ALLHAT-feats}
\end{minipage}%
\begin{minipage}{0.5\textwidth}
\small
    \centering
    \resizebox{1\textwidth}{!}{
    \begin{tabular}{|r|l|}
    \hline
    \textbf{Name} & \textbf{Description}\\
    \hline
    \textbf{\texttt{hxmi}}& History of MI\\
    \textbf{\texttt{age}}& Age upon entry\\
    \textbf{\texttt{dbp\_stand}}& Standing diastolic BP\\
    \textbf{\texttt{sbp\_stand}}& Standing systolic BP\\
    \textbf{\texttt{sex}}& Sex\\
    \textbf{\texttt{asp}}& Aspirin use\\
    \textbf{\texttt{smkcat}}& Cigarette smoking category\\
    \textbf{\texttt{betab}}& Beta blocker use\\
    \textbf{\texttt{ccb}}& Calcium blocker use\\
    \textbf{\texttt{hxhtn}}& History of hypertension requiring tx\\
    \textbf{\texttt{insulin}}& Insulin use\\
    \textbf{\texttt{weight}}& Weight (kg) upon entry\\
    \textbf{\texttt{bmi}}& BMI upon entry\\
    \textbf{\texttt{qabn}}& Abnormal Q-Wave\\
    \textbf{\texttt{trig}}& Triglycerides (mg/dl) upon entry\\
    \textbf{\texttt{dmdur}}& Duration of diabetes mellitus\\
    \textbf{\texttt{ablvef}}& Left ventricular ejection fraction \textless 50\% \\
    \textbf{\texttt{race}}& Race\\
    \textbf{\texttt{priorrev}}& Prior revascularization\\
    \textbf{\texttt{hxcva}}& Cerebrovascular accident\\
    \textbf{\texttt{screat}}& Serum creatinine (mg/dl)\\
    \textbf{\texttt{hmg}}& Statin\\
    \textbf{\texttt{hxhypo}}& History of hypoglycemic episode\\
    \textbf{\texttt{hba1c}}& Hemoglobin A1c(\%)\\
    \textbf{\texttt{priorstent}}& Prior stent\\
    \textbf{\texttt{spotass}}& Serum Potassium(mEq/L)\\
    \textbf{\texttt{hispanic}}& Hispanic ethnicity\\
    \textbf{\texttt{tchol}}& Total Cholesterol\\
    \textbf{\texttt{hdl}}& HDL Cholesterol\\
    \textbf{\texttt{insul\_circ}}& Circulating insulin (IU/ml)\\
    \textbf{\texttt{tzd}}& Thiazolidinedione\\
    \textbf{\texttt{ldl}}& LDL Cholesterol\\
    \textbf{\texttt{tabn}}& Abnormal T-waves\\
    \textbf{\texttt{nsgn}}& Nonsublingual nitrate\\
    \textbf{\texttt{sulf}}& Sulfonylurea\\
    \textbf{\texttt{hxchf}}& Histoty of congestive heart failure req tx\\
    \textbf{\texttt{arb}}& Angiotensin receptor blocker\\
    \textbf{\texttt{acr}}& Urine albumin/creatinine ratio mg/g\\
    \textbf{\texttt{diur}}& Diuretic\\
    \textbf{\texttt{apa}}& Anti-platelet\\
    \textbf{\texttt{hxchl}}& Hypercholesterolemia req tx\\
    \textbf{\texttt{acei}}& ACE inhibitor\\
    \textbf{\texttt{abilow}}& Low ABI (\textless = 0.9)\\
    \textbf{\texttt{biguanide}}& Biguanide\\
    \textbf{\texttt{stabn}}& Abnormal ST depression\\
    \hline
    \end{tabular}}
    \captionof{table}{\small List of confounding variables used for experiments involving the \textbf{BARI2D} dataset.}
    \label{tab:BARI2D-feats}
\end{minipage}

\begin{table}[!htb]
    \centering
    \begin{minipage}{0.5\textwidth}
\small
\centering
\textbf{ALLHAT}
\resizebox{1\textwidth}{!}{
    \begin{tabular}{|r|l|}
    \hline
    \textbf{Name} & \textbf{Description}\\
    \hline
    \textbf{\texttt{BV2SBP}}& Baseline Seated Diastolic Pressure\\
    \textbf{\texttt{BLGFR}}& Baseline est Glomerular Filteration Rate\\
    \textbf{\texttt{BLMEDS}}& Antihypertensive Treatment\\
    \textbf{\texttt{CURSMOKE}}& Current Smoking Status\\
    \textbf{\texttt{SEX}}& Sex of Participant\\
    \hline
    \end{tabular}}
\end{minipage}%
\begin{minipage}{0.5\textwidth}
\small
\centering
\textbf{BARI2D}
\resizebox{1\textwidth}{!}{
    \begin{tabular}{|r|l|}
    \hline
    \textbf{Name} & \textbf{Description}\\
    \hline
    \textbf{\texttt{age}}& Age upon entry\\
    \textbf{\texttt{asp}}& Aspirin use\\
    \textbf{\texttt{hxhtn}}& History of hypertension requiring tx\\
    \textbf{\texttt{hxchl}}& Hypercholesterolemia req tx\\
    \textbf{\texttt{priorstent}}& Prior stent\\
    \hline
    \end{tabular}}
\end{minipage}
\caption{List of selected features with sparsity level: \textbf{$||\bm{\theta}||_0 \leq 5$}}
    \label{tab:5feats}
\end{table}

\vspace{3em} 

\begin{table}[!htb]
    \centering
\begin{minipage}{0.5\textwidth}
\small
\centering
\textbf{ALLHAT}
\resizebox{1\textwidth}{!}{
    \begin{tabular}{|r|l|}
    \hline
    \textbf{Name} & \textbf{Description}\\
    \hline
    \textbf{\texttt{BV2SBP}}& Baseline Seated Diastolic Pressure\\
    \textbf{\texttt{BLGFR}}& Baseline est Glomerular Filtration Rate\\
    \textbf{\texttt{BLMEDS}}& Antihypertensive Treatment\\
    \textbf{\texttt{CURSMOKE}}& Current Smoking Status\\
    \textbf{\texttt{SEX}}& Sex of Participant\\
    \textbf{\texttt{ASPIRIN}}& Aspirin Use\\
    \textbf{\texttt{ACHOL}}& Total Cholesterol\\
    \textbf{\texttt{BLWGT}}& Weight upon entry\\
    \textbf{\texttt{BMI}}& Body mass index upon entry\\
    \textbf{\texttt{OASCVD}}& Other atherosclerotic cardiovascular disease\\
    \hline
    \end{tabular}
    }
\end{minipage}%
\begin{minipage}{0.5\textwidth}
\small
\centering
\textbf{BARI2D}
\resizebox{0.9\textwidth}{!}{
    \begin{tabular}{|r|l|}
    \hline
    \textbf{Name} & \textbf{Description}\\
    \hline
    \textbf{\texttt{age}}& Age upon entry\\
    \textbf{\texttt{asp}}& Aspirin use\\
    \textbf{\texttt{hxhtn}}& History of hypertension requiring tx\\
    \textbf{\texttt{hxchl}}& Hypercholesterolemia req tx\\
    \textbf{\texttt{priorstent}}& Prior stent\\
    \textbf{\texttt{acei}}& ACE Inhibitor\\
    \textbf{\texttt{acr}}& Urine albumin/creatinine ratio mg/g\\
        \textbf{\texttt{insul\_circ}}& Circulating insulin\\
    \textbf{\texttt{screat}}& Serum creatinine (mg/dl)\\
    \textbf{\texttt{tchol}}& Total Cholesterol\\

    \hline
    \end{tabular}
    }
\end{minipage}
\caption{List of selected features with sparsity level: \textbf{$||\bm{\theta}||_0 \leq 10$}}
    \label{tab:10feats}
\end{table}

\section{Derivation of the Inference Algorithm}

\label{apx:posterior}

\noindent \textbf{Censored Instances}:
Note that in the case of the censored instances we will condition on the thresholded survival $(T>\vu)$. The the posterior counts thus reduce to:
\begin{align}
 \nonumber \bm{\gamma}^{k} &= \mathbb{P}(Z=k|X=\vx, A=\va, T>\vu) \\
  &= \frac{\mathbb{P}(T>\vt|Z=\vk, X=\vx, A=\va)\bm{p}(Z=\vk|X=\vx)}{\sum_k \mathbb{P}(T>\vt|Z=\vk, X=\vx, A=\va)\mathbb{P}(Z=\vk|X=\vx)}
  \label{eqn:censored}
\end{align}
$$ \text{Here, } \mathbb{P}(T>\vt| Z=\vk, X=\vx, A=\va) \\ \nonumber = \exp \big(- \bm{\Lambda}(t)\big)^{\bm{h}(\vx, \va, k)}$$

\noindent \textbf{Uncensored Instances} The posteriors are $\bm{\gamma}^{k} = \bm{p}_{\bm \theta}(Z=k|X=\vx, T = \vu) $, \\

\noindent Posteriors for the uncensored data are more involved and involve the base hazard $\bm{\lambda}_0(\cdot)$. Posteriors for uncensored data are independent of the base hazard function, $\bm{\lambda}_0(\cdot)$ as,

\begin{align}
 \bm{\gamma}^{k}  = \frac{\cancel{\bm\lambda_0(u)} \bm{h}_k(\vx,\bm{a})  \mS_0(u_i) ^{\bm{h}_k(\vx,\bm{a})  }} { \mathop{\sum}\limits_{k} \cancel{\bm\lambda_0(u)} \bm{h}_k(\vx,\bm{a})  \nonumber \mS_0(u) ^{\bm{h}_k(\vx,\bm{a})  } } = \frac{\bm{h}_k(\vx,\bm{a})  \mS_0(u_i) ^{\bm{h}_k(\vx,\bm{a})  }} { \mathop{\sum}\limits_{k} \bm{h}_k(\vx,\bm{a})  \mS_0(u_i) ^{\bm{h}_k(\vx,\bm{a})  }} \label{eqn:uncensored2}
\end{align}

\noindent Combining Equations \ref{eqn:censored} and  \ref{eqn:uncensored2} we arrive at the following estimate for the posterior counts

\begin{align}
 \nonumber \bm{\gamma}^{k} &= \widehat{\mathbb{P}}(Z=k|X=\vx, A=\va, \vu) \\
 \nonumber &= \frac{\mathbb{P}(\vu|Z=\vk, X=\vx, A=\va)\mathbb{P}(Z=\vk|X=\vx)}{\sum_k \mathbb{P}(\vu|Z=\vk, X=\vx, A=\va)\mathbb{P}(Z=\vk|X=\vx)}\\
 &= \frac{  {\vh(\vx, \va, \vk)}^{\delta_i} \widehat{\mS}_0( \vu )^{\vh(\vx, \va, \vk)} \exp(\bm{\theta_k}^\top \bm{x})}{\mathop{\sum}_{j\in \mathcal{Z}} {\vh(\vx, \va, \vj)}^{\delta_i}\widehat{\mS}_0( \vu )^{\vh(\vx, \va, \vj)} \exp(\bm{\theta_j}^\top \bm{x})} .
\end{align}

\newpage

\section{Additional Results}

\label{apx:addresults}
\begin{figure}[!htbp]
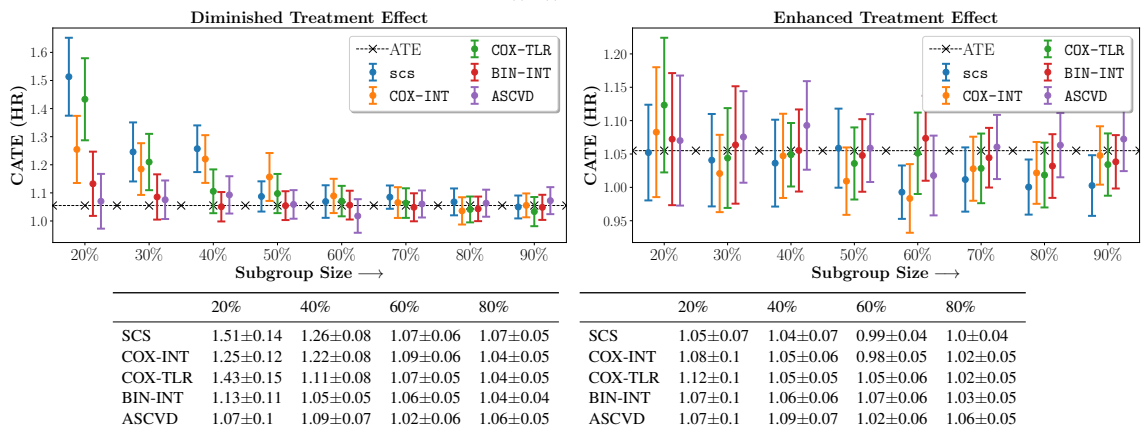
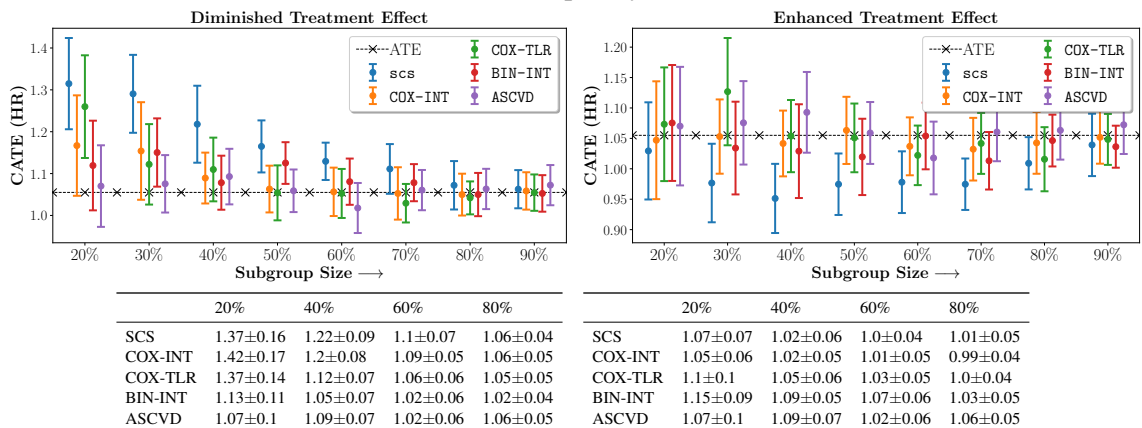

\centering
    \textbf{$||\bm{\theta}||_0 \leq 5$}
    \includegraphics[width=0.5\textwidth]{figures/results/allhat1/allhat1_HR_0_5.pdf}%
    \includegraphics[width=0.5\textwidth]{figures/results/allhat1/allhat1_HR_1_5.pdf}\\
    \begin{minipage}{1\textwidth}
    \centering
    \resizebox{0.8\columnwidth}{!}{%
    \begin{tabular}{lllll}
        \toprule {} &           20\% &           40\% &           60\% &           80\% \\
        \midrule
        SCS     &  1.31$\pm$0.11 &  1.22$\pm$0.09 &  1.13$\pm$0.04 &  1.07$\pm$0.06 \\
        COX-INT &  1.17$\pm$0.12 &  1.09$\pm$0.06 &  1.06$\pm$0.06 &  1.05$\pm$0.05 \\
        COX-TLR &  1.26$\pm$0.12 &  1.11$\pm$0.08 &  1.05$\pm$0.06 &  1.04$\pm$0.04 \\
        BIN-INT &  1.12$\pm$0.11 &  1.08$\pm$0.06 &  1.08$\pm$0.06 &  1.05$\pm$0.05 \\
        ASCVD   &   1.07$\pm$0.1 &  1.09$\pm$0.07 &  1.02$\pm$0.06 &  1.06$\pm$0.05 \\
        \bottomrule
        \end{tabular}
        \quad
        \begin{tabular}{lllll}
        \toprule
        {} &           20\% &           40\% &           60\% &           80\% \\
        \midrule
        SCS     &  1.03$\pm$0.08 &  0.95$\pm$0.06 &  0.98$\pm$0.05 &  1.01$\pm$0.04 \\
        COX-INT &   1.05$\pm$0.1 &  1.04$\pm$0.05 &  1.04$\pm$0.05 &  1.04$\pm$0.05 \\
        COX-TLR &  1.07$\pm$0.09 &  1.05$\pm$0.06 &  1.02$\pm$0.05 &  1.02$\pm$0.05 \\
        BIN-INT &   1.08$\pm$0.1 &  1.03$\pm$0.08 &  1.05$\pm$0.05 &  1.05$\pm$0.04 \\
        ASCVD   &   1.07$\pm$0.1 &  1.09$\pm$0.07 &  1.02$\pm$0.06 &  1.06$\pm$0.05 \\
        \bottomrule
        \end{tabular}
        }
        \vspace{1em} 
        \end{minipage}

    \textbf{$||\bm{\theta}||_0 \leq 10$}
    \includegraphics[width=0.5\textwidth]{figures/results/allhat1/allhat1_HR_0_10.pdf}%
    \includegraphics[width=0.5\textwidth]{figures/results/allhat1/allhat1_HR_1_10.pdf}
    
    \begin{minipage}{1\textwidth}
        \centering
        \resizebox{0.8\columnwidth}{!}{%
            \begin{tabular}{lllll}
\toprule
{} &           20\% &           40\% &           60\% &           80\% \\
\midrule
SCS     &  1.51$\pm$0.14 &  1.26$\pm$0.08 &  1.07$\pm$0.06 &  1.07$\pm$0.05 \\
COX-INT &  1.25$\pm$0.12 &  1.22$\pm$0.08 &  1.09$\pm$0.06 &  1.04$\pm$0.05 \\
COX-TLR &  1.43$\pm$0.15 &  1.11$\pm$0.08 &  1.07$\pm$0.05 &  1.04$\pm$0.05 \\
BIN-INT &  1.13$\pm$0.11 &  1.05$\pm$0.05 &  1.06$\pm$0.05 &  1.04$\pm$0.04 \\
ASCVD   &   1.07$\pm$0.1 &  1.09$\pm$0.07 &  1.02$\pm$0.06 &  1.06$\pm$0.05 \\
\bottomrule
\end{tabular}
        \quad
        \begin{tabular}{lllll}
\toprule
{} &           20\% &           40\% &           60\% &           80\% \\
\midrule
SCS     &  1.05$\pm$0.07 &  1.04$\pm$0.07 &  0.99$\pm$0.04 &   1.0$\pm$0.04 \\
COX-INT &   1.08$\pm$0.1 &  1.05$\pm$0.06 &  0.98$\pm$0.05 &  1.02$\pm$0.05 \\
COX-TLR &   1.12$\pm$0.1 &  1.05$\pm$0.05 &  1.05$\pm$0.06 &  1.02$\pm$0.05 \\
BIN-INT &   1.07$\pm$0.1 &  1.06$\pm$0.06 &  1.07$\pm$0.06 &  1.03$\pm$0.05 \\
ASCVD   &   1.07$\pm$0.1 &  1.09$\pm$0.07 &  1.02$\pm$0.06 &  1.06$\pm$0.05 \\
\bottomrule
\end{tabular}
        
        }
        \vspace{1em} 
        \end{minipage}\\

    No Sparsity\\
   \includegraphics[width=0.5\textwidth]{figures/results/allhat1/allhat1_HR_0_20.pdf}%
    \includegraphics[width=0.5\textwidth]{figures/results/allhat1/allhat1_HR_1_20.pdf}
    
    \begin{minipage}{1\textwidth}
        \centering
        \resizebox{0.8\columnwidth}{!}{%
            \begin{tabular}{lllll}
\toprule
{} &           20\% &           40\% &           60\% &           80\% \\
\midrule
SCS     &  1.37$\pm$0.16 &  1.22$\pm$0.09 &   1.1$\pm$0.07 &  1.06$\pm$0.04 \\
COX-INT &  1.42$\pm$0.17 &   1.2$\pm$0.08 &  1.09$\pm$0.05 &  1.06$\pm$0.05 \\
COX-TLR &  1.37$\pm$0.14 &  1.12$\pm$0.07 &  1.06$\pm$0.06 &  1.05$\pm$0.05 \\
BIN-INT &  1.13$\pm$0.11 &  1.05$\pm$0.07 &  1.02$\pm$0.06 &  1.02$\pm$0.04 \\
ASCVD   &   1.07$\pm$0.1 &  1.09$\pm$0.07 &  1.02$\pm$0.06 &  1.06$\pm$0.05 \\
\bottomrule
\end{tabular}
        \quad
        \begin{tabular}{lllll}
\toprule
{} &           20\% &           40\% &           60\% &           80\% \\
\midrule
SCS     &  1.07$\pm$0.07 &  1.02$\pm$0.06 &   1.0$\pm$0.04 &  1.01$\pm$0.05 \\
COX-INT &  1.05$\pm$0.06 &  1.02$\pm$0.05 &  1.01$\pm$0.05 &  0.99$\pm$0.04 \\
COX-TLR &    1.1$\pm$0.1 &  1.05$\pm$0.06 &  1.03$\pm$0.05 &   1.0$\pm$0.04 \\
BIN-INT &  1.15$\pm$0.09 &  1.09$\pm$0.05 &  1.07$\pm$0.06 &  1.03$\pm$0.05 \\
ASCVD   &   1.07$\pm$0.1 &  1.09$\pm$0.07 &  1.02$\pm$0.06 &  1.06$\pm$0.05 \\
\bottomrule
\end{tabular}
        
        }
        \end{minipage}\\
    
    \caption{\small Conditional Average Treatment Effect in Hazard Ratio versus subgroup size for the latent phenogroups extracted from the \textbf{ALLHAT} study.}
    \label{fig:allhathr}

\end{figure}

\begin{figure}
\centering
    
    \textbf{$||\bm{\theta}||_0 \leq 5$}
    \includegraphics[width=0.5\textwidth]{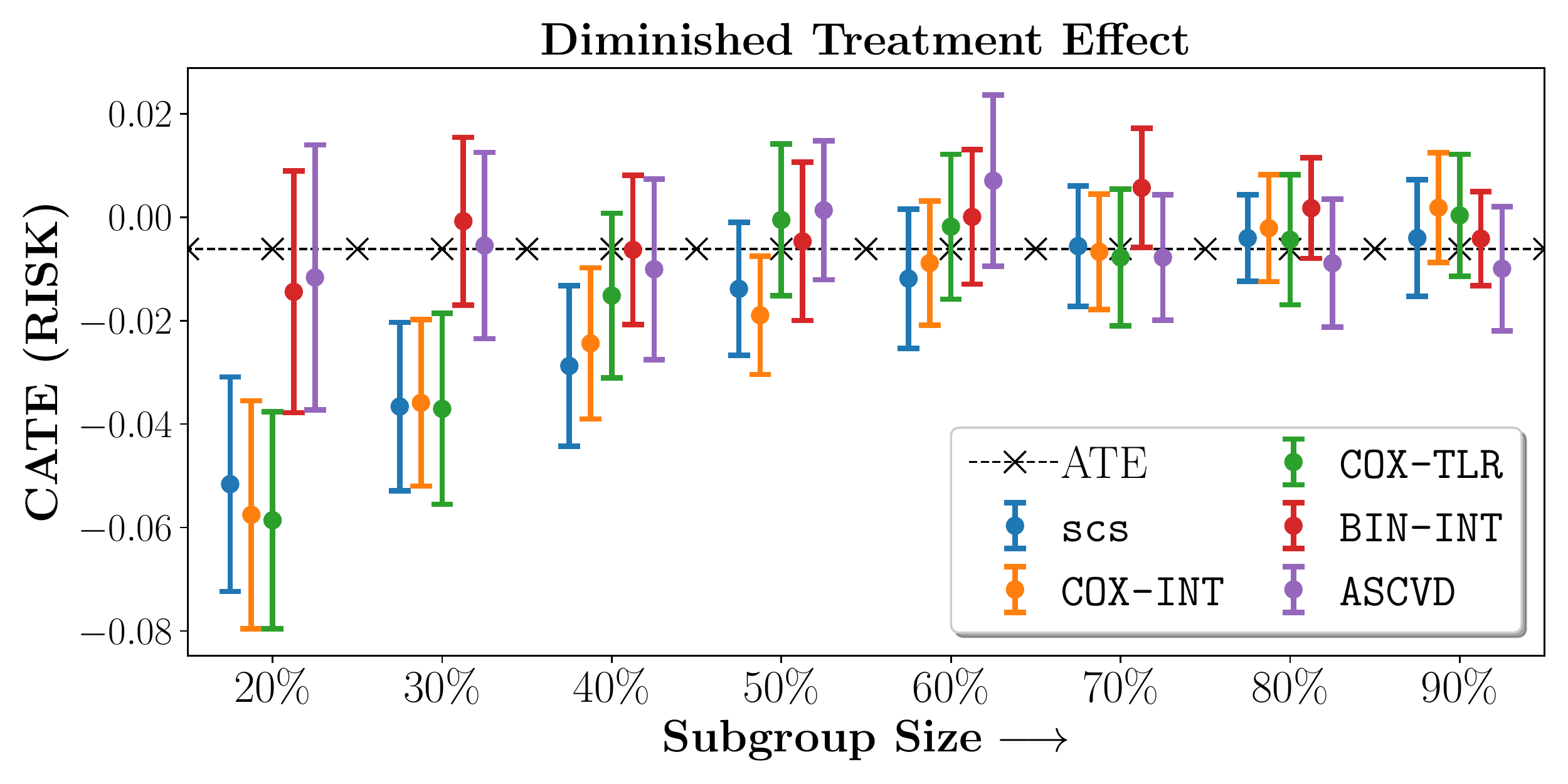}%
    \includegraphics[width=0.5\textwidth]{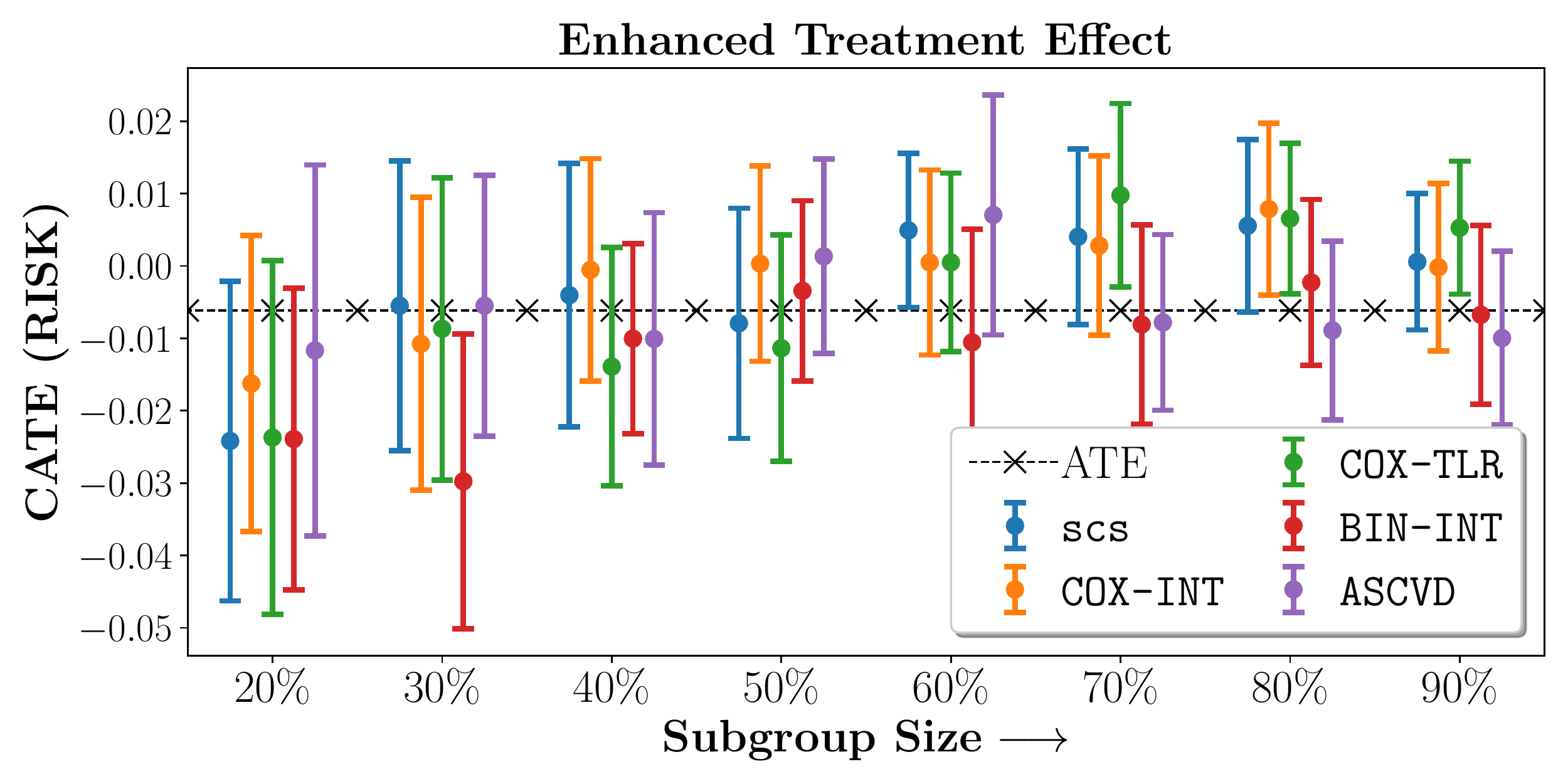}
    \begin{minipage}{1\textwidth}
        \centering
        \resizebox{0.8\columnwidth}{!}{%
           \begin{tabular}{lllll}
\toprule
{} &            20\% &            40\% &            60\% &            80\% \\
\midrule
SCS     &  -0.05$\pm$0.02 &  -0.03$\pm$0.02 &  -0.01$\pm$0.01 &   -0.0$\pm$0.01 \\
COX-INT &  -0.06$\pm$0.02 &  -0.02$\pm$0.01 &  -0.01$\pm$0.01 &   -0.0$\pm$0.01 \\
COX-TLR &  -0.06$\pm$0.02 &  -0.02$\pm$0.02 &   -0.0$\pm$0.01 &   -0.0$\pm$0.01 \\
BIN-INT &  -0.01$\pm$0.02 &  -0.01$\pm$0.01 &    0.0$\pm$0.01 &    0.0$\pm$0.01 \\
ASCVD   &  -0.01$\pm$0.03 &  -0.01$\pm$0.02 &   0.01$\pm$0.02 &  -0.01$\pm$0.01 \\
\bottomrule
\end{tabular}
        \quad
       \begin{tabular}{lllll}
\toprule
{} &            20\% &            40\% &            60\% &            80\% \\
\midrule
SCS     &  -0.02$\pm$0.02 &   -0.0$\pm$0.02 &    0.0$\pm$0.01 &   0.01$\pm$0.01 \\
COX-INT &  -0.02$\pm$0.02 &   -0.0$\pm$0.02 &    0.0$\pm$0.01 &   0.01$\pm$0.01 \\
COX-TLR &  -0.02$\pm$0.02 &  -0.01$\pm$0.02 &    0.0$\pm$0.01 &   0.01$\pm$0.01 \\
BIN-INT &  -0.02$\pm$0.02 &  -0.01$\pm$0.01 &  -0.01$\pm$0.02 &   -0.0$\pm$0.01 \\
ASCVD   &  -0.01$\pm$0.03 &  -0.01$\pm$0.02 &   0.01$\pm$0.02 &  -0.01$\pm$0.01 \\
\bottomrule
\end{tabular}
        
        }
        \vspace{1em} 
        \end{minipage}\\
        
    \textbf{$||\bm{\theta}||_0 \leq 10$}
    \includegraphics[width=0.5\textwidth]{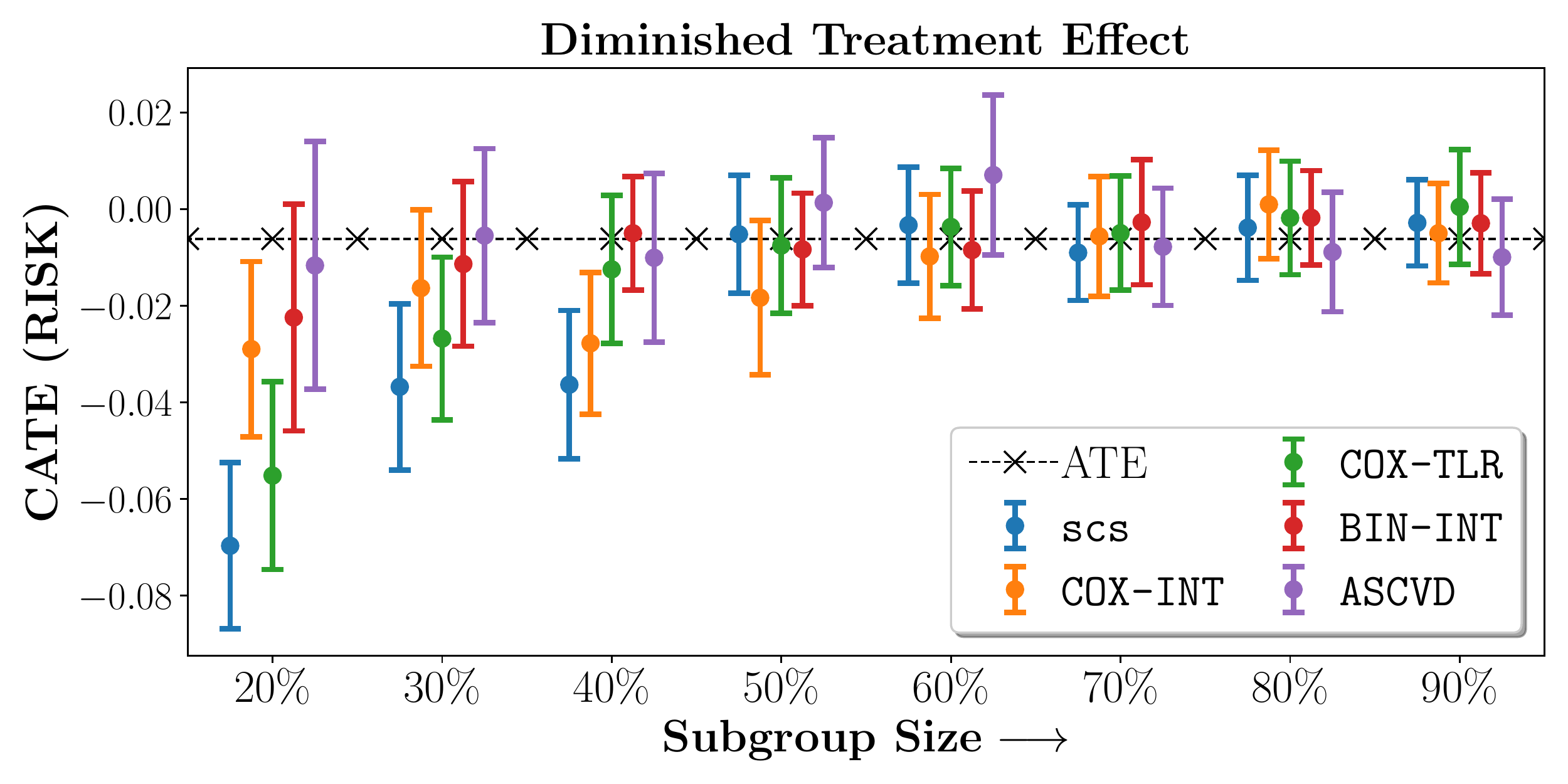}%
    \includegraphics[width=0.5\textwidth]{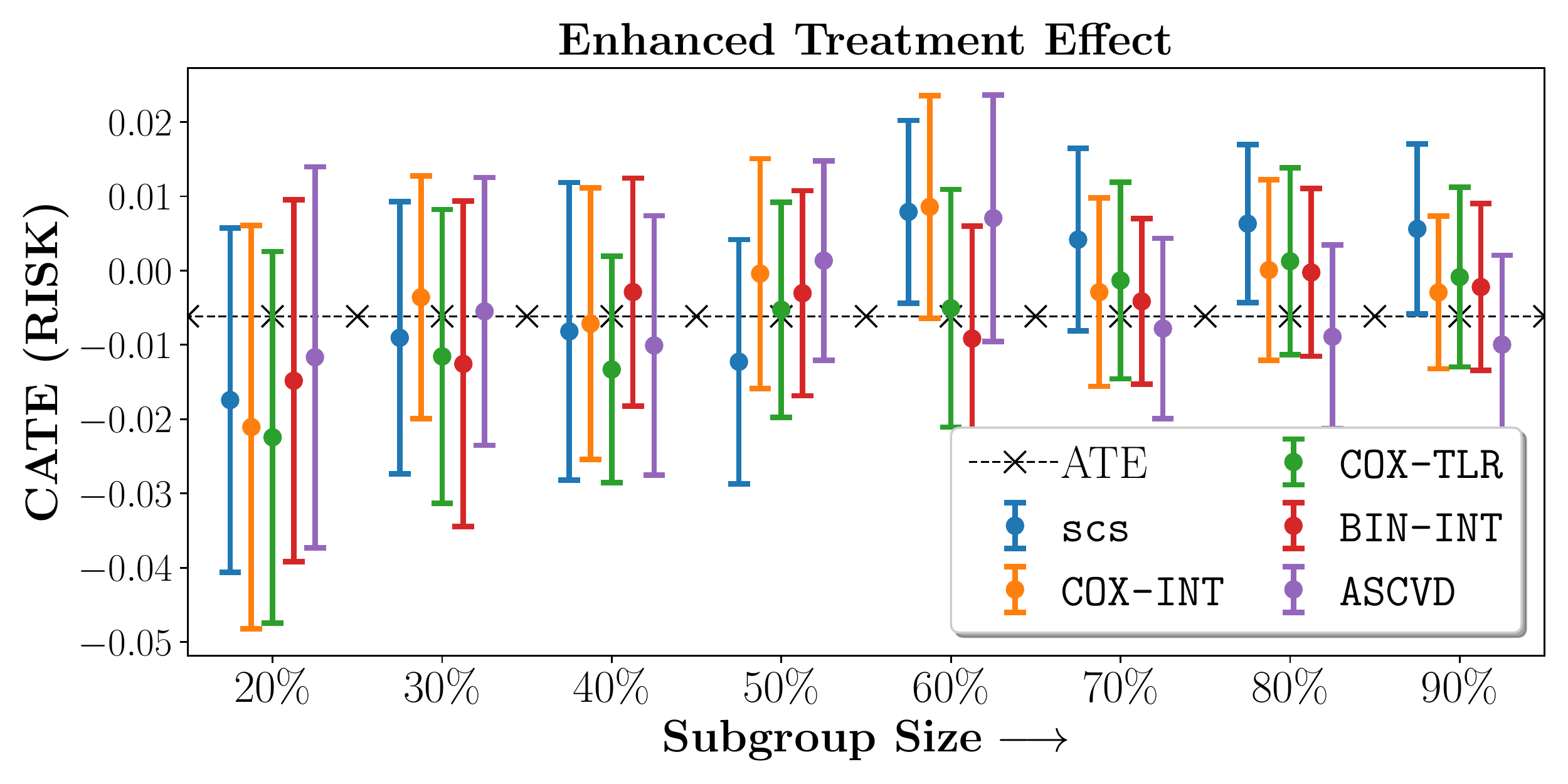}
    
    \begin{minipage}{1\textwidth}
        \centering
        \resizebox{0.8\columnwidth}{!}{%
            \begin{tabular}{lllll}
\toprule
{} &            20\% &            40\% &            60\% &            80\% \\
\midrule
SCS     &  -0.07$\pm$0.02 &  -0.04$\pm$0.02 &   -0.0$\pm$0.01 &   -0.0$\pm$0.01 \\
COX-INT &  -0.03$\pm$0.02 &  -0.03$\pm$0.01 &  -0.01$\pm$0.01 &    0.0$\pm$0.01 \\
COX-TLR &  -0.06$\pm$0.02 &  -0.01$\pm$0.02 &   -0.0$\pm$0.01 &   -0.0$\pm$0.01 \\
BIN-INT &  -0.02$\pm$0.02 &  -0.01$\pm$0.01 &  -0.01$\pm$0.01 &   -0.0$\pm$0.01 \\
ASCVD   &  -0.01$\pm$0.03 &  -0.01$\pm$0.02 &   0.01$\pm$0.02 &  -0.01$\pm$0.01 \\
\bottomrule
\end{tabular}
        \quad
        \begin{tabular}{lllll}
\toprule
{} &            20\% &            40\% &            60\% &            80\% \\
\midrule
SCS     &  -0.02$\pm$0.02 &  -0.01$\pm$0.02 &   0.01$\pm$0.01 &   0.01$\pm$0.01 \\
COX-INT &  -0.02$\pm$0.03 &  -0.01$\pm$0.02 &   0.01$\pm$0.01 &    0.0$\pm$0.01 \\
COX-TLR &  -0.02$\pm$0.03 &  -0.01$\pm$0.02 &  -0.01$\pm$0.02 &    0.0$\pm$0.01 \\
BIN-INT &  -0.01$\pm$0.02 &   -0.0$\pm$0.02 &  -0.01$\pm$0.02 &   -0.0$\pm$0.01 \\
ASCVD   &  -0.01$\pm$0.03 &  -0.01$\pm$0.02 &   0.01$\pm$0.02 &  -0.01$\pm$0.01 \\
\bottomrule
\end{tabular}
        
        }
        \vspace{1em} 
        \end{minipage}\\

    No Sparsity\\
   \includegraphics[width=0.5\textwidth]{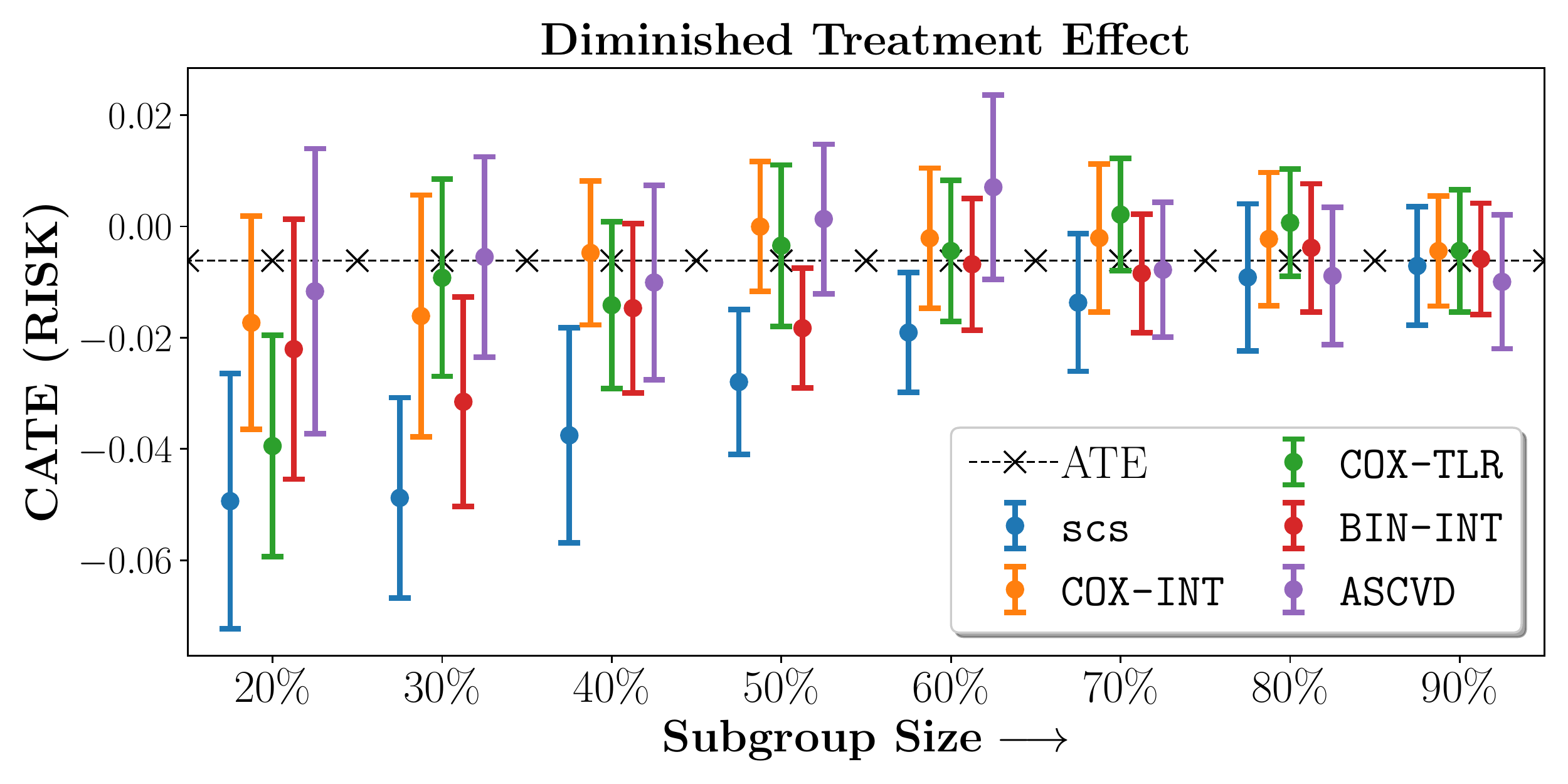}%
    \includegraphics[width=0.5\textwidth]{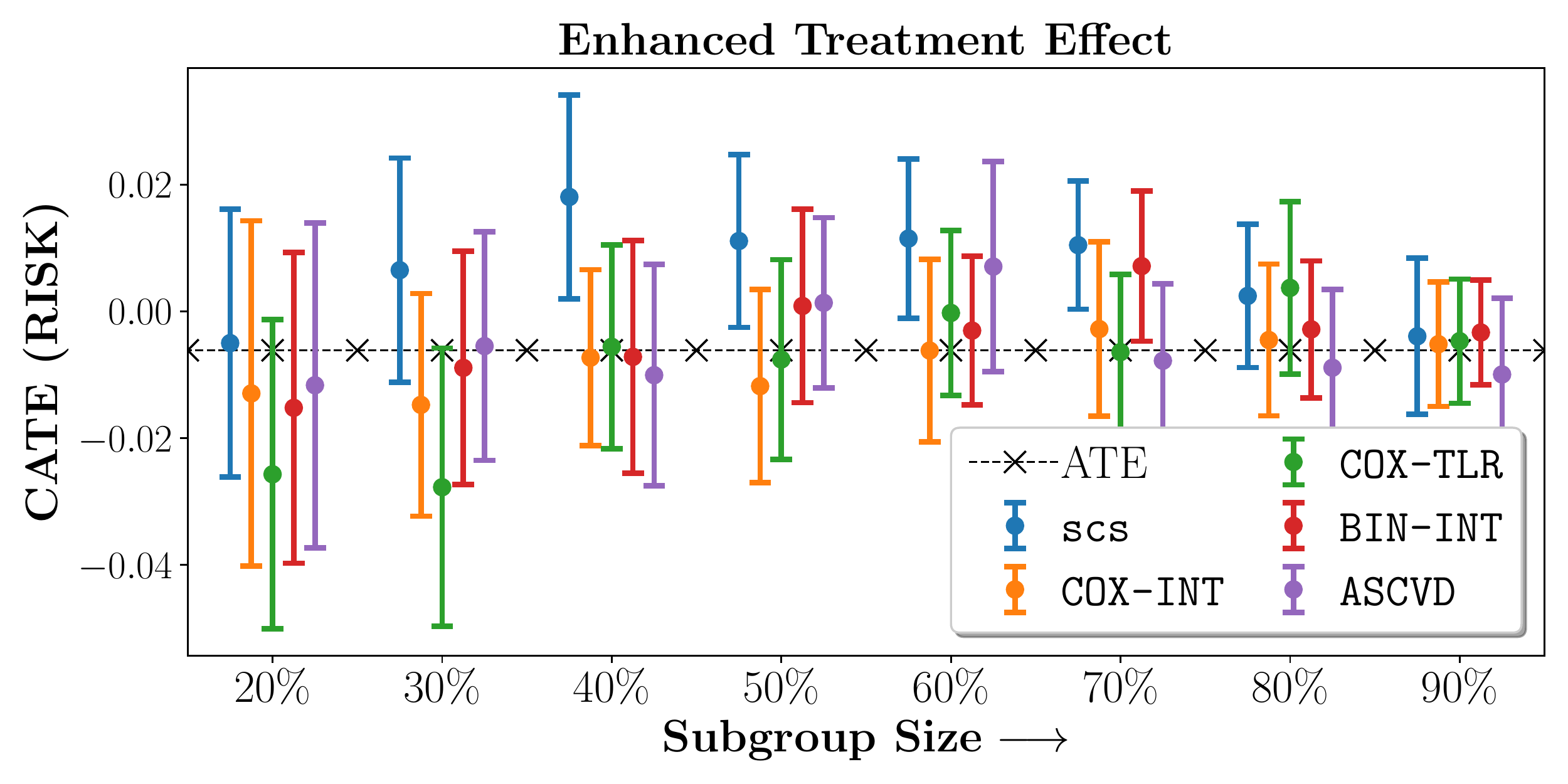}
\begin{minipage}{1\textwidth}
\centering
\resizebox{0.8\columnwidth}{!}{%
\begin{tabular}{lllll}
\toprule
{} &            20\% &            40\% &            60\% &            80\% \\
\midrule
SCS     &  -0.05$\pm$0.02 &  -0.04$\pm$0.02 &  -0.02$\pm$0.01 &  -0.01$\pm$0.01 \\
COX-INT &  -0.02$\pm$0.02 &   -0.0$\pm$0.01 &   -0.0$\pm$0.01 &   -0.0$\pm$0.01 \\
COX-TLR &  -0.04$\pm$0.02 &  -0.01$\pm$0.01 &   -0.0$\pm$0.01 &    0.0$\pm$0.01 \\
BIN-INT &  -0.02$\pm$0.02 &  -0.01$\pm$0.02 &  -0.01$\pm$0.01 &   -0.0$\pm$0.01 \\
ASCVD   &  -0.01$\pm$0.03 &  -0.01$\pm$0.02 &   0.01$\pm$0.02 &  -0.01$\pm$0.01 \\
\bottomrule
\end{tabular}%
        \begin{tabular}{lllll}
\toprule
{} &            20\% &            40\% &            60\% &            80\% \\
\midrule
SCS     &  -0.01$\pm$0.02 &   0.02$\pm$0.02 &   0.01$\pm$0.01 &    0.0$\pm$0.01 \\
COX-INT &  -0.01$\pm$0.03 &  -0.01$\pm$0.01 &  -0.01$\pm$0.01 &   -0.0$\pm$0.01 \\
COX-TLR &  -0.03$\pm$0.02 &  -0.01$\pm$0.02 &   -0.0$\pm$0.01 &    0.0$\pm$0.01 \\
BIN-INT &  -0.02$\pm$0.02 &  -0.01$\pm$0.02 &   -0.0$\pm$0.01 &   -0.0$\pm$0.01 \\
ASCVD   &  -0.01$\pm$0.03 &  -0.01$\pm$0.02 &   0.01$\pm$0.02 &  -0.01$\pm$0.01 \\
\bottomrule
\end{tabular}
        
        }
        \vspace{1em} 
        \end{minipage}\\
    
    \caption{\small  Conditional Average Treatment Effect in Risk versus subgroup size for the latent phenogroups extracted from the \textbf{ALLHAT} study.}
    \label{fig:allhatrisk}

\end{figure}

\begin{figure}
\centering
    
    \textbf{$||\bm{\theta}||_0 \leq 5$}
    \includegraphics[width=0.5\textwidth]{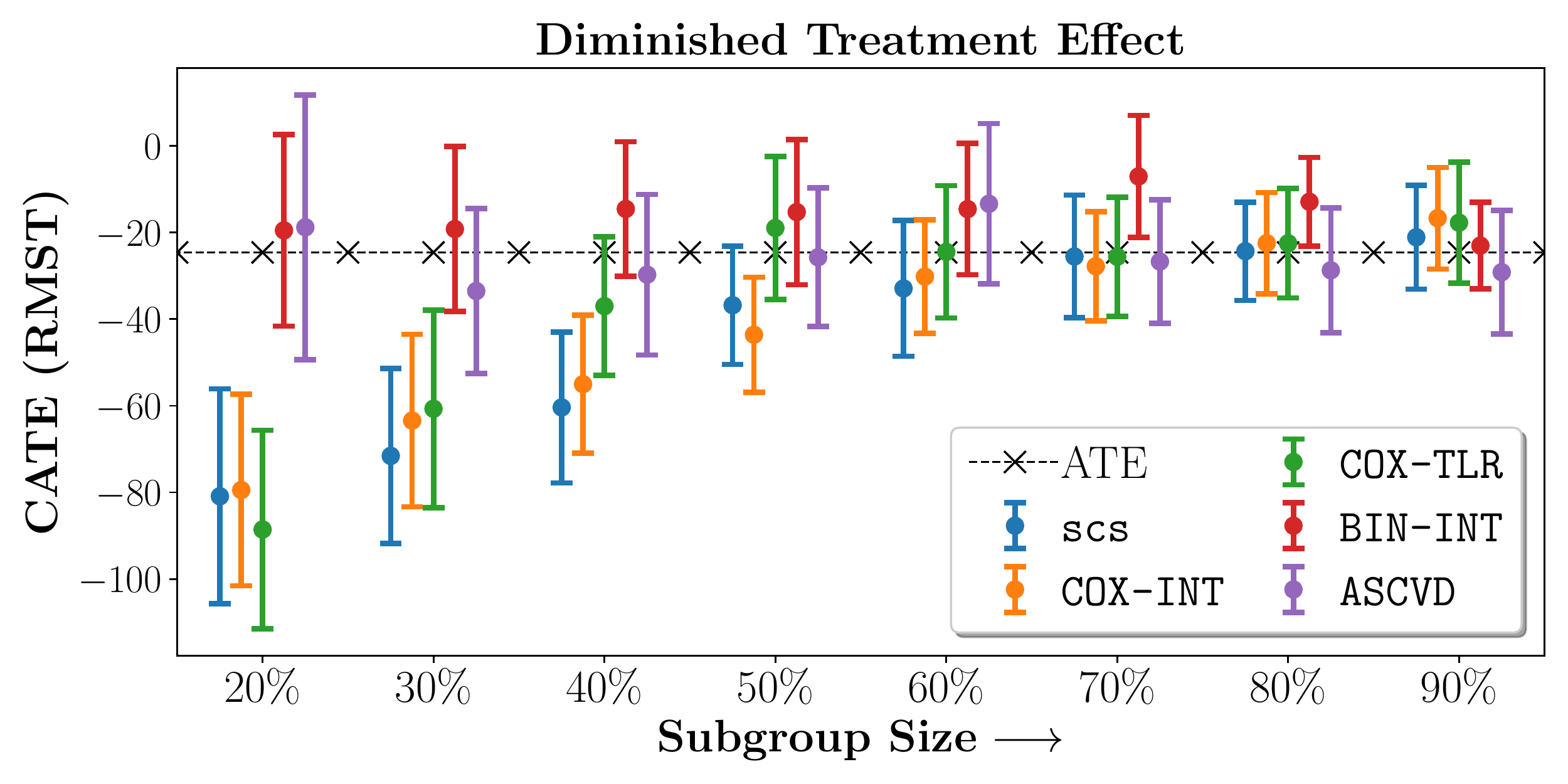}%
    \includegraphics[width=0.5\textwidth]{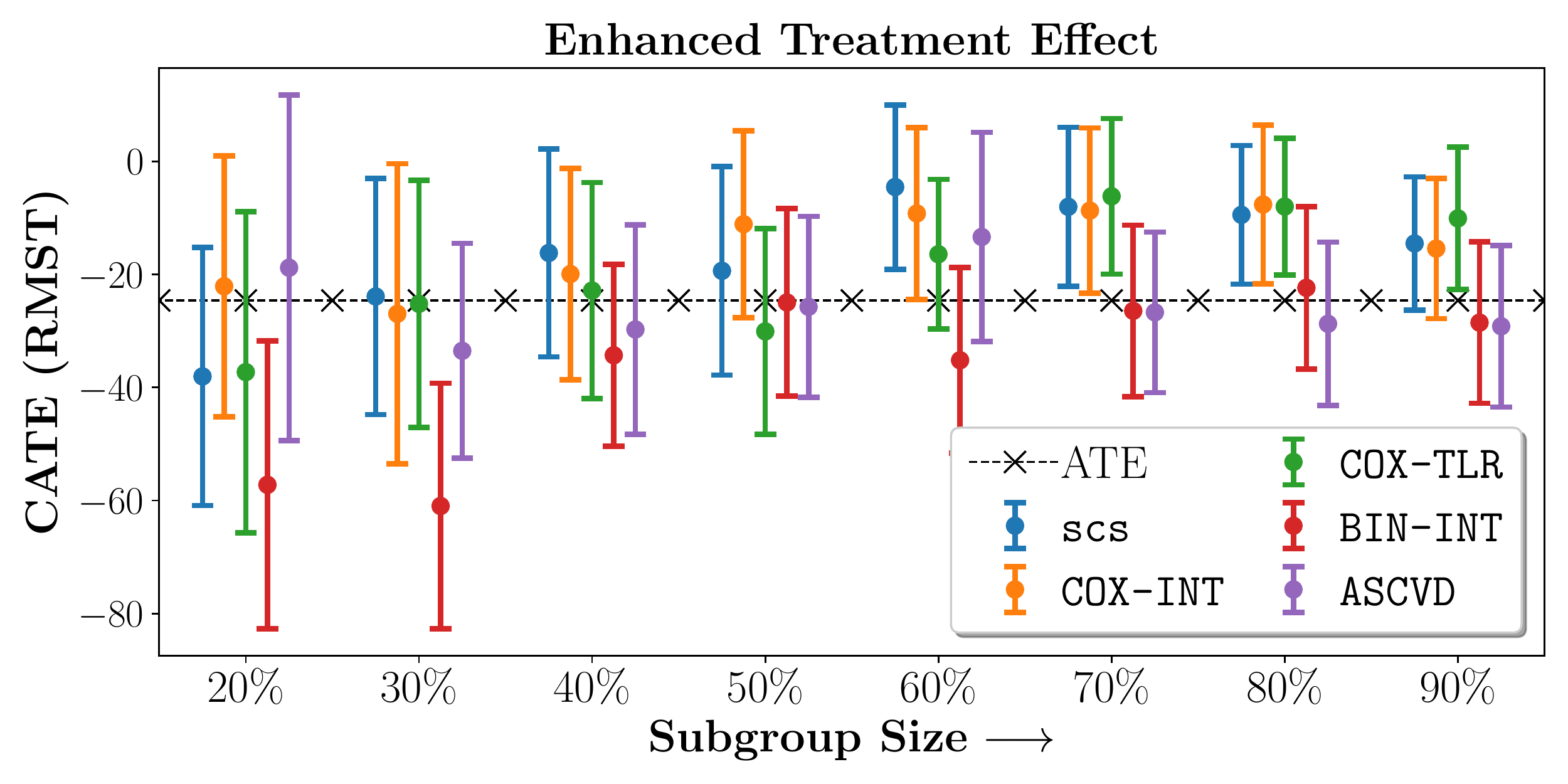}
    \begin{minipage}{1\textwidth}
        \centering
        \resizebox{0.8\columnwidth}{!}{%
            \begin{tabular}{lllll}
\toprule
{} &              20\% &              40\% &              60\% &              80\% \\
\midrule
SCS     &  -80.91$\pm$24.81 &  -60.39$\pm$17.44 &  -32.92$\pm$15.67 &  -24.33$\pm$11.35 \\
COX-INT &   -79.47$\pm$22.1 &  -55.03$\pm$15.93 &   -30.2$\pm$13.11 &  -22.46$\pm$11.68 \\
COX-TLR &  -88.58$\pm$22.92 &  -37.01$\pm$15.97 &  -24.48$\pm$15.29 &  -22.49$\pm$12.61 \\
BIN-INT &  -19.52$\pm$22.12 &  -14.56$\pm$15.51 &  -14.59$\pm$15.14 &  -12.92$\pm$10.25 \\
ASCVD   &  -18.81$\pm$30.57 &  -29.74$\pm$18.53 &  -13.37$\pm$18.52 &  -28.73$\pm$14.45 \\
\bottomrule
\end{tabular}
        \quad
        \begin{tabular}{lllll}
\toprule
{} &              20\% &              40\% &              60\% &              80\% \\
\midrule
SCS     &  -38.04$\pm$22.85 &   -16.19$\pm$18.4 &   -4.54$\pm$14.54 &   -9.46$\pm$12.27 \\
COX-INT &  -22.11$\pm$23.05 &  -19.93$\pm$18.72 &   -9.21$\pm$15.22 &   -7.63$\pm$14.01 \\
COX-TLR &  -37.29$\pm$28.39 &   -22.86$\pm$19.1 &  -16.42$\pm$13.25 &    -8.0$\pm$12.11 \\
BIN-INT &  -57.22$\pm$25.47 &   -34.3$\pm$16.07 &  -35.18$\pm$16.41 &  -22.37$\pm$14.36 \\
ASCVD   &  -18.81$\pm$30.57 &  -29.74$\pm$18.53 &  -13.37$\pm$18.52 &  -28.73$\pm$14.45 \\
\bottomrule
\end{tabular}
        
        }
        \vspace{1em} 
        \end{minipage}\\
        
    \textbf{$||\bm{\theta}||_0 \leq 10$}
    \includegraphics[width=0.5\textwidth]{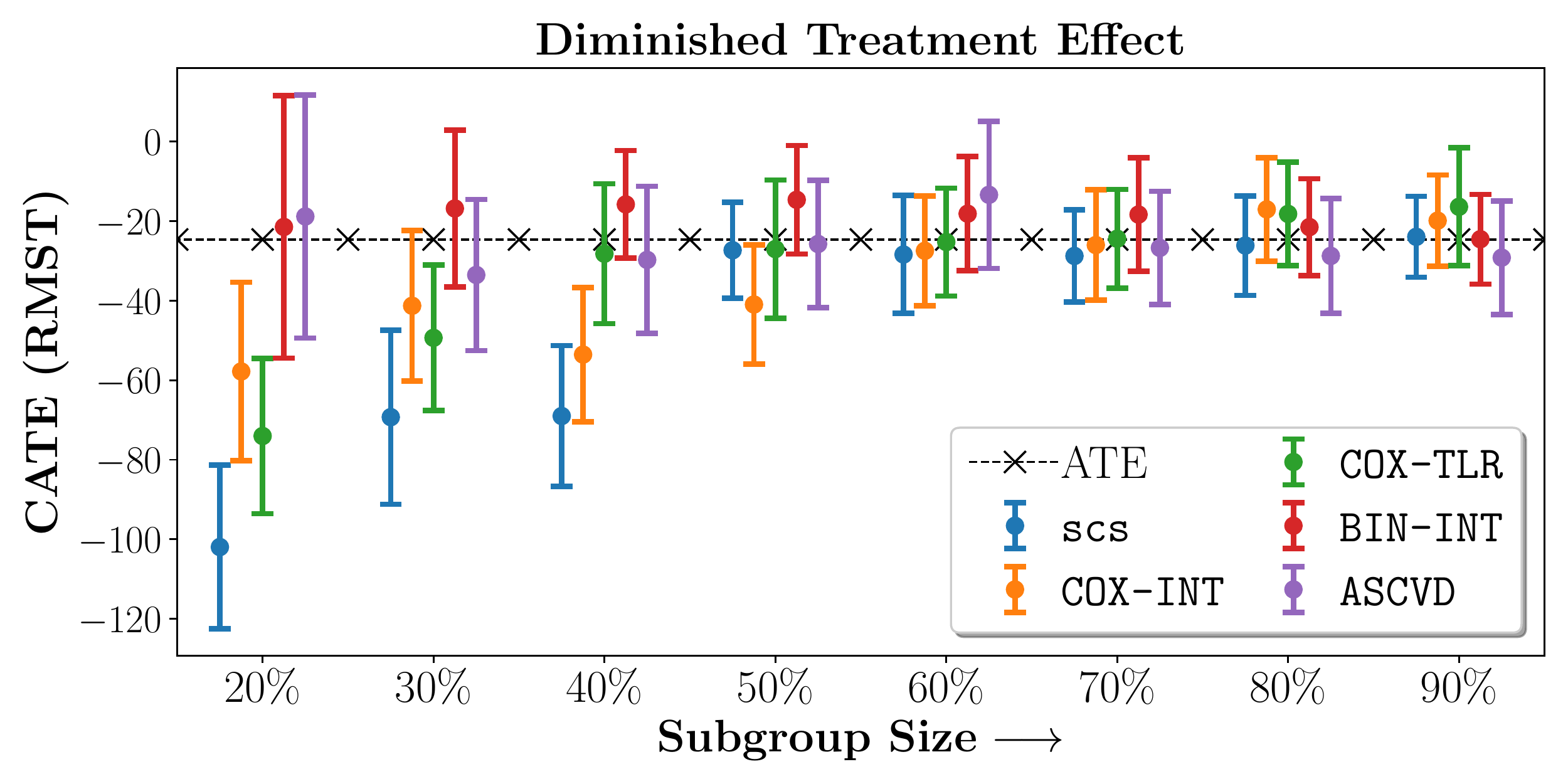}%
    \includegraphics[width=0.5\textwidth]{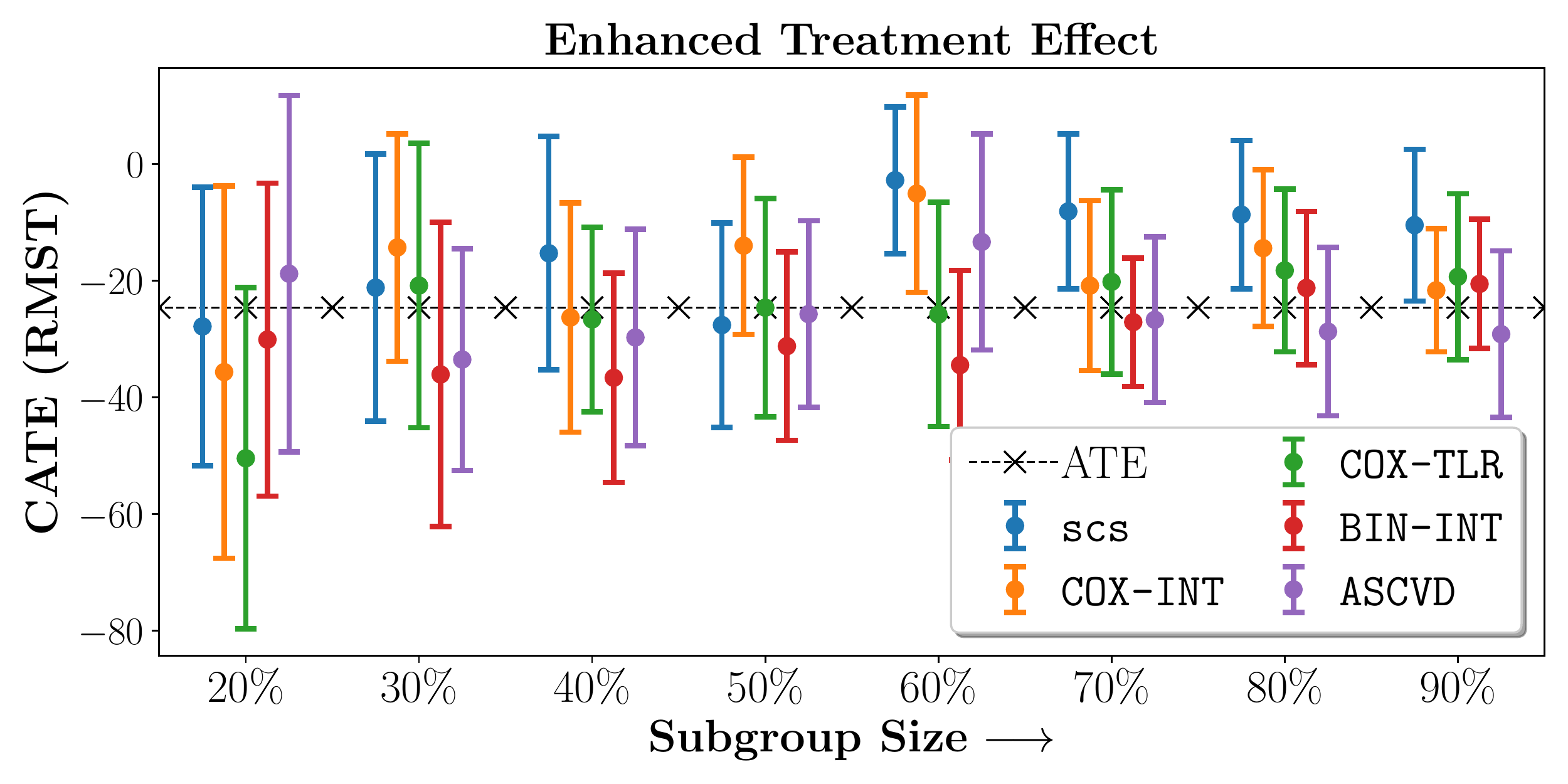}
    
    \begin{minipage}{1\textwidth}
        \centering
        \resizebox{0.8\columnwidth}{!}{%
            \begin{tabular}{lllll}
\toprule
{} &               20\% &              40\% &              60\% &              80\% \\
\midrule
SCS     &  -101.97$\pm$20.57 &  -69.01$\pm$17.72 &  -28.38$\pm$14.82 &  -26.13$\pm$12.52 \\
COX-INT &    -57.8$\pm$22.46 &  -53.56$\pm$16.89 &  -27.46$\pm$13.79 &  -17.04$\pm$13.01 \\
COX-TLR &   -74.04$\pm$19.52 &   -28.19$\pm$17.6 &  -25.25$\pm$13.57 &  -18.17$\pm$13.06 \\
BIN-INT &    -21.45$\pm$33.0 &  -15.78$\pm$13.54 &  -18.12$\pm$14.34 &  -21.48$\pm$12.17 \\
ASCVD   &   -18.81$\pm$30.57 &  -29.74$\pm$18.53 &  -13.37$\pm$18.52 &  -28.73$\pm$14.45 \\
\bottomrule
\end{tabular}
        \quad
        \begin{tabular}{lllll}
\toprule
{} &              20\% &              40\% &              60\% &              80\% \\
\midrule
SCS     &  -27.85$\pm$23.85 &    -15.3$\pm$20.0 &   -2.78$\pm$12.58 &   -8.69$\pm$12.72 \\
COX-INT &  -35.66$\pm$31.92 &  -26.32$\pm$19.65 &   -5.07$\pm$16.91 &  -14.43$\pm$13.43 \\
COX-TLR &  -50.44$\pm$29.24 &   -26.65$\pm$15.8 &   -25.8$\pm$19.21 &  -18.25$\pm$13.94 \\
BIN-INT &   -30.1$\pm$26.83 &  -36.65$\pm$17.93 &  -34.47$\pm$16.25 &  -21.25$\pm$13.13 \\
ASCVD   &  -18.81$\pm$30.57 &  -29.74$\pm$18.53 &  -13.37$\pm$18.52 &  -28.73$\pm$14.45 \\
\bottomrule
\end{tabular}
        
        }
        \vspace{1em} 
        \end{minipage}\\

    No Sparsity\\
   \includegraphics[width=0.5\textwidth]{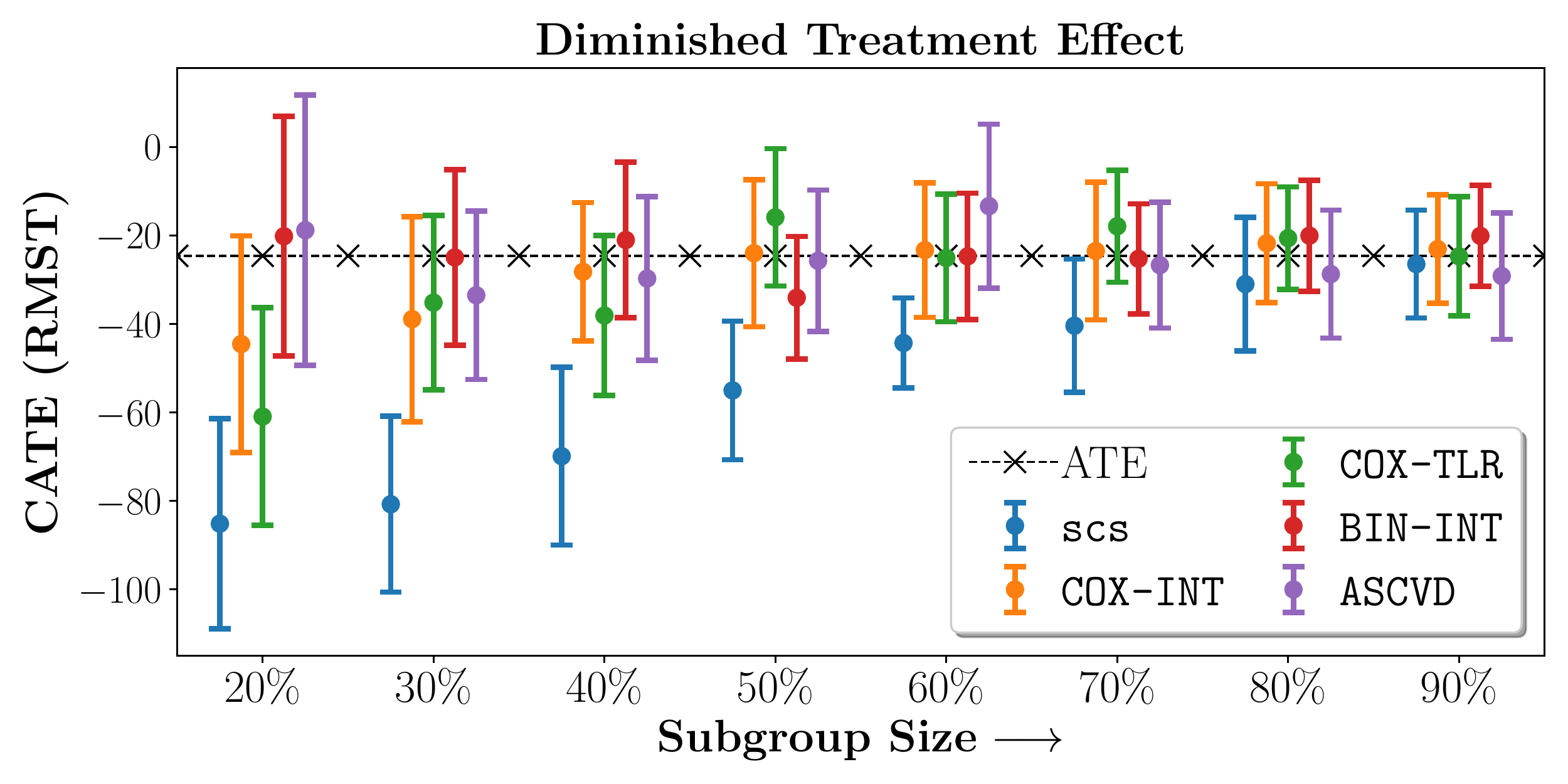}%
    \includegraphics[width=0.5\textwidth]{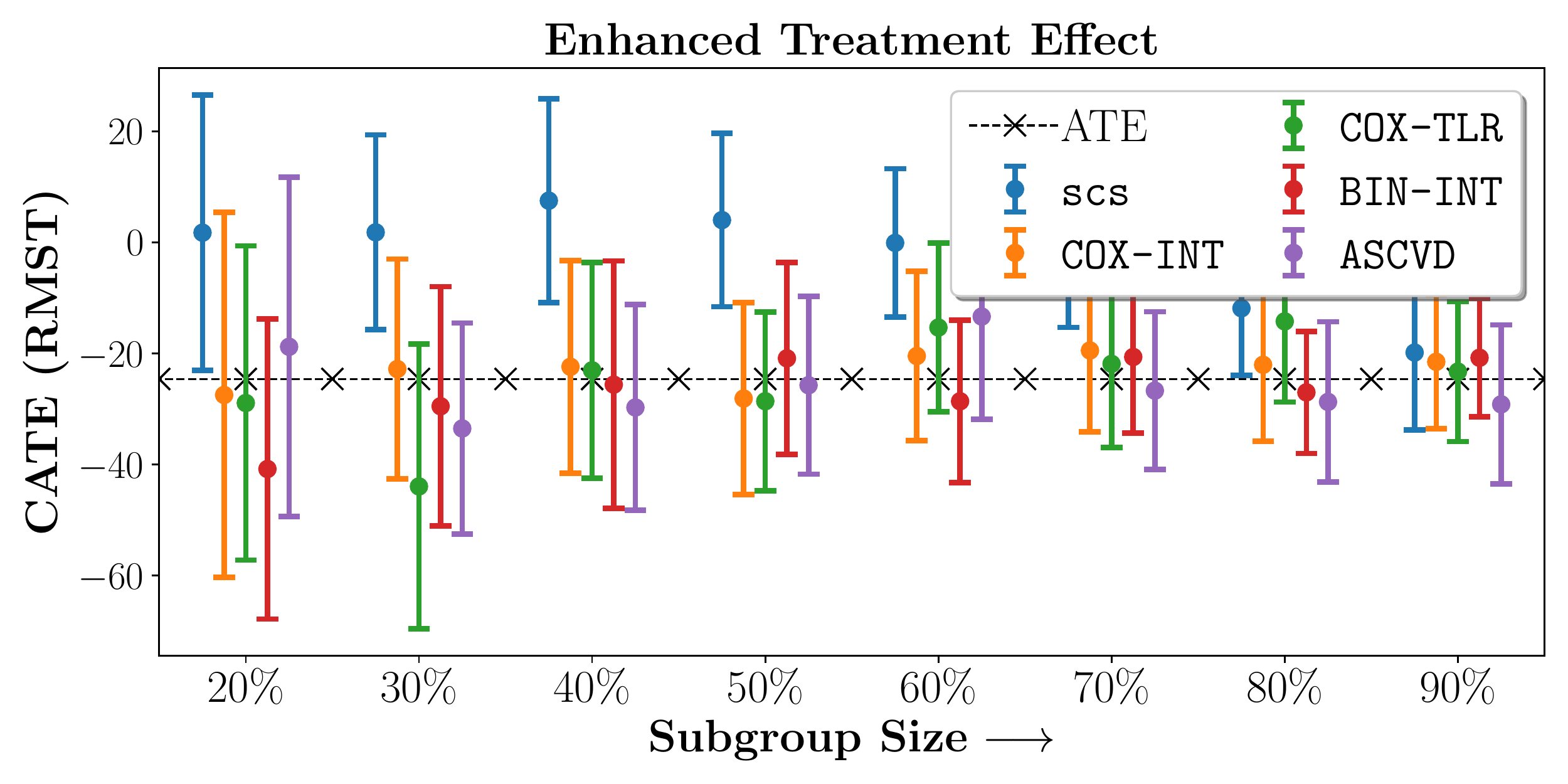}
    
    \begin{minipage}{1\textwidth}
        \centering
        \resizebox{0.8\columnwidth}{!}{%
            \begin{tabular}{lllll}
\toprule
{} &              20\% &              40\% &              60\% &              80\% \\
\midrule
SCS     &  -85.16$\pm$23.76 &    -69.9$\pm$20.1 &   -44.31$\pm$10.2 &  -31.01$\pm$15.13 \\
COX-INT &  -44.56$\pm$24.49 &  -28.22$\pm$15.64 &   -23.32$\pm$15.2 &  -21.78$\pm$13.45 \\
COX-TLR &  -60.94$\pm$24.58 &  -38.11$\pm$18.07 &  -25.09$\pm$14.44 &  -20.61$\pm$11.56 \\
BIN-INT &  -20.17$\pm$27.07 &  -21.04$\pm$17.57 &  -24.72$\pm$14.28 &  -20.06$\pm$12.56 \\
ASCVD   &  -18.81$\pm$30.57 &  -29.74$\pm$18.53 &  -13.37$\pm$18.52 &  -28.73$\pm$14.45 \\
\bottomrule
\end{tabular}
        \quad
        \begin{tabular}{lllll}
\toprule
{} &              20\% &              40\% &              60\% &              80\% \\
\midrule
SCS     &    1.74$\pm$24.81 &     7.5$\pm$18.37 &    -0.1$\pm$13.34 &  -11.88$\pm$12.05 \\
COX-INT &  -27.49$\pm$32.86 &  -22.41$\pm$19.13 &  -20.47$\pm$15.24 &  -22.05$\pm$13.78 \\
COX-TLR &  -28.94$\pm$28.29 &  -23.05$\pm$19.46 &  -15.34$\pm$15.18 &  -14.23$\pm$14.52 \\
BIN-INT &  -40.82$\pm$27.02 &  -25.63$\pm$22.29 &  -28.62$\pm$14.63 &  -27.02$\pm$11.01 \\
ASCVD   &  -18.81$\pm$30.57 &  -29.74$\pm$18.53 &  -13.37$\pm$18.52 &  -28.73$\pm$14.45 \\
\bottomrule
\end{tabular}
        
        }
        \vspace{1em} 
        \end{minipage}\\
    
    \caption{\small  Conditional Average Treatment Effect in Restricted Mean Survival Time versus subgroup size for the latent phenogroups extracted from the \textbf{ALLHAT} study.}
    \label{fig:allhatrmst}

\end{figure}

\begin{figure}
    \centering
    
    \textbf{$||\bm{\theta}||_0 \leq 5$}
    \includegraphics[width=0.5\textwidth]{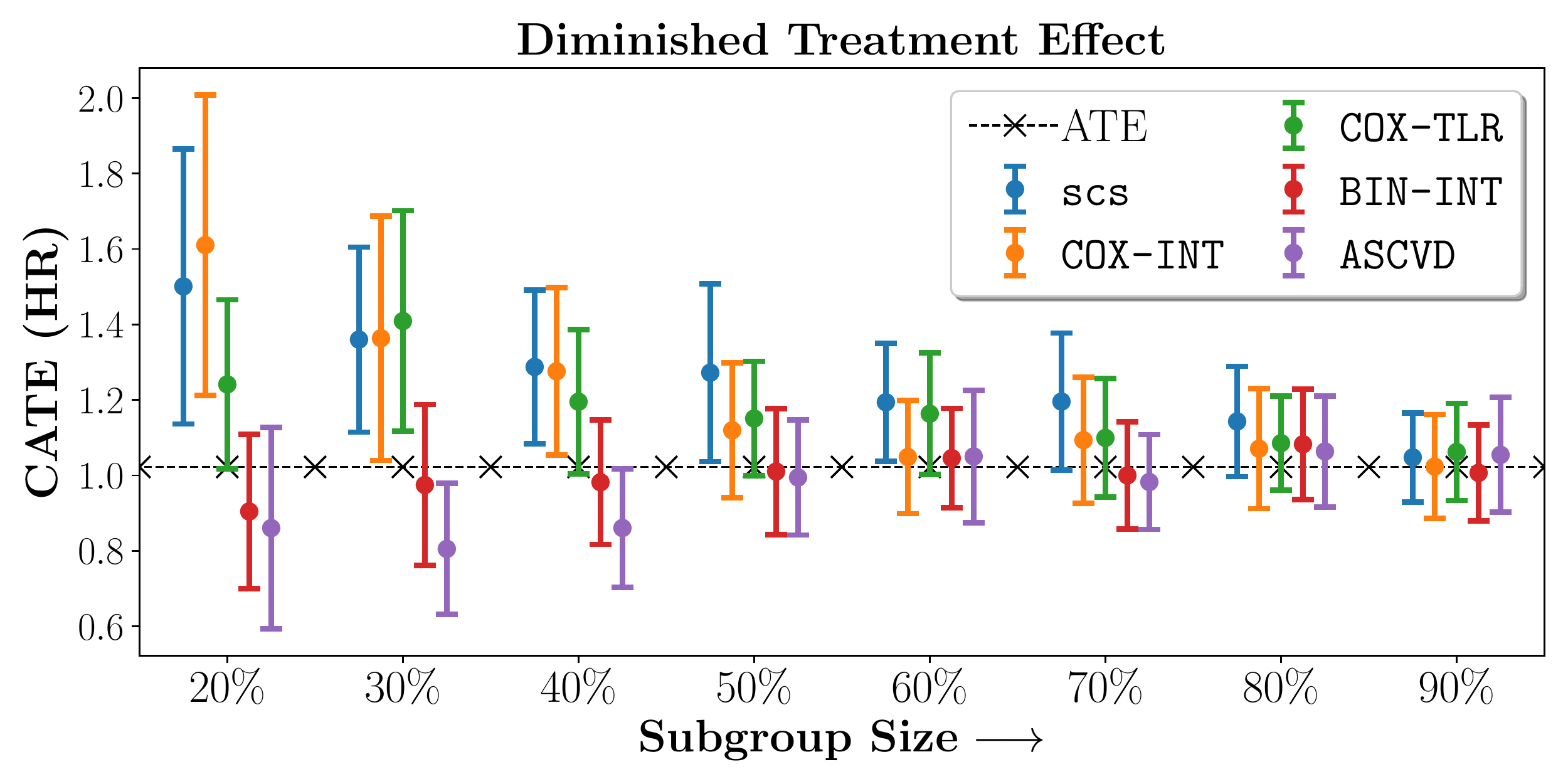}%
    \includegraphics[width=0.5\textwidth]{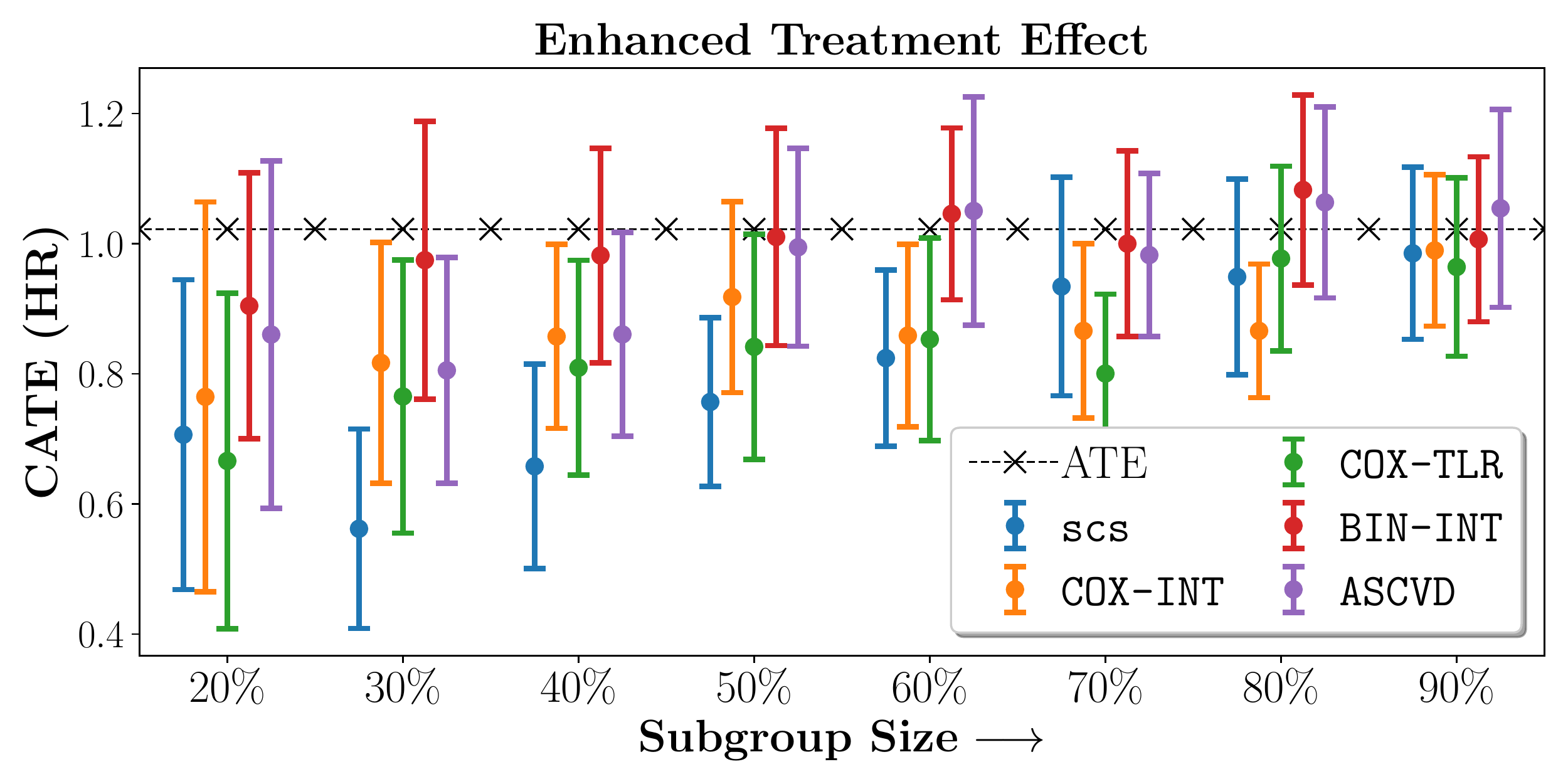}
    \begin{minipage}{1\textwidth}
        \centering
        \resizebox{0.8\columnwidth}{!}{%
            \begin{tabular}{lllll}
\toprule
{} &           20\% &           40\% &           60\% &           80\% \\
\midrule
SCS     &   1.5$\pm$0.36 &   1.29$\pm$0.2 &  1.19$\pm$0.16 &  1.14$\pm$0.15 \\
COX-INT &   1.61$\pm$0.4 &  1.28$\pm$0.22 &  1.05$\pm$0.15 &  1.07$\pm$0.16 \\
COX-TLR &  1.24$\pm$0.22 &   1.2$\pm$0.19 &  1.16$\pm$0.16 &  1.09$\pm$0.13 \\
BIN-INT &    0.9$\pm$0.2 &  0.98$\pm$0.16 &  1.05$\pm$0.13 &  1.08$\pm$0.15 \\
ASCVD   &  0.86$\pm$0.27 &  0.86$\pm$0.16 &  1.05$\pm$0.18 &  1.06$\pm$0.15 \\
\bottomrule
\end{tabular}
        \quad
        \begin{tabular}{lllll}
\toprule
{} &           20\% &           40\% &           60\% &           80\% \\
\midrule
SCS     &  0.71$\pm$0.24 &  0.66$\pm$0.16 &  0.82$\pm$0.14 &  0.95$\pm$0.15 \\
COX-INT &   0.76$\pm$0.3 &  0.86$\pm$0.14 &  0.86$\pm$0.14 &   0.87$\pm$0.1 \\
COX-TLR &  0.67$\pm$0.26 &  0.81$\pm$0.17 &  0.85$\pm$0.16 &  0.98$\pm$0.14 \\
BIN-INT &    0.9$\pm$0.2 &  0.98$\pm$0.16 &  1.05$\pm$0.13 &  1.08$\pm$0.15 \\
ASCVD   &  0.86$\pm$0.27 &  0.86$\pm$0.16 &  1.05$\pm$0.18 &  1.06$\pm$0.15 \\
\bottomrule
\end{tabular}
        
        }
        \vspace{1em} 
        \end{minipage}\\
        
    \textbf{$||\bm{\theta}||_0 \leq 10$}
    \includegraphics[width=0.5\textwidth]{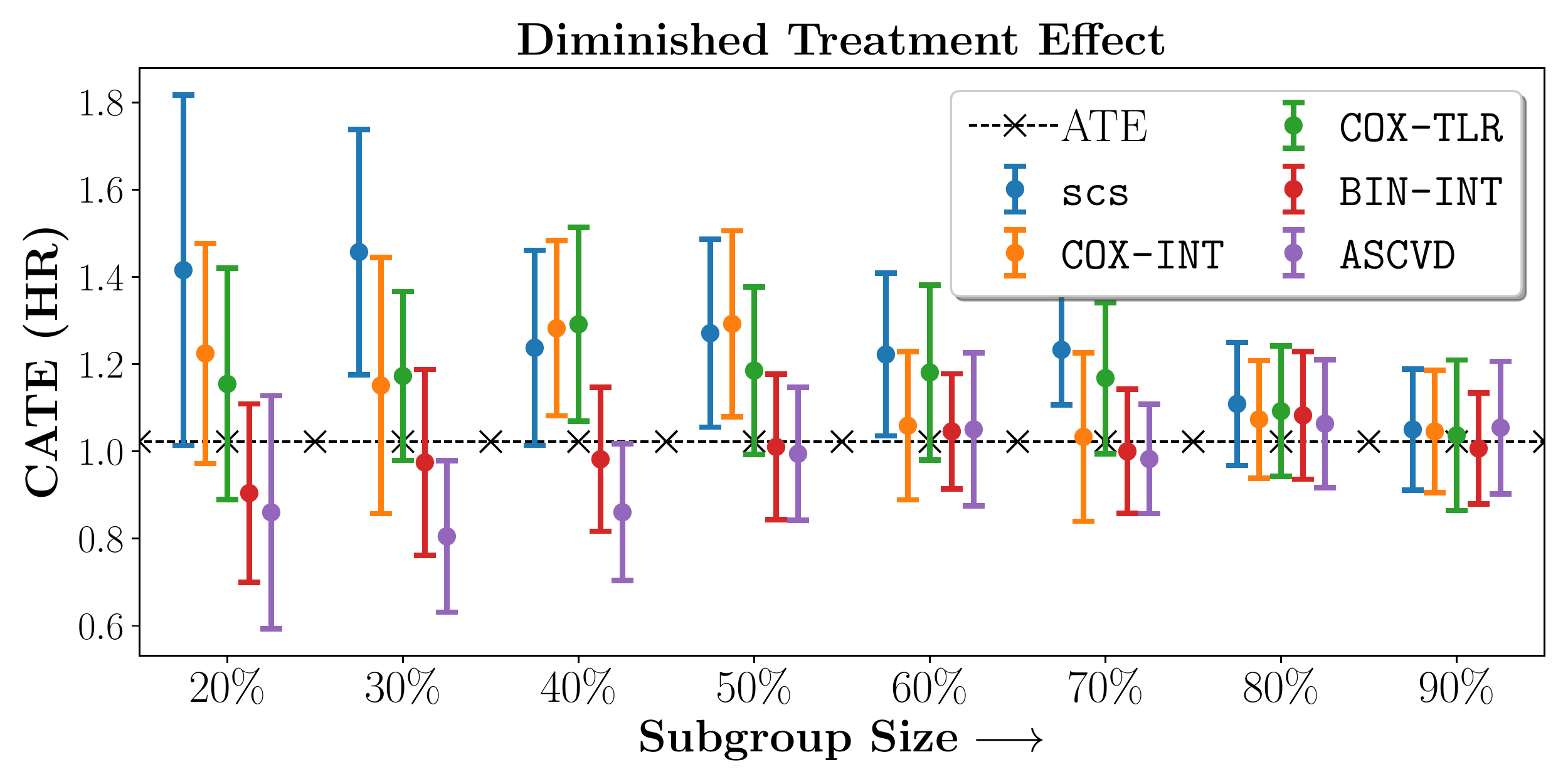}%
    \includegraphics[width=0.5\textwidth]{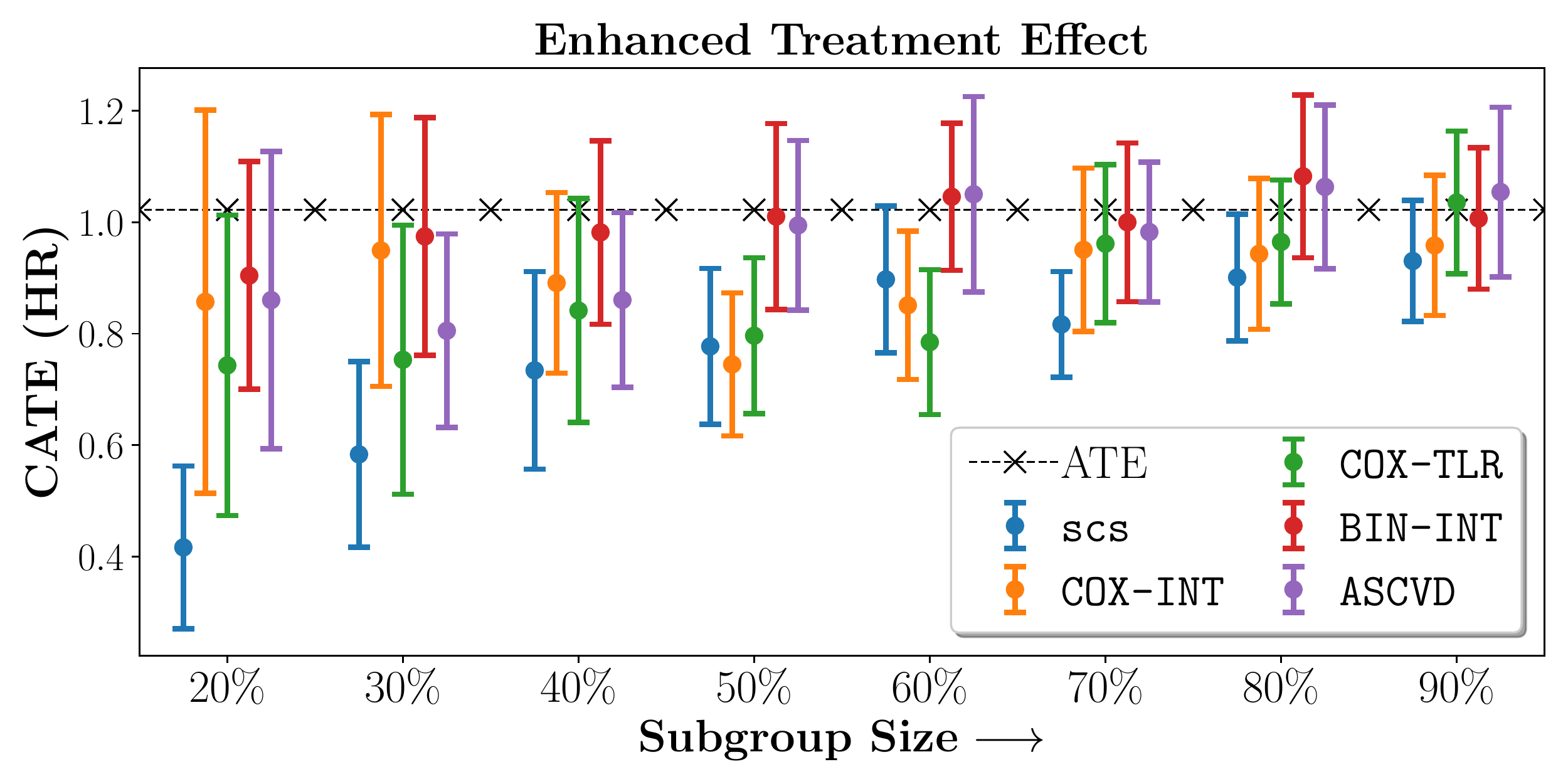}
    
    \begin{minipage}{1\textwidth}
        \centering
        \resizebox{0.8\columnwidth}{!}{%
           \begin{tabular}{lllll}
\toprule
{} &           20\% &           40\% &           60\% &           80\% \\
\midrule
scs     &   1.42$\pm$0.4 &  1.24$\pm$0.22 &  1.22$\pm$0.19 &  1.11$\pm$0.14 \\
COX-INT &  1.22$\pm$0.25 &   1.28$\pm$0.2 &  1.06$\pm$0.17 &  1.07$\pm$0.13 \\
COX-TLR &  1.15$\pm$0.27 &  1.29$\pm$0.22 &   1.18$\pm$0.2 &  1.09$\pm$0.15 \\
BIN-INT &    0.9$\pm$0.2 &  0.98$\pm$0.16 &  1.05$\pm$0.13 &  1.08$\pm$0.15 \\
ASCVD   &  0.86$\pm$0.27 &  0.86$\pm$0.16 &  1.05$\pm$0.18 &  1.06$\pm$0.15 \\
\bottomrule
\end{tabular}
        \quad
        \begin{tabular}{lllll}
\toprule
{} &           20\% &           40\% &           60\% &           80\% \\
\midrule
SCS     &   1.21$\pm$0.3 &  1.35$\pm$0.22 &  1.38$\pm$0.25 &   1.1$\pm$0.16 \\
COX-INT &  1.37$\pm$0.31 &  1.29$\pm$0.25 &  1.15$\pm$0.18 &   1.1$\pm$0.14 \\
COX-TLR &   1.34$\pm$0.4 &  1.18$\pm$0.26 &  1.16$\pm$0.17 &  1.05$\pm$0.14 \\
BIN-INT &    0.9$\pm$0.2 &  0.98$\pm$0.16 &  1.05$\pm$0.13 &  1.08$\pm$0.15 \\
ASCVD   &  0.86$\pm$0.27 &  0.86$\pm$0.16 &  1.05$\pm$0.18 &  1.06$\pm$0.15 \\
\bottomrule
\end{tabular}
        
        }
        \vspace{1em} 
        \end{minipage}\\

    No Sparsity\\
   \includegraphics[width=0.5\textwidth]{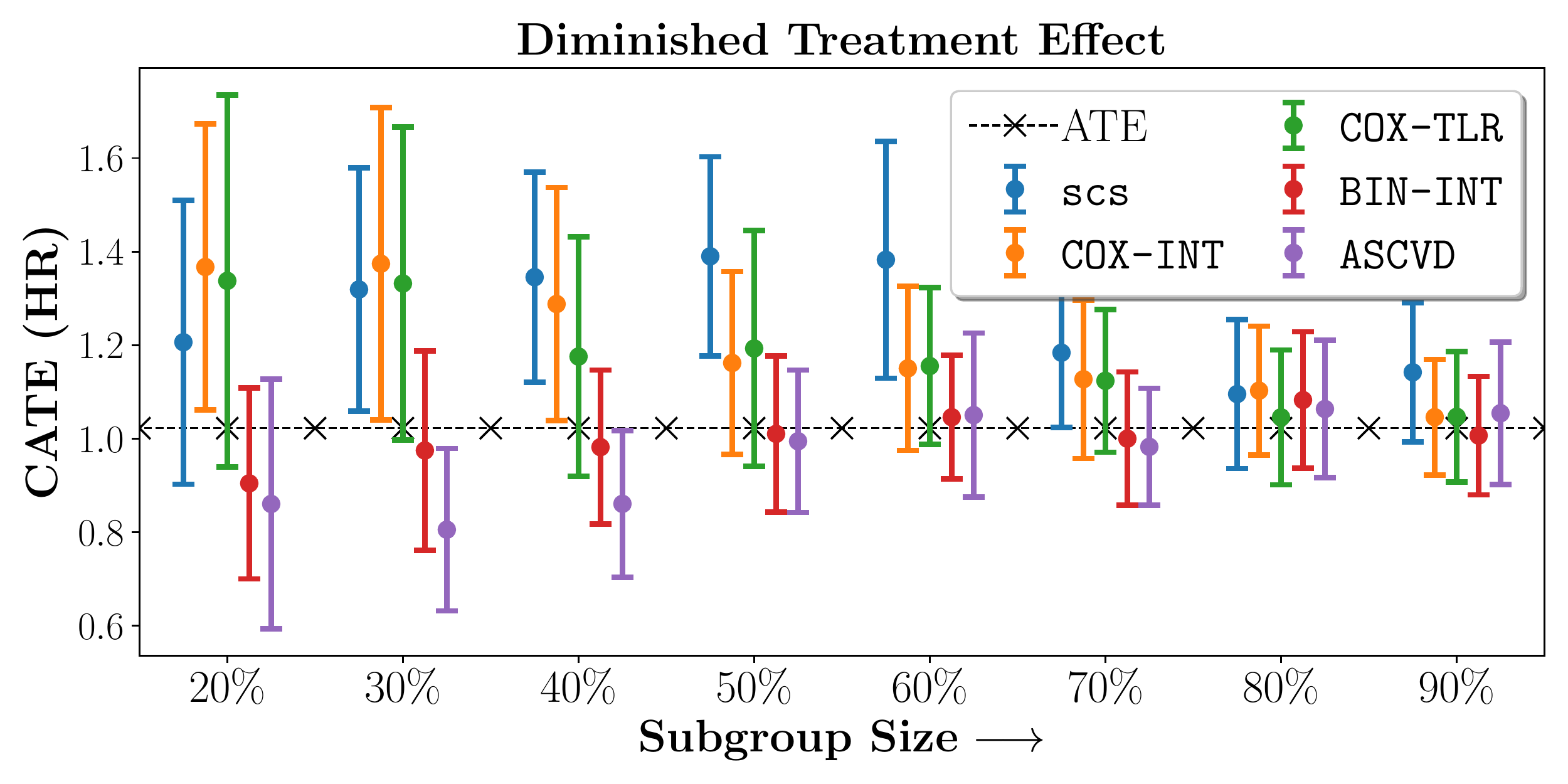}%
    \includegraphics[width=0.5\textwidth]{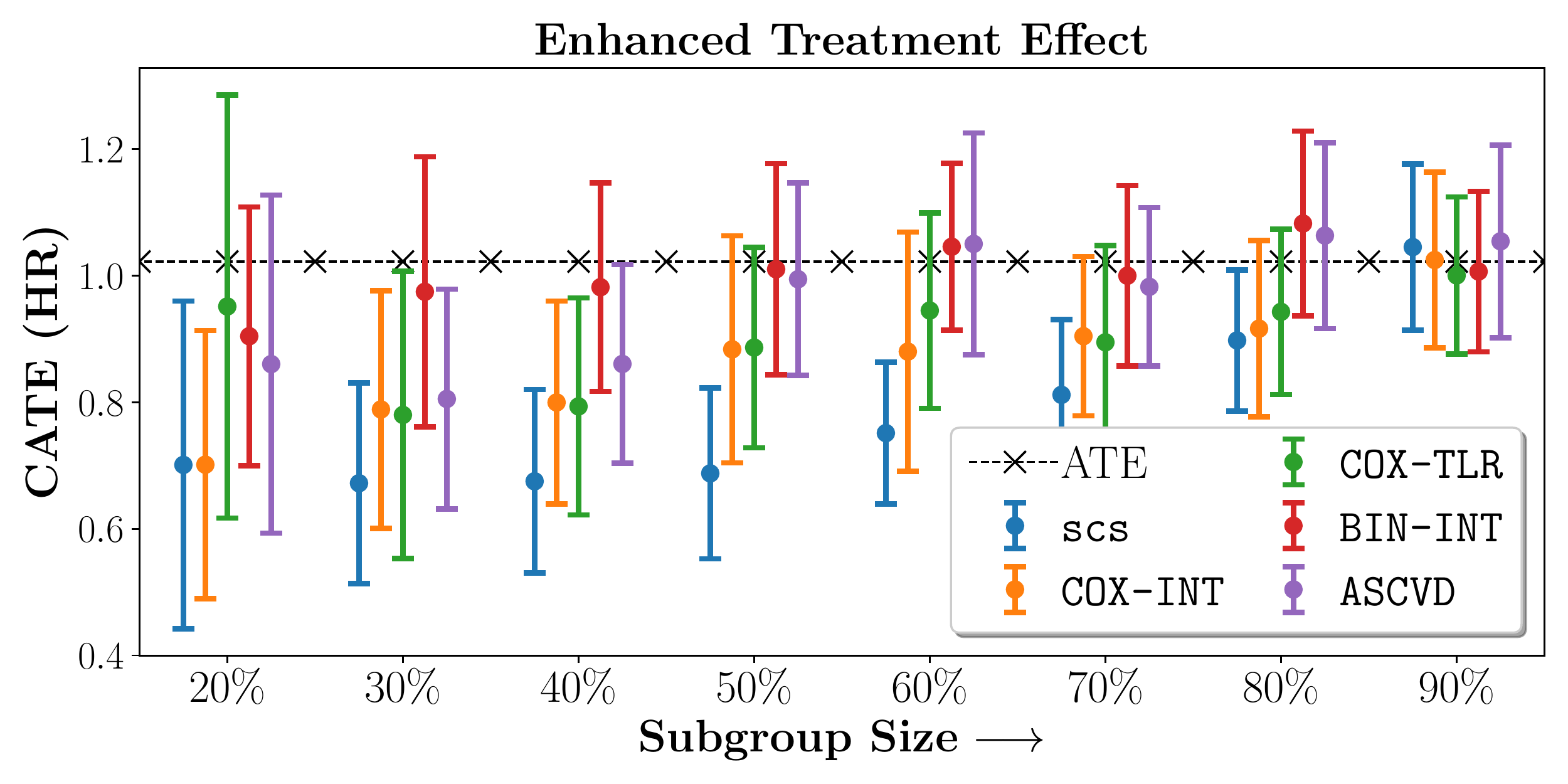}
    
    \begin{minipage}{1\textwidth}
        \centering
        \resizebox{0.8\columnwidth}{!}{%
            \begin{tabular}{lllll}
\toprule
{} &           20\% &           40\% &           60\% &           80\% \\
\midrule
scs     &   1.21$\pm$0.3 &  1.35$\pm$0.22 &  1.38$\pm$0.25 &   1.1$\pm$0.16 \\
COX-INT &  1.37$\pm$0.31 &  1.29$\pm$0.25 &  1.15$\pm$0.18 &   1.1$\pm$0.14 \\
COX-TLR &   1.34$\pm$0.4 &  1.18$\pm$0.26 &  1.16$\pm$0.17 &  1.05$\pm$0.14 \\
BIN-INT &    0.9$\pm$0.2 &  0.98$\pm$0.16 &  1.05$\pm$0.13 &  1.08$\pm$0.15 \\
ASCVD   &  0.86$\pm$0.27 &  0.86$\pm$0.16 &  1.05$\pm$0.18 &  1.06$\pm$0.15 \\
\bottomrule
\end{tabular}
        \quad
        \begin{tabular}{lllll}
\toprule
{} &           20\% &           40\% &           60\% &           80\% \\
\midrule
SCS     &   0.7$\pm$0.26 &  0.68$\pm$0.15 &  0.75$\pm$0.11 &   0.9$\pm$0.11 \\
COX-INT &   0.7$\pm$0.21 &   0.8$\pm$0.16 &  0.88$\pm$0.19 &  0.92$\pm$0.14 \\
COX-TLR &  0.95$\pm$0.33 &  0.79$\pm$0.17 &  0.95$\pm$0.15 &  0.94$\pm$0.13 \\
BIN-INT &    0.9$\pm$0.2 &  0.98$\pm$0.16 &  1.05$\pm$0.13 &  1.08$\pm$0.15 \\
ASCVD   &  0.86$\pm$0.27 &  0.86$\pm$0.16 &  1.05$\pm$0.18 &  1.06$\pm$0.15 \\
\bottomrule
\end{tabular}
        
        }
        \vspace{1em} 
        \end{minipage}\\
    
    \caption{\small  Conditional Average Treatment Effect in Hazard Ratio versus subgroup size for the latent phenogroups extracted from the \textbf{BARI 2D} study.}
    \label{fig:bari2dhr}

\end{figure}

\begin{figure}
    \centering
    
    \textbf{$||\bm{\theta}||_0 \leq 5$}
    \includegraphics[width=0.5\textwidth]{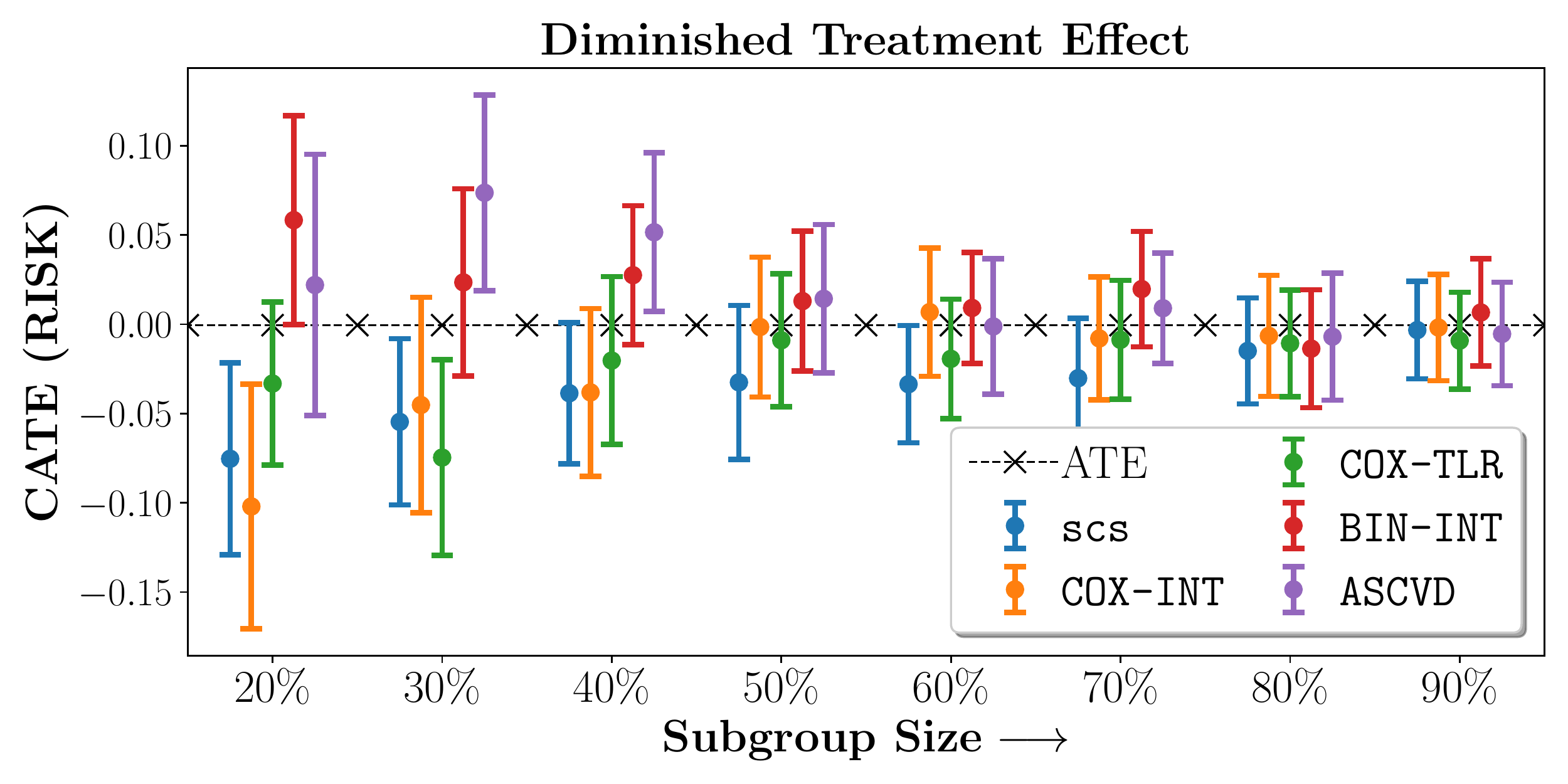}%
    \includegraphics[width=0.5\textwidth]{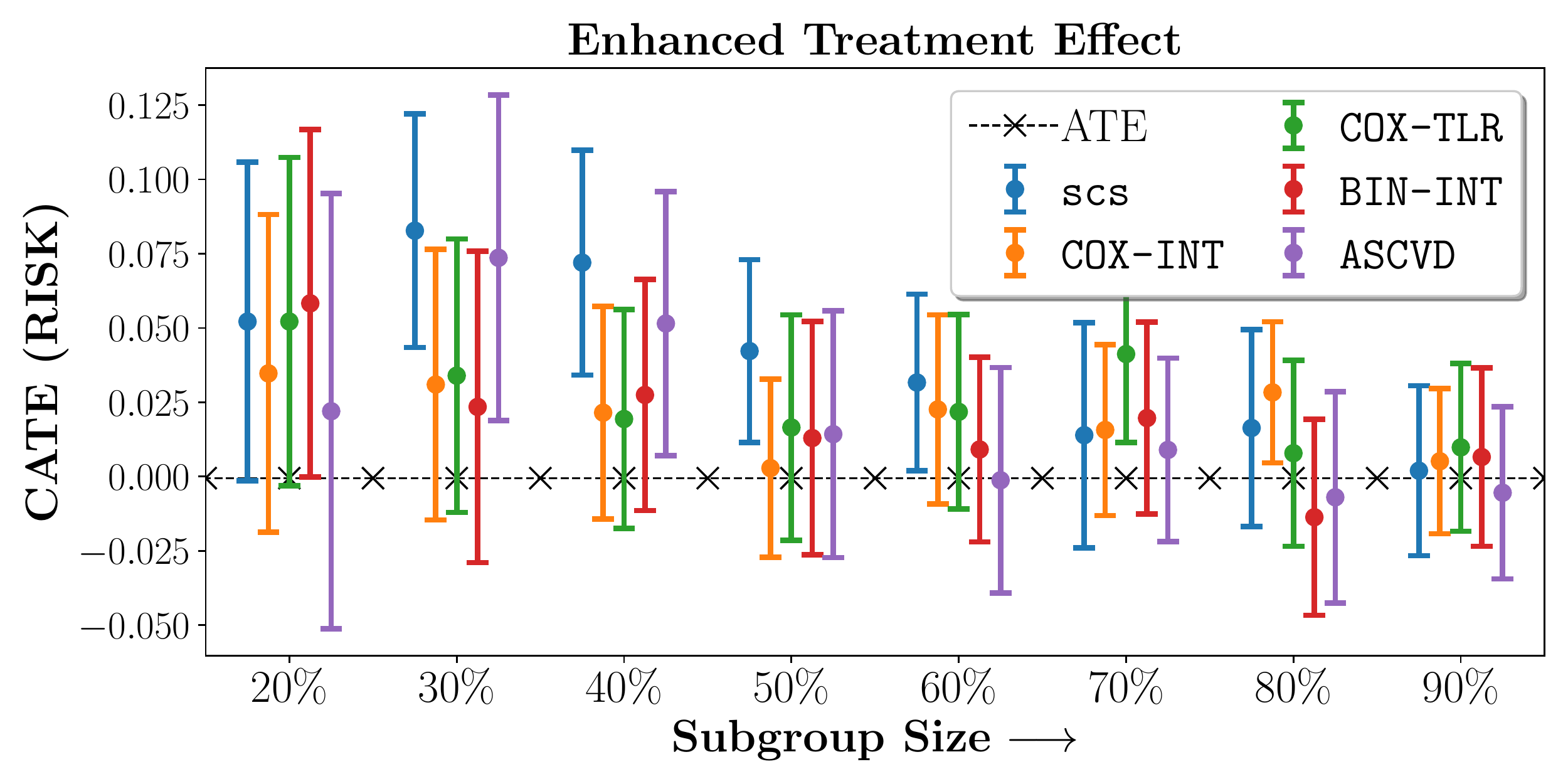}
    \begin{minipage}{1\textwidth}
        \centering
        \resizebox{0.8\columnwidth}{!}{%
            \begin{tabular}{lllll}
\toprule
{} &            20\% &            40\% &            60\% &            80\% \\
\midrule
SCS     &  -0.08$\pm$0.05 &  -0.04$\pm$0.04 &  -0.03$\pm$0.03 &  -0.01$\pm$0.03 \\
COX-INT &   -0.1$\pm$0.07 &  -0.04$\pm$0.05 &   0.01$\pm$0.04 &  -0.01$\pm$0.03 \\
COX-TLR &  -0.03$\pm$0.05 &  -0.02$\pm$0.05 &  -0.02$\pm$0.03 &  -0.01$\pm$0.03 \\
BIN-INT &   0.06$\pm$0.06 &   0.03$\pm$0.04 &   0.01$\pm$0.03 &  -0.01$\pm$0.03 \\
ASCVD   &   0.02$\pm$0.07 &   0.05$\pm$0.04 &   -0.0$\pm$0.04 &  -0.01$\pm$0.04 \\
\bottomrule
\end{tabular}
        \quad
        \begin{tabular}{lllll}
\toprule
{} &           20\% &           40\% &           60\% &            80\% \\
\midrule
SCS     &  0.05$\pm$0.05 &  0.07$\pm$0.04 &  0.03$\pm$0.03 &   0.02$\pm$0.03 \\
COX-INT &  0.03$\pm$0.05 &  0.02$\pm$0.04 &  0.02$\pm$0.03 &   0.03$\pm$0.02 \\
COX-TLR &  0.05$\pm$0.06 &  0.02$\pm$0.04 &  0.02$\pm$0.03 &   0.01$\pm$0.03 \\
BIN-INT &  0.06$\pm$0.06 &  0.03$\pm$0.04 &  0.01$\pm$0.03 &  -0.01$\pm$0.03 \\
ASCVD   &  0.02$\pm$0.07 &  0.05$\pm$0.04 &  -0.0$\pm$0.04 &  -0.01$\pm$0.04 \\
\bottomrule
\end{tabular}
        
        }
        \vspace{1em} 
        \end{minipage}\\
        
    \textbf{$||\bm{\theta}||_0 \leq 10$}
    \includegraphics[width=0.5\textwidth]{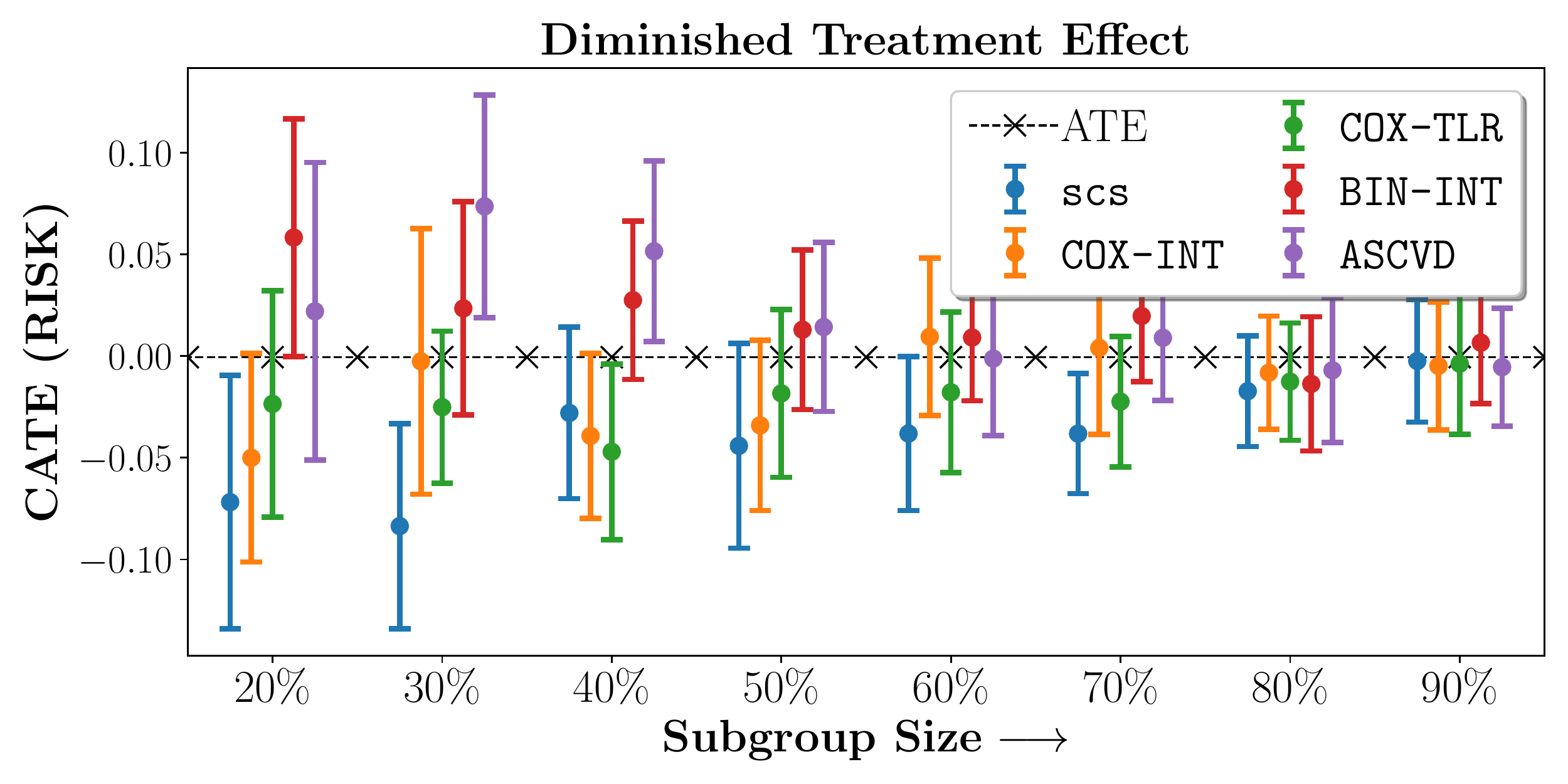}%
    \includegraphics[width=0.5\textwidth]{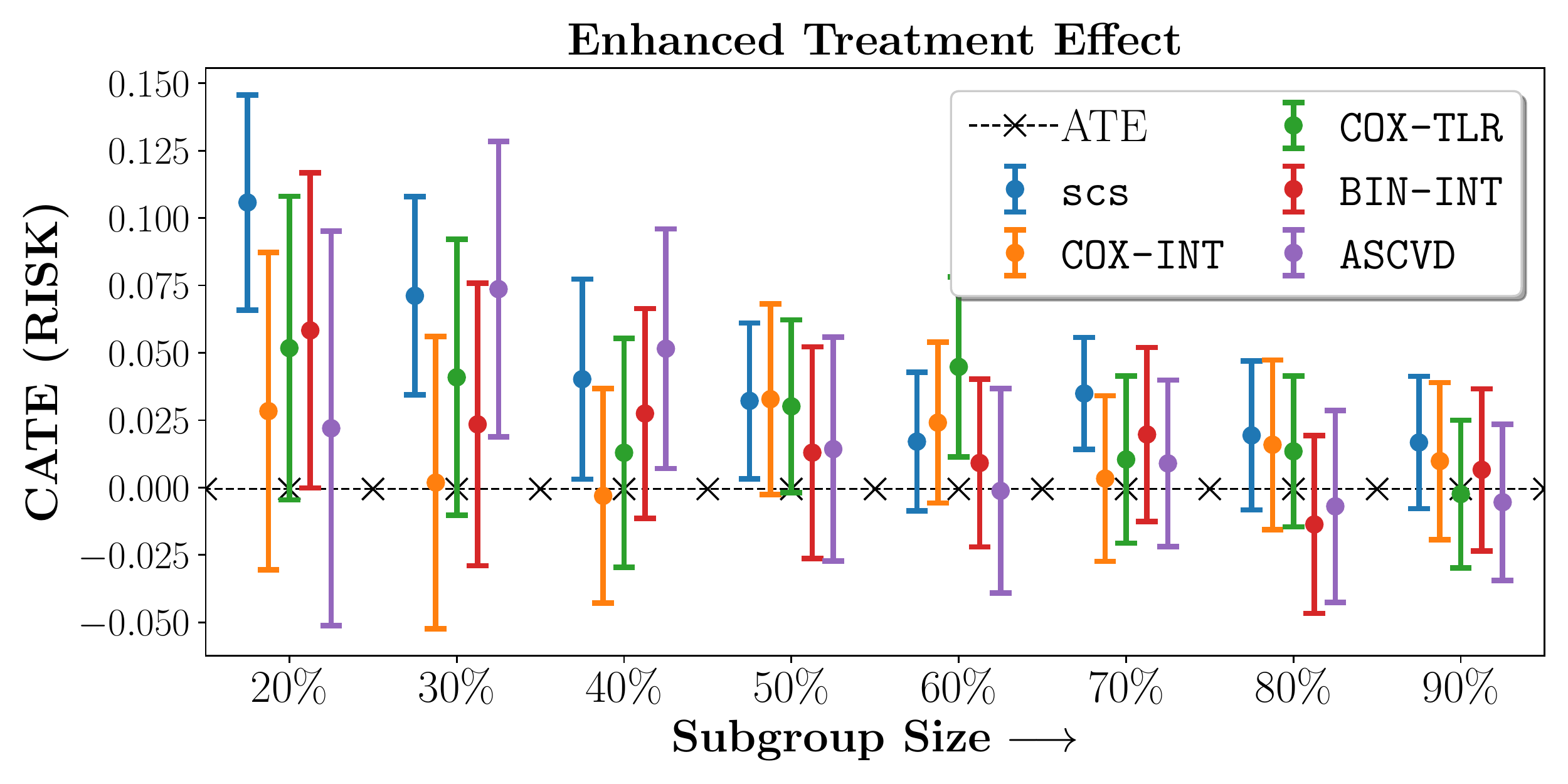}
    
    \begin{minipage}{1\textwidth}
        \centering
        \resizebox{0.8\columnwidth}{!}{%
            \begin{tabular}{lllll}
\toprule
{} &            20\% &            40\% &            60\% &            80\% \\
\midrule
SCS     &  -0.07$\pm$0.06 &  -0.03$\pm$0.04 &  -0.04$\pm$0.04 &  -0.02$\pm$0.03 \\
COX-INT &  -0.05$\pm$0.05 &  -0.04$\pm$0.04 &   0.01$\pm$0.04 &  -0.01$\pm$0.03 \\
COX-TLR &  -0.02$\pm$0.06 &  -0.05$\pm$0.04 &  -0.02$\pm$0.04 &  -0.01$\pm$0.03 \\
BIN-INT &   0.06$\pm$0.06 &   0.03$\pm$0.04 &   0.01$\pm$0.03 &  -0.01$\pm$0.03 \\
ASCVD   &   0.02$\pm$0.07 &   0.05$\pm$0.04 &   -0.0$\pm$0.04 &  -0.01$\pm$0.04 \\
\bottomrule
\end{tabular}
        \quad
       \begin{tabular}{lllll}
\toprule
{} &            20\% &            40\% &            60\% &            80\% \\
\midrule
SCS     &  -0.07$\pm$0.06 &  -0.03$\pm$0.04 &  -0.04$\pm$0.04 &  -0.02$\pm$0.03 \\
COX-INT &  -0.05$\pm$0.05 &  -0.04$\pm$0.04 &   0.01$\pm$0.04 &  -0.01$\pm$0.03 \\
COX-TLR &  -0.02$\pm$0.06 &  -0.05$\pm$0.04 &  -0.02$\pm$0.04 &  -0.01$\pm$0.03 \\
BIN-INT &   0.06$\pm$0.06 &   0.03$\pm$0.04 &   0.01$\pm$0.03 &  -0.01$\pm$0.03 \\
ASCVD   &   0.02$\pm$0.07 &   0.05$\pm$0.04 &   -0.0$\pm$0.04 &  -0.01$\pm$0.04 \\
\bottomrule
\end{tabular}
        
        }
        \vspace{1em} 
        \end{minipage}\\

    No Sparsity\\
   \includegraphics[width=0.5\textwidth]{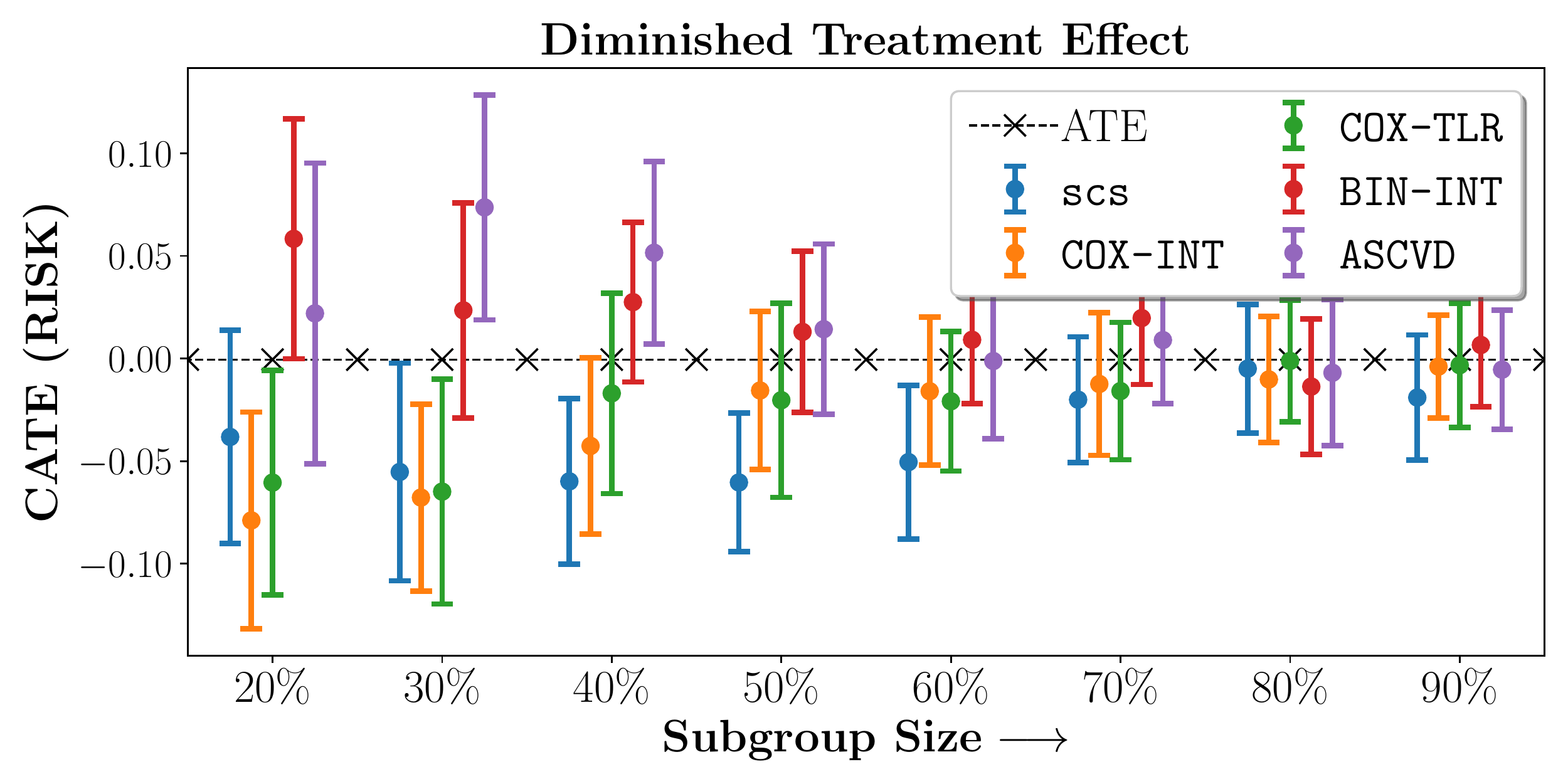}%
    \includegraphics[width=0.5\textwidth]{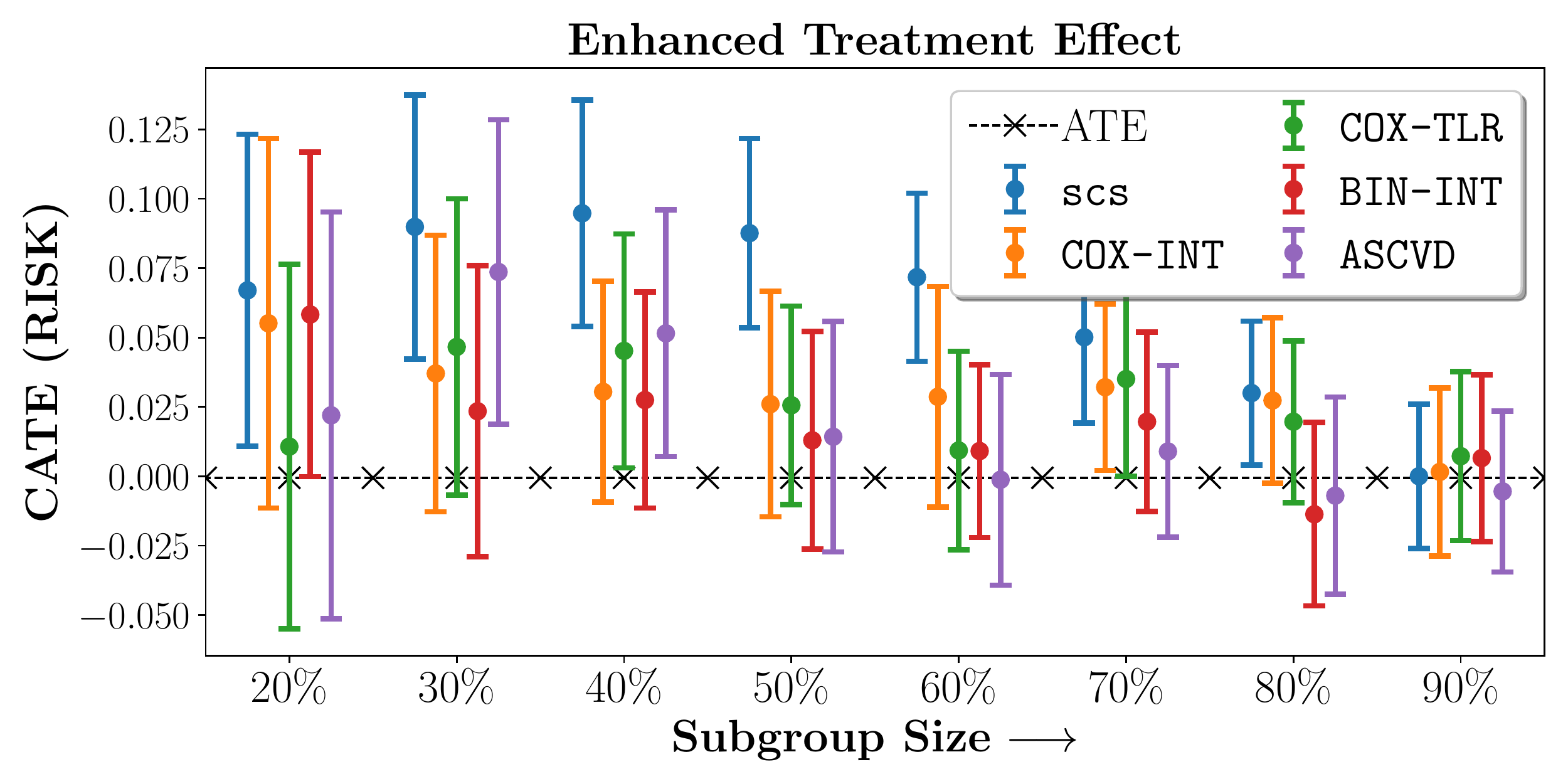}
    
    \begin{minipage}{1\textwidth}
        \centering
        \resizebox{0.8\columnwidth}{!}{%
        \begin{tabular}{lllll}
\toprule
{} &           20\% &           40\% &           60\% &            80\% \\
\midrule
SCS     &  0.07$\pm$0.06 &  0.09$\pm$0.04 &  0.07$\pm$0.03 &   0.03$\pm$0.03 \\
COX-INT &  0.06$\pm$0.07 &  0.03$\pm$0.04 &  0.03$\pm$0.04 &   0.03$\pm$0.03 \\
COX-TLR &  0.01$\pm$0.07 &  0.05$\pm$0.04 &  0.01$\pm$0.04 &   0.02$\pm$0.03 \\
BIN-INT &  0.06$\pm$0.06 &  0.03$\pm$0.04 &  0.01$\pm$0.03 &  -0.01$\pm$0.03 \\
ASCVD   &  0.02$\pm$0.07 &  0.05$\pm$0.04 &  -0.0$\pm$0.04 &  -0.01$\pm$0.04 \\
\bottomrule
\end{tabular}

        \quad
        \begin{tabular}{lllll}
\toprule
{} &           20\% &           40\% &           60\% &            80\% \\
\midrule
SCS     &  0.07$\pm$0.06 &  0.09$\pm$0.04 &  0.07$\pm$0.03 &   0.03$\pm$0.03 \\
COX-INT &  0.06$\pm$0.07 &  0.03$\pm$0.04 &  0.03$\pm$0.04 &   0.03$\pm$0.03 \\
COX-TLR &  0.01$\pm$0.07 &  0.05$\pm$0.04 &  0.01$\pm$0.04 &   0.02$\pm$0.03 \\
BIN-INT &  0.06$\pm$0.06 &  0.03$\pm$0.04 &  0.01$\pm$0.03 &  -0.01$\pm$0.03 \\
ASCVD   &  0.02$\pm$0.07 &  0.05$\pm$0.04 &  -0.0$\pm$0.04 &  -0.01$\pm$0.04 \\
\bottomrule
\end{tabular}

        }
        \vspace{1em} 
        \end{minipage}\\
    
    \caption{Conditional Average Treatment Effect in Risk versus subgroup size for the latent phenogroups extracted from the \textbf{BARI 2D} study.}
    \label{fig:bari2drisk}

\end{figure}

\begin{figure}
    \centering
    
    \textbf{$||\bm{\theta}||_0 \leq 5$}
    \includegraphics[width=0.5\textwidth]{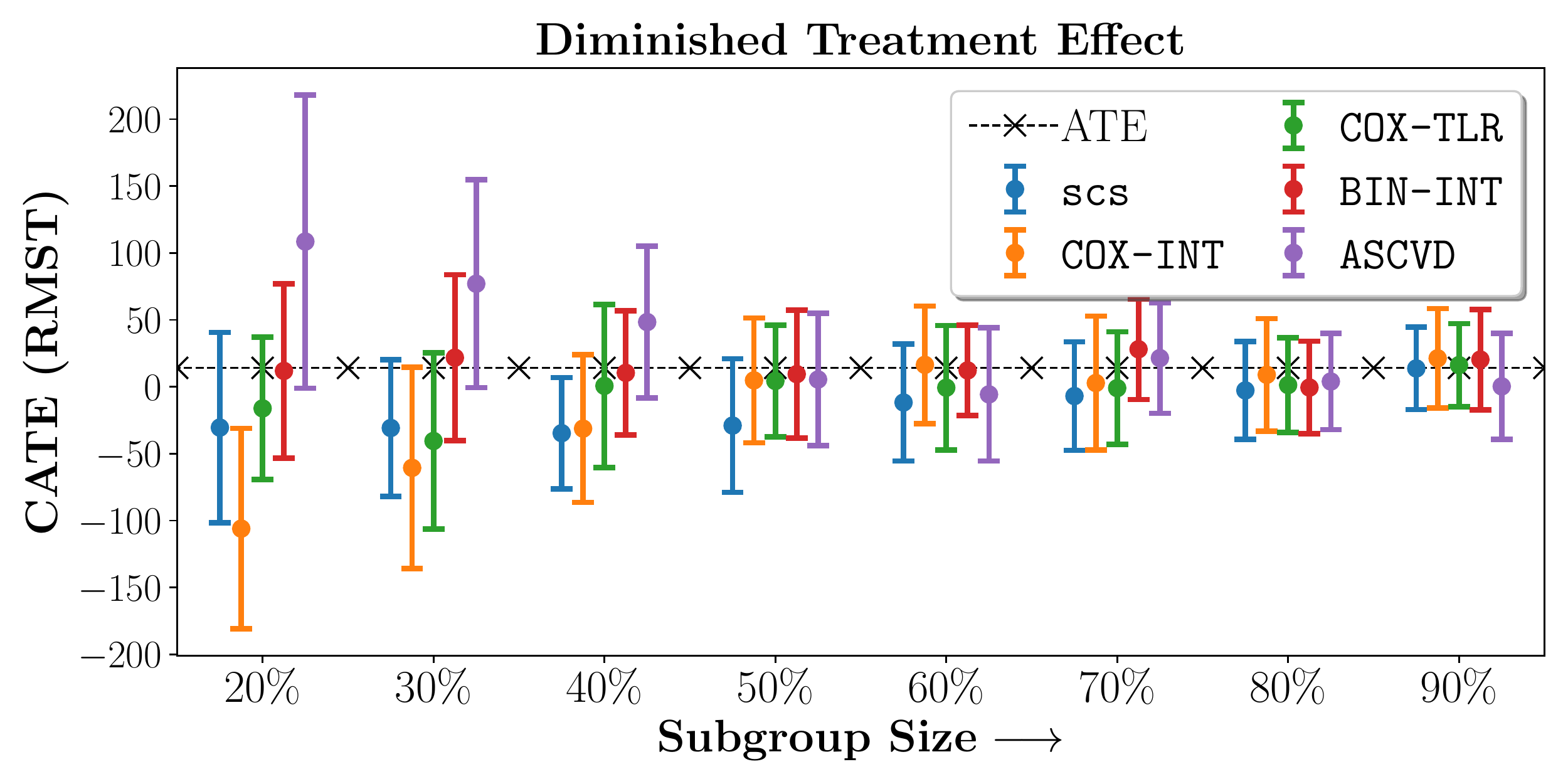}%
    \includegraphics[width=0.5\textwidth]{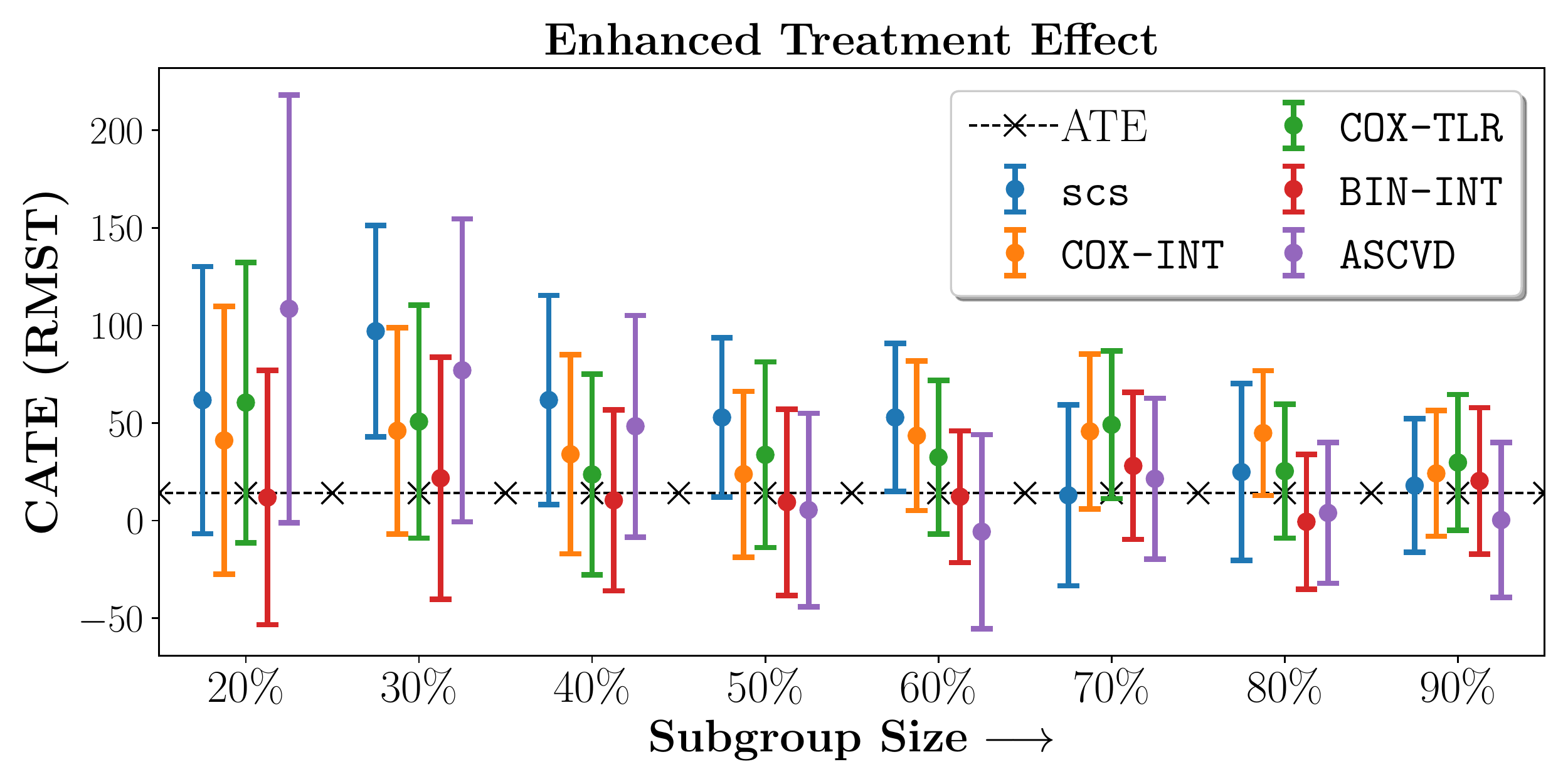}
    \begin{minipage}{1\textwidth}
        \centering
        \resizebox{0.8\columnwidth}{!}{%
            \begin{tabular}{lllll}
\toprule
{} &               20\% &              40\% &              60\% &             80\% \\
\midrule
SCS     &   -30.51$\pm$71.06 &  -34.71$\pm$41.55 &  -11.77$\pm$43.75 &  -2.73$\pm$36.52 \\
COX-INT &   -106.06$\pm$74.9 &   -31.3$\pm$55.18 &   16.38$\pm$43.89 &   8.93$\pm$42.04 \\
COX-TLR &   -16.18$\pm$53.18 &    0.65$\pm$60.94 &    -0.7$\pm$46.54 &   1.29$\pm$35.33 \\
BIN-INT &    11.84$\pm$65.18 &   10.39$\pm$46.33 &    12.23$\pm$33.8 &    -0.6$\pm$34.6 \\
ASCVD   &  108.54$\pm$109.62 &   48.36$\pm$56.77 &    -5.7$\pm$49.77 &   3.93$\pm$36.04 \\
\bottomrule
\end{tabular}
        \quad
        \begin{tabular}{lllll}
\toprule
{} &               20\% &             40\% &             60\% &             80\% \\
\midrule
SCS     &    61.73$\pm$68.48 &  61.76$\pm$53.61 &  52.87$\pm$37.98 &  24.86$\pm$45.35 \\
COX-INT &    41.07$\pm$68.66 &   33.99$\pm$51.0 &  43.48$\pm$38.35 &  44.84$\pm$31.96 \\
COX-TLR &    60.48$\pm$71.86 &  23.63$\pm$51.48 &  32.45$\pm$39.39 &  25.35$\pm$34.38 \\
BIN-INT &    11.84$\pm$65.18 &  10.39$\pm$46.33 &   12.23$\pm$33.8 &    -0.6$\pm$34.6 \\
ASCVD   &  108.54$\pm$109.62 &  48.36$\pm$56.77 &   -5.7$\pm$49.77 &   3.93$\pm$36.04 \\
\bottomrule
\end{tabular}
        
        }
        \vspace{1em} 
        \end{minipage}\\
        
    \textbf{$||\bm{\theta}||_0 \leq 10$}
    \includegraphics[width=0.5\textwidth]{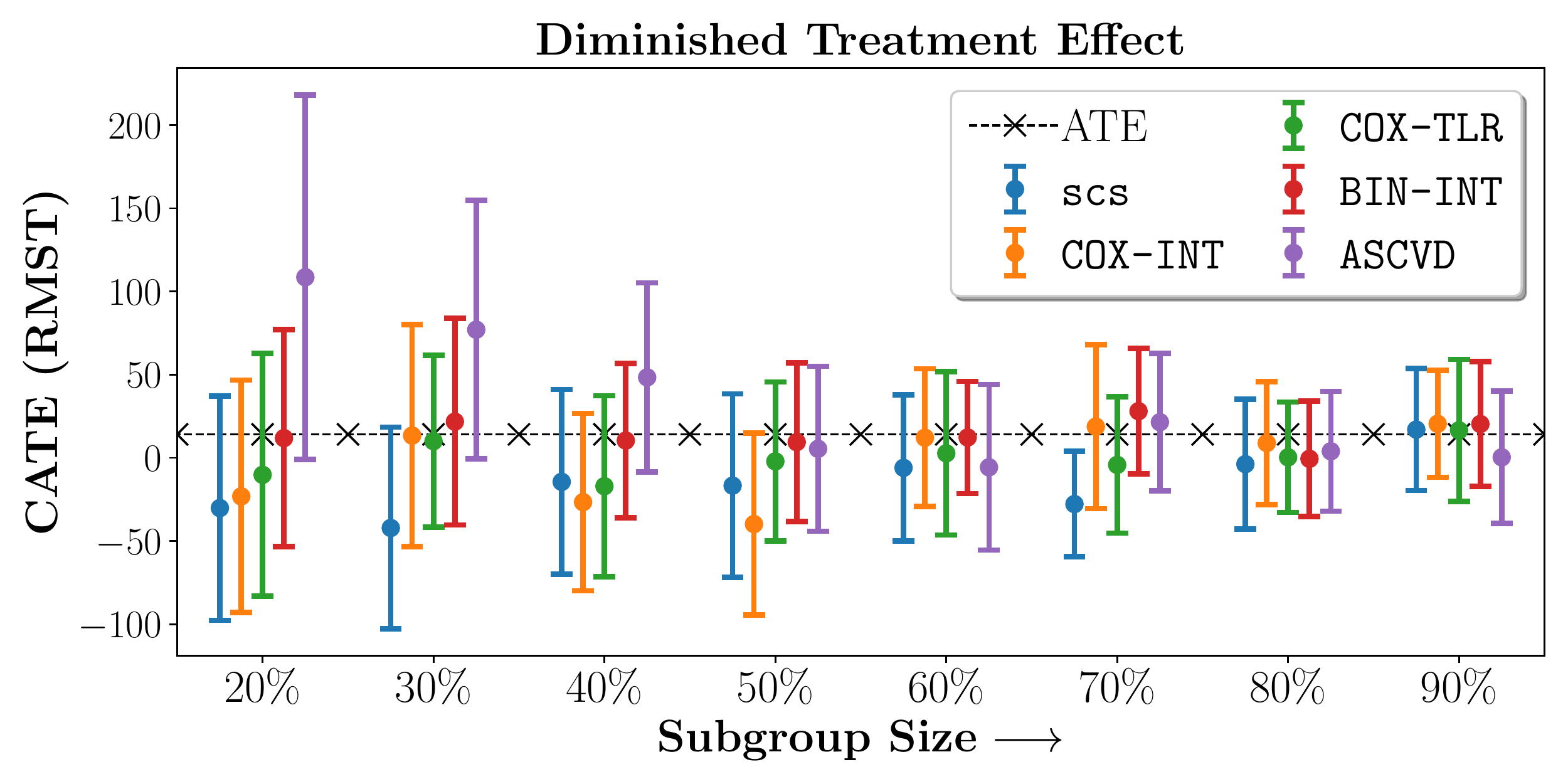}%
    \includegraphics[width=0.5\textwidth]{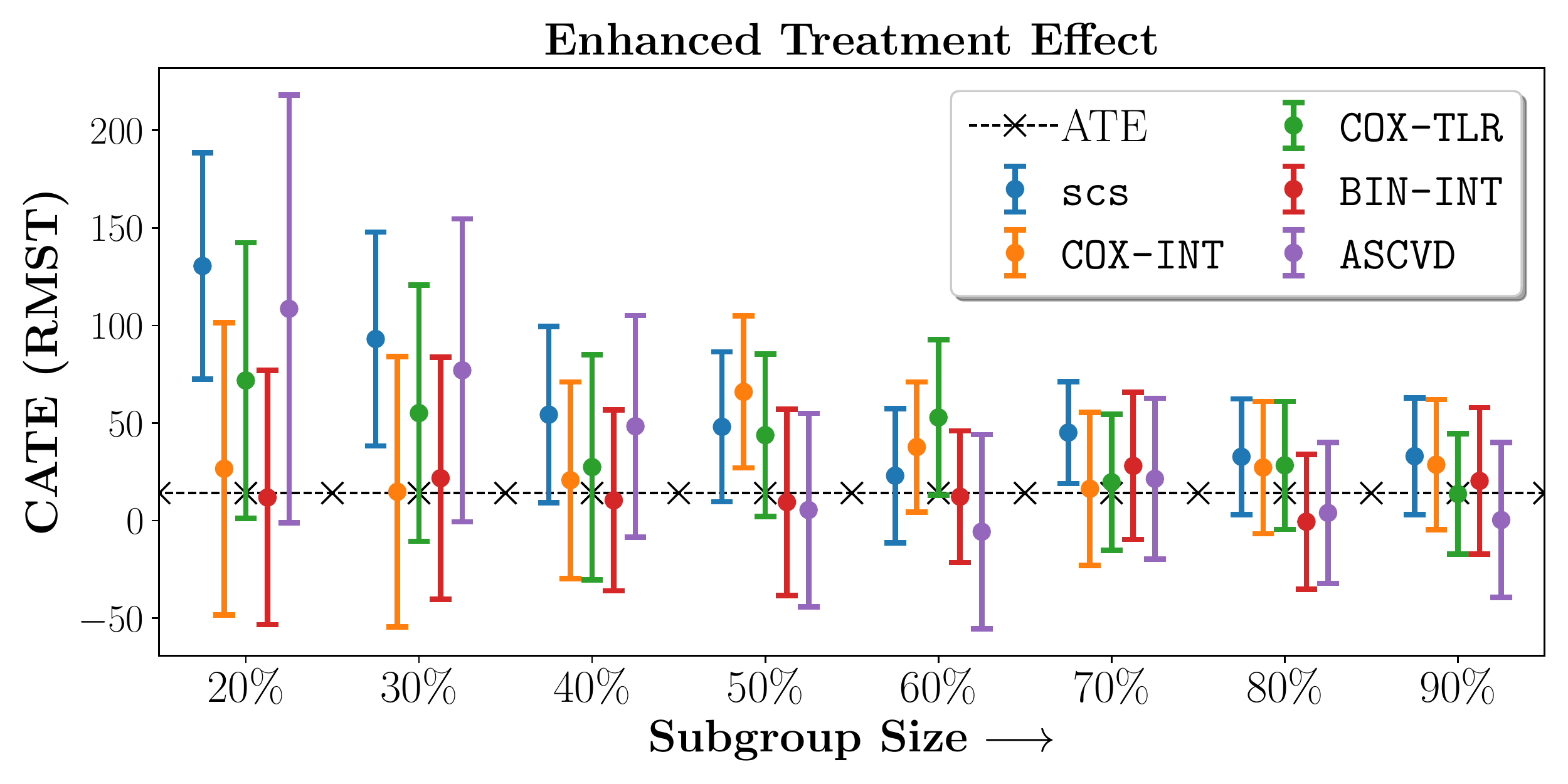}
    
    \begin{minipage}{1\textwidth}
        \centering
        \resizebox{0.8\columnwidth}{!}{%
            \begin{tabular}{lllll}
\toprule
{} &               20\% &              40\% &             60\% &            80\% \\
\midrule
SCS     &    -30.2$\pm$67.44 &  -14.45$\pm$55.58 &  -6.03$\pm$43.99 &  -3.8$\pm$39.02 \\
COX-INT &   -23.18$\pm$69.92 &  -26.64$\pm$53.34 &  12.09$\pm$41.39 &  8.87$\pm$36.87 \\
COX-TLR &   -10.22$\pm$73.05 &  -17.07$\pm$54.39 &    2.7$\pm$49.05 &  0.31$\pm$33.17 \\
BIN-INT &    11.84$\pm$65.18 &   10.39$\pm$46.33 &   12.23$\pm$33.8 &   -0.6$\pm$34.6 \\
ASCVD   &  108.54$\pm$109.62 &   48.36$\pm$56.77 &   -5.7$\pm$49.77 &  3.93$\pm$36.04 \\
\bottomrule
\end{tabular}
        \quad
        \begin{tabular}{lllll}
\toprule
{} &               20\% &             40\% &             60\% &             80\% \\
\midrule
SCS     &   130.48$\pm$58.03 &   54.3$\pm$45.22 &   23.0$\pm$34.47 &  32.71$\pm$29.64 \\
COX-INT &    26.58$\pm$74.93 &   20.7$\pm$50.38 &  37.69$\pm$33.38 &  27.18$\pm$33.95 \\
COX-TLR &    71.81$\pm$70.62 &  27.41$\pm$57.71 &  52.85$\pm$39.85 &  28.33$\pm$32.79 \\
BIN-INT &    11.84$\pm$65.18 &  10.39$\pm$46.33 &   12.23$\pm$33.8 &    -0.6$\pm$34.6 \\
ASCVD   &  108.54$\pm$109.62 &  48.36$\pm$56.77 &   -5.7$\pm$49.77 &   3.93$\pm$36.04 \\
\bottomrule
\end{tabular}
        
        }
        \vspace{1em} 
        \end{minipage}\\

    No Sparsity\\
   \includegraphics[width=0.5\textwidth]{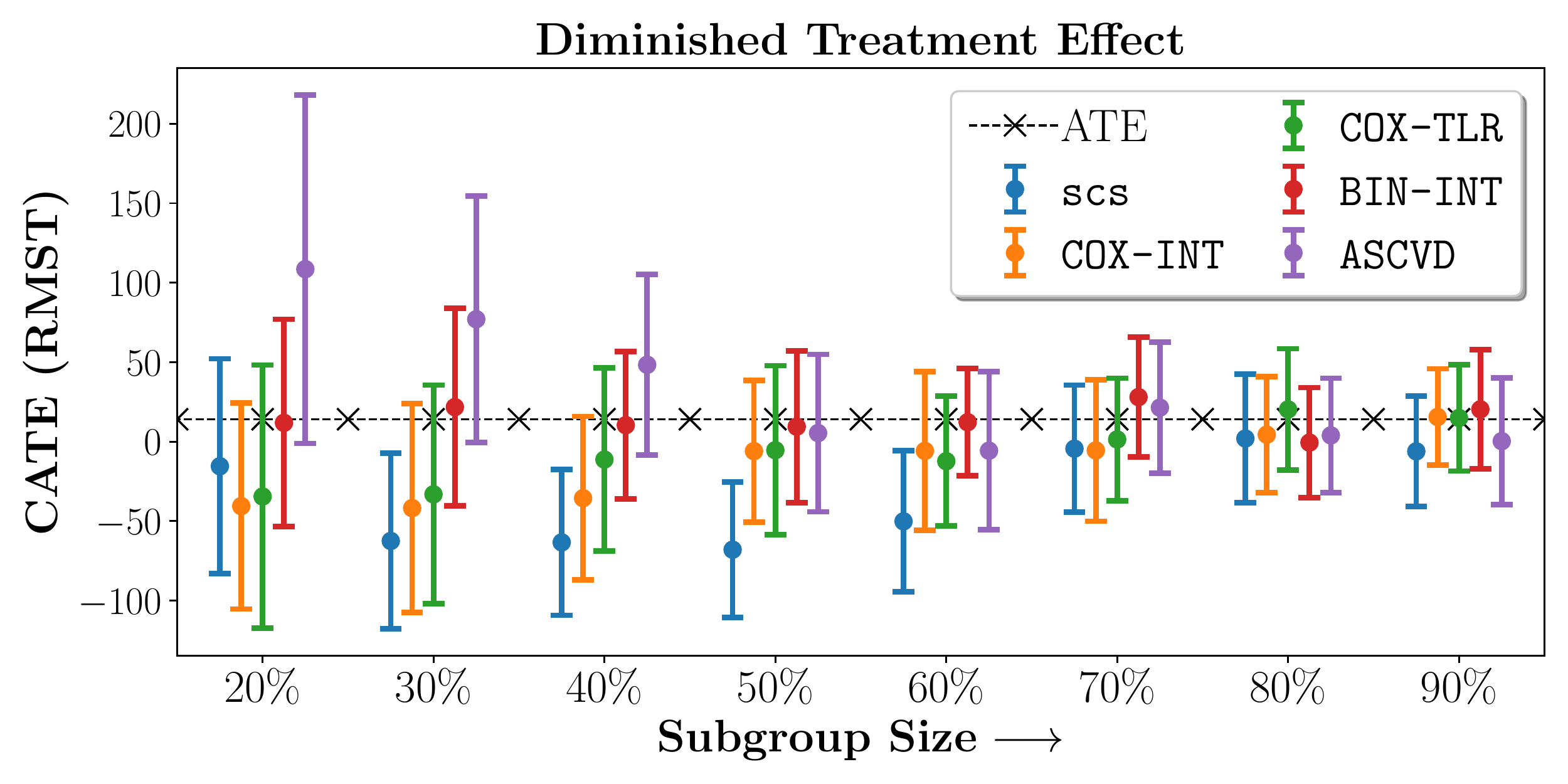}%
    \includegraphics[width=0.5\textwidth]{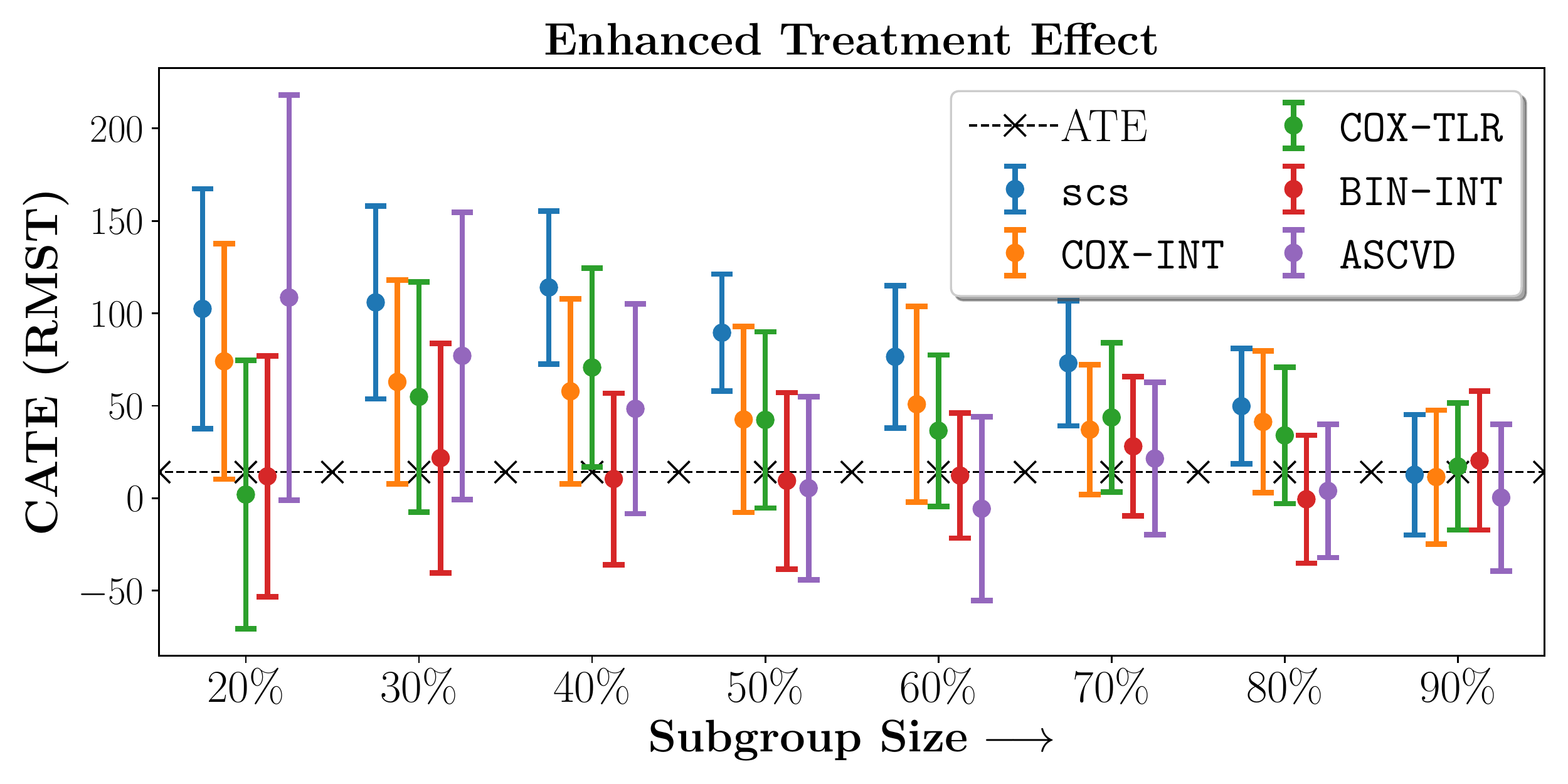}
    
    \begin{minipage}{1\textwidth}
        \centering
        \resizebox{0.8\columnwidth}{!}{%
            \begin{tabular}{lllll}
\toprule
{} &               20\% &              40\% &              60\% &            80\% \\
\midrule
SCS     &   -15.46$\pm$67.61 &  -63.38$\pm$45.83 &  -50.14$\pm$44.42 &  1.99$\pm$40.43 \\
COX-INT &    -40.54$\pm$64.8 &   -35.58$\pm$51.4 &   -5.82$\pm$49.97 &  4.42$\pm$36.41 \\
COX-TLR &   -34.51$\pm$82.71 &  -11.28$\pm$57.62 &   -12.3$\pm$40.85 &  20.28$\pm$38.2 \\
BIN-INT &    11.84$\pm$65.18 &   10.39$\pm$46.33 &    12.23$\pm$33.8 &   -0.6$\pm$34.6 \\
ASCVD   &  108.54$\pm$109.62 &   48.36$\pm$56.77 &    -5.7$\pm$49.77 &  3.93$\pm$36.04 \\
\bottomrule
\end{tabular}
        \quad
        \begin{tabular}{lllll}
\toprule
{} &               20\% &              40\% &             60\% &             80\% \\
\midrule
SCS     &    102.43$\pm$64.9 &  113.99$\pm$41.35 &  76.49$\pm$38.53 &  49.76$\pm$31.25 \\
COX-INT &    74.04$\pm$63.72 &   57.82$\pm$50.04 &   50.8$\pm$52.89 &  41.27$\pm$38.38 \\
COX-TLR &     1.96$\pm$72.63 &    70.7$\pm$53.71 &  36.51$\pm$40.91 &   33.98$\pm$36.9 \\
BIN-INT &    11.84$\pm$65.18 &   10.39$\pm$46.33 &   12.23$\pm$33.8 &    -0.6$\pm$34.6 \\
ASCVD   &  108.54$\pm$109.62 &   48.36$\pm$56.77 &   -5.7$\pm$49.77 &   3.93$\pm$36.04 \\
\bottomrule
\end{tabular}
        
        }
        \vspace{1em} 
        \end{minipage}\\
    
    \caption{\small  Conditional Average Treatment Effect in Restricted Mean Survival Time versus subgroup size for the latent phenogroups extracted from the \textbf{BARI 2D} study.}
    \label{fig:bari2drmst}

\end{figure}

{ 
Figures \ref{fig:allhathr}, \ref{fig:allhatrisk}, \ref{fig:allhatrmst} present tabulated metrics on \textbf{ALLHAT} with Hazard Ratio, Risk Difference and Restricted Mean Survival Time respectively.
\noindent 
Figures \ref{fig:bari2dhr}, \ref{fig:bari2drisk}, \ref{fig:bari2drmst} present tabulated metrics \textbf{BARI2D} with Hazard Ratio, Risk Difference and Restricted Mean Survival Time metrics respectively. 
}

\end{document}

%% file: math_commands.tex
\usepackage{amsmath,amsfonts,bm}

\def\eqref#1{equation~\ref{#1}}

\def\1{\bm{1}}

\def\eps{{\epsilon}}

\def\va{{\bm{a}}}

\def\vh{{\bm{h}}}

\def\vj{{\bm{j}}}
\def\vk{{\bm{k}}}

\def\vt{{\bm{t}}}
\def\vu{{\bm{u}}}

\def\vw{{\bm{w}}}
\def\vx{{\bm{x}}}
\def\vy{{\bm{y}}}

\def\mA{{\bm{A}}}
\def\mB{{\bm{B}}}

\def\mS{{\bm{S}}}

\DeclareMathAlphabet{\mathsfit}{\encodingdefault}{\sfdefault}{m}{sl}
\SetMathAlphabet{\mathsfit}{bold}{\encodingdefault}{\sfdefault}{bx}{n}

